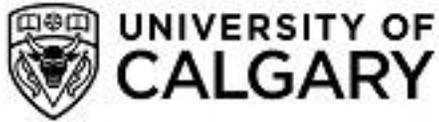

# VLSI-Inspired Methods for Student Learning Community Creation and Refinement

Cao, Sheng Lun

UNIVERSITY OF CALGARY

VLSI-Inspired Methods for Student Learning Community Creation and Refinement

by

Sheng Lun (Christine) Cao

A THESIS

SUBMITTED TO THE FACULTY OF GRADUATE STUDIES

IN PARTIAL FULFILMENT OF THE REQUIREMENTS FOR THE

DEGREE OF MASTER OF SCIENCE

GRADUATE PROGRAM IN ELECTRICAL ENGINEERING

CALGARY, ALBERTA

AUGUST, 2021

# Abstract

The unprecedented global pandemic COVID-19 significantly disrupted how educational contents are delivered in academic institutions, rapidly accelerating the adoption of online and blended learning. This thesis explores the creation and refinement of optimized student learning communities as a mean to support online and blended learning in the pandemic and post-pandemic setting.

Students enrolled in courses at a university can be modelled as an enrollment network similar to a circuit netlist. Learning communities are created by clustering students into groups, optimizing for maximum internal connection to support student learning, and minimum external connection to reduce disease transmission. Three VLSI-based clustering algorithms: Hyperedge Coarsening, Modified Hyperedge Coarsening, and Best Choice, are modified to cluster student enrollment networks. Further experimentations are conducted for Best Choice to fine-tune its clustering parameters.

The learning communities created by the clustering algorithms are further refined by the Simulated Annealing algorithm using the same optimization criteria. Experiments are performed to fine-tune the algorithmic parameters of Simulated Annealing. The Learning Community Creation and Refinement Framework combines all three stages of network modeling, learning community creation, and learning community refinement.

The proposed framework is tested on both the 3$^{rd}$ year Electrical Engineering Fall 2020 enrollment dataset and a very large Fall 2020 and Winter 2021 enrollment dataset. Best Choice performed the best among the clustering algorithms, capable of creating learning communities for the optimization criteria for a given maximum cluster size. Simulated Annealing is able to refine the clustering results by significantly increase cluster quality. The framework is capable of creating and refining learning communities for both the small and the large enrollment networks, but it is better suited for creating tailored learning communities at a program level. Future work, including creating student learning communities based on other optimization criteria, should be explored.

# Acknowledgements

I would like to express my deepest appreciation and gratitude towards my co-supervisors Dr. Laleh Behjat and Dr. Erin Gibbs Van Brunschot. Without their scholarship, mentorship, guidance, and support over the last three years, this work would not be possible. Your patience in guiding me through the haze of ideation, and your feedback through different stages of my research and writing, were invaluable in me reaching where I am today. I am enormously grateful especially to Dr. Behjat, for seeing the potential in a confused and lost undergrad, and humouring my request to "chat and find out more about grad school". It has truly been an honour working with you both, and I consider these years an experience for a lifetime.

To my examination committee, Dr. Vassil Dimitrov and Dr. Mohammad Moshirpour, thank you for your insightful questions and feedback on my research during the memorable defence. To my Cool Lab labmates, thank you for hanging out with my fish and I over these years, and giving me great feedback on my presentations. I especially want to thank Aysa, Erfan, and Renan for explaining the different algorithms to me, and the Engineering Education group (Robyn, Kat, Jason, and Kirill) for bouncing ideas and proofreading my work. I also want to thank the Conjoint Faculties Research Ethics Board (CFREB) for approving this research, and Jeff Lee at the Office of the Registrar for releasing my research data.

Last but not least, I want to thank my parents, for always unconditionally loving me, supporting my career, and always reminding me that I am good enough. I love you mom and dad!

# Dedication

我把此论文献给我亲爱的姥姥。有您的爱我什么都能做到。

(*I dedicate this thesis to my beloved maternal grandmother. I can accomplish anything with your love.*)

# Table of Contents

# List of Tables

# List of Figures

# List of Symbols, Abbreviations, and Nomenclature

**SYMBOLS AND ACRONYMS**

| | | |
|---|---|---|
| **A** | | *Adjacency Matrix* |
| | **A$_{ij}$** | *Matrix Element of Adjacency Matrix* |
| **ASEE** | | *American Society for Engineering Education* |
| **BC** | | *Best Choice* |
| | **$a(\cdot)$** | *Area Function* |
| | **$d(\cdot)$** | Clustering Score Function |
| | **e** | Hyperedge |
| | **PQ** | Priority Queue |
| | **u** | Cell |
| | **v** | *Closest Cell to u* |
| | **$w_e$** | Weight of Hyperedge |
| **BNDE** | | *Nursing, Direct Entry for High School Applicants* |
| **BNDH** | | *Nursing, Degree Holder* |
| **BNTR** | | *Nursing, Transfer Student* |
| **BUSI** | | *Business* |
| **C** | | *Connectivity Matrix* |
| | **C$_{ij}$** | *Matrix Element of Connectivity Matrix* |
| **CFREB** | | *Conjoint Faculties Research Ethics Board* |
| **CoI** | | *Community of Inquiry Framework* |
| **COVID-19** | | *SARS-CoV-2* |
| **EC** | | *Edge Coarsening* |

| **ECON** | *Economics* |
|---|---|
| **EDA** | *Electronic Design Automation* |
| **ENEL** | *Electrical Engineering* |
|     **ENCM 369** | *Computer Organization* |
|     **ENEL 300** | *Electrical and Computer Engineering Professional Skills* |
|     **ENEL 327** | *Signals and Transforms* |
|     **ENEL 343** | *Circuits II* |
|     **ENEL 361** | *Electronic Devices and Materials* |
| **ENER** | *Energy Engineering* |
| **ENSF** | *Software Engineering* |
| **F2F** | *Face-to-Face* |
| **FM** | *Fiduccia-Mattheyses* |
| **G** | *Graph* |
|     **E** | *Edges* |
|     **V** | *Vertices* |
| **GENL** | *Business* |
| **HC** | *Hyperedge Coarsening* |
| **IC** | *Integrated Circuits* |
| **ICT** | *Information and Communication Technology* |
| **ILS** | *Integrated Learning Stream* |
| **LC / LCs** | *Learning Community / Communities* |
| **MHC** | *Modified Hyperedge Coarsening* |
| **MENG** | *Mechanical Engineering* |

| | | |
|---|---|---|
| **NTSC** | | *Natural Sciences* |
| **PLC / PLCs** | | *Professional Learning Community / Communities* |
| **POLI** | | *Political Sciences* |
| **SA** | | *Simulated Annealing* |
| | $(a_i, b_i)$ | *Selected Pair* |
| | **AP** | *Acceptance Probability* |
| | $\alpha$ | *Cooling Rate* |
| | **Cost** | *Cost Value* |
| | **COST**($\cdot$) | *Cost Function* |
| | $\Delta$**Cost** | *Change in Cost* |
| | $\overline{\Delta\text{Cost}}$ | *Average Change in Cost* |
| | **Cost**$_{\text{current}}$ | *Cost of the Current Solution* |
| | **Cost**$_{\text{initial}}$ | *Cost of Initial Solution* |
| | **Cost**$_{\text{trial}}$ | *Cost of the Trial Solution* |
| | $\overline{\text{Cost}_{\text{trial}}}$ | *Average Cost After the First Perturbation* |
| | $d_l$ | *Linear Clustering Score Function* |
| | $d_n$ | *Nonlinear Clustering Score Function* |
| | **e** | *Euler's Number* |
| | $e^{\frac{-\Delta\text{Cost}}{T}}$ | *Boltzmann Acceptance Criterion* |
| | **i** | *Current Iteration at Each Temperature* |
| | $i_T$ | *Maximum Number of Iterations at Each Temperature* |
| | **N** | *Number of Initial Perturbations for Average Initial Cost* |
| | **PERTURB**($\cdot$) | *Perturbation Function* |
| **RANDOM(·)** | | *Random Function* |
| **r** | | *Random Value* |
| **SELECT_PAIR(·)** | | *Select Swapping Pair Function* |
| $S_{E_j}$ | | *Total Number of External Cluster Connections* |
| $S_{I_j}$ | | *Total Number of Internal Cluster Connections* |
| $S_T$ | | *Total Cluster Quality Score* |
| **T** | | *Temperature* |
| $T_0$ | | *Initial Temperature* |
| $T_{current}$ | | *Current Temperature* |
| $T_{min}$ | | *Minimum Temperature* |
| $T_{next}$ | | *Next Temperature* |
| **VLSI** | | *Very Large-Scale Integration* |

## **TERMS**

*Cell*

*Chip Planning*

*Clock Tree Synthesis*

*Clustering*

*Monte Carlo Simulation*

*Multilevel Partitioning*

*Net*

*Netlist*

*Partitioning*

# Chapter 1 : Introduction

In March of 2020, the novel virus SARS-CoV-2 (COVID-19) was declared a global pandemic by the World Health Organization [1]. Due to the strict social distancing measures enacted to preserve public health and safety, academic institutions were forced to shift away from traditional course delivery inside classrooms and lecture halls. Millions of students from across the globe, from the United States [2], to Saudi Arabia [3], to Zambia [4], were impacted by the need to receive their education through online learning.

More than one year later, as the vaccination rollout for COVID-19 steadily improves, academic institutions can begin to explore what a physical return to campus could look like for courses that spent the last year online. Research on the impact of COVID-19 on higher education is still emerging, but the growing trend towards online learning and blended learning may have been accelerated by the pandemic. This thesis discusses the creation of optimized learning communities (LCs) as a mean to support online and blended learning in a pandemic and post-pandemic setting.

The remainder of this chapter is organized as follows: Section 1.1 discusses the motivation for the research presented in this thesis. In Section 1.2, the major research contributions of this thesis are outlined. Section 1.3 presents an outline for the remainder of the thesis. Finally, Section 1.4 includes a statement of ethics approval regarding this research.

## 1.1 Motivation

The shift to online learning during the beginning of COVID-19 was largely unplanned. In the year since, there have been some opportunities for students to return in campus, based both on school re-opening data [5] and anecdotal evidence at the University of Calgary, where a small percentage of undergraduate courses continued in-person. In online learning, students experiences social isolation leading to loneliness and demotivation in their studies [6]. If students return to campus in the same fashion as pre-pandemic, there are concerns around health and safety, and whether social distancing rules [7] can be complied with.

Learning communities (LCs) are a viable method to address these concerns, as they can provide a sense of community for students in online learning, while grouping students together to minimize disease transmission for in-person components. As academic institutions slowly re-open, LCs can be utilised to provide support for a range of course content delivery modes. However, before LCs can be utilised effectively, they must be created effectively.

The goal of this thesis is to provide a method to optimally create and refine LCs, based on students' course enrollment in a given institution. These LCs should be created to maximize the amount of time students spend within the LC, and minimize cross-LC contact, in order to adhere to public health and safety guidelines.

Methodological inspiration for creating optimized LC configurations comes from algorithms used to design Very Large-Scale Integrated (VLSI) circuits. Students enrolled in an institution, connected by their common courses, forms a complex network, much like advanced circuits. Therefore, algorithms designed to cluster and refine complex circuits should theoretically also operate well on complex student networks with modification.

## 1.2 Research Contributions

The research in this thesis aims to provide a methodology to create and refine optimal LCs for student, based on VLSI algorithms. The main contributions of this thesis are as follows:

- Propose a method of configuring students enrolled in courses into a circuit-like network
- Modify VLSI clustering algorithms for application in student LC creation, based on a set clustering criterion
- Modify a global optimization algorithm to refine student LC clustering results, based on a set refinement criterion
- Experimentally tune parameters for both the clustering and optimization algorithms
- Propose an operational framework for optimized LC creation and refinement, from student enrollment data to refined LCs, and implement the framework on a large student enrollment dataset

The methodology of these contributions is outlined in Chapter 4, and results and implications from experimentations are outlined in Chapter 5.

## 1.3 Thesis Organization

The remainder of this thesis is organized as follows:

Chapter 2: In this chapter, the background on course content delivery formats is introduced. The goal of this chapter is to contextualize different formats of content delivery, including online learning, blended learning, and learning communities. A case study from Electrical & Software

Engineering at the University of Calgary is presented, demonstrating how these types of content delivery can be operationalized.

Chapter 3: In this chapter, the background information on the VLSI algorithms used in this thesis is presented. The goal of this chapter is to provide an explanation on all of the algorithms used in this thesis. The relevant algorithms discussed includes Edge Coarsening, Hyperedge Coarsening, Modified Hyperedge Coarsening, Best Choice, and Simulated Annealing.

Chapter 4: In this chapter, the methodology of the research contributions is outlined. First, student networks creation is explored. Proposed modifications to three clustering algorithms (Hyperedge Coarsening, Modified Hyperedge Coarsening, and Best Choice) are presented. Experiments are designed to determine ideal clustering parameters for Best Choice. Modifications to Simulated Annealing as a clustering refinement method is presented, and experiments to fine-tune Simulated Annealing parameters are designed. Finally, the operational framework for optimized LC creation is proposed.

Chapter 5: In this chapter, the clustering and experiment results are presented. These results include clustering outcomes using Hyperedge Coarsening, Modified Hyperedge Coarsening, and Best Choice, as well as experiment results for fine-tuning clustering parameters in Best Choice. Refined clusters and experiment results for fine-tuning Simulated Annealing parameters are presented. Finally, the proposed operational framework is applied to an extended student enrollment dataset from Fall 2020 and Winter 2021 terms.

Chapter 6: In this chapter, the conclusion of this thesis is presented, including a summary of research contributions, as well as a discussion of future work of this research.

## 1.4 Ethics Approval

The research discussed in this thesis has been approved by the Conjoint Faculties Research Ethics Board (CFREB). The data used in this thesis have been anonymized and released by the Office of the Registrar at the University of Calgary.

# Chapter 2 : Background in Course Content Delivery Formats

## 2.1 Introduction

The sudden onset of the global COVID-19 pandemic brought forth rapid changes to how the entire world operate, communicate, and collaborate. Universities across the world, from the United States [2] to Zambia [4] have been forced to move away from face-to-face (F2F) courses to online and remote learning formats. The goal of this chapter is to contextualize how non-F2F educational content delivery operates, the benefits and drawbacks these types of content delivery for students, and how benefits can be augmented through learning communities (LCs).

The remainder of this chapter is organized as follows: in Section 2.2, the basics of online learning, including synchronous and asynchronous learning, is outlined. Section 2.3 highlights the current use and future viability of blended learning. Section 2.4 contextualizes social presence and LCs, as well as concepts closely related to LCs, such as "community" and "cohort". Section 2.5 presents a case study of a pilot project from Electrical and Software Engineering at the University of Calgary, which implemented the concepts outlined in Sections 2.2 to 2.4.

Some details outlined in Section 2.5, especially pertaining to the Winter terms of 2020 and 2021, are anecdotal and unpublished in literature. These details are obtained through first-person recollection, as Christine was a Teaching Assistant for the project in both the 2020 and 2021 terms.

## 2.2 Online Learning

Long before COVID-19 changed how university courses are taught around the world, online learning has been quietly proliferating across higher education over the last two decades. In 1995, online learning was simply equivalent to using web-based learning management systems or uploading PDFs and texts online [8]. Online learning has since become more viable, thanks to the rapid growth of both Information and Communication Technology (ICT) [9] and higher education institutions' investment in educational technology [10]. In the US, 14.3% of students (2,902,756) were enrolled in online learning in the Fall of 2015 [11], and by 2018 that number has risen to about 6.3 million students [12]. Even prior to COVID-19, scholars predicted that the steady growth of online learning would reach the mainstream by 2025 [13].

Despite the popularity of this learning format, there exist a level of confusion and debate around how to define the term "online learning". The term has been considered synonymous with web-based education, e-learning [14], and distance learning [15]. A review of 46 definitions of this concept [12] concluded that it should be defined as:

> "… education being delivered in an online environment through the use of the internet for teaching and learning. This includes online learning on the part of the students that is not dependent on their physical or virtual co-location. The teaching content is delivered online and the instructors develop teaching modules that enhance learning and interactivity in the synchronous or asynchronous environment".

The goal of online learning is to free students from restrictions in geographic proximity, allowing for freer student-student and student-instructor interactions [16]. As referenced in the previous definition, there are two main modes of online learning: synchronous and asynchronous learning.

**2.2.1 Synchronous Learning**

Synchronous learning takes place in real-time, where the sender and receiver both have to be present for communication to take place [17]. An example of synchronous learning is live lectures using audio/video conferencing tools and sharing of whiteboards. Due to the in-time content delivery, many educators feel synchronous learning is perhaps the best approximation of teaching in a traditional classroom [18]. Students engaged in this type of online learning benefit from a stable mean of communication, tend to stay on task, feel a greater sense of participation, and tend to experience higher task and course completion rates [19][20]. In addition, a live synchronous lecture can help students quickly clarify problems, decrease a sense of isolation, and improve their sense of community [21][22]. Students also benefit from the interactive learning environment and active class participation [6].

However, synchronous learning is not without difficulties. Barriers to technological access is often a challenge for students enrolled in synchronous learning, as students may have slow broadband access [6] and students who rely on public-access technology (i.e. university computer labs or public libraries) may have find difficulties with participating in synchronous lectures [23].

At the onset of COVID-19, many educators transitioned their courses to live synchronous lectures using web conferencing tools like Zoom [18][24][25], likely due to the fact that the unfamiliarity of online learning led educators to default to replicating the F2F practices they are familiar with [26]. During these synchronous lectures, they often went overtime [25], saw more students multitasking and distracted [6], thus rendering participation difficult due to external situation factors, and leaving students feeling fatigued and frustrated due to the long sessions [18].

**2.2.2 Asynchronous Learning**

The other main delivery mode for online learning is asynchronous learning, a time-delayed or time-deferred computer mediated mode of delivery [17]. The sender and receiver do not have to be present at the same time for communication to occur, and this is exemplified by recorded lectures and discussion forums. Some scholars have noted that asynchronous learning seems to be more widely preferred [27], while others believe that students do not have a preference between synchronous and asynchronous teaching formats [23].

Asynchronous learning offers students flexibility to study at their own pace, as they do not have to be online at the same time as the instructor [28][29]. Students can also take time to reflect on the lessons to learn flexibly [30], express their viewpoints more freely on online discussion boards, easing the pressure they feel to immediately respond to a question or comment [31], and increasing their overall satisfaction [32]. This mode of learning allows students to exercise self-control in arranging their own schedule, and self-direction to limitlessly review lecture content [6][33], thus allowing them to learn independent of time and space. Therefore, this type of learning may be easier for students who experience intense business or family lives [34].

Much like synchronous learning, asynchronous learning also faces a slew of difficulties for students. Students in this learning environment experience social isolation from their peers, leading to demotivation, loneliness, and lowered commitment to their online courses [6][27][35]. In addition, students in asynchronous learning also noted higher workload, evaluation pressure, lack of real-time instructor feedback, and technology issues as barriers they faced in asynchronous learning [6][32].

Courses that transitioned to asynchronous learning during COVID-19 were able to maintain the goal of connecting and engaging students during the pandemic [18]. Some have also found that students must be given the opportunity to work collaboratively in this situation, to prevent their feeling of loneliness [33].

**2.3 Blended Learning**

On the spectrum between traditional F2F teaching and online learning, there exist the education model of blended learning, a thoughtful combination of F2F and technology-mediated online instruction [36][37]. Blended learning can also be known as flexible, hybrid, flipped, mixed-mode, multi-mode, or inverted learning, based on how the F2F and online components are integrated [36][38] [39]. Higher education has increasingly utilized blended learning; if it is done well, it can take advantage of both traditional and online teaching approaches [40][41]. Blended learning adoption has been carried out across the world, from Canada to Malaysia [37].

Despite pedagogical advantages, students are less satisfied and less motivated in online learning due to lack of interactions [42]. Conversely, F2F learning has the built-in advantages of learning in a social environment, which facilitates the exchange of ideas [43]. By mixing both types of learning, blended learning can enhance the student experience by overcoming the shortcomings of both F2F and online learning. Blended learning enables students to be more involved and more enthused in the learning process, thus improving their perseverance and commitment [44]. Blended learning can also provide pedagogical productivity, knowledge access, cost efficiency, collective collaboration, and may address student attendance problems [45].

Blended learning has been implemented in engineering's introduction to programming [42] and engineering education [46]. Even prior to COVID-19, educators have regarded blended learning as "the new normal" due to its high adoption rate, popularity, and perceived benefits [47]. As the world transitions to post-COVID-19, the trend towards adopting blended learning will continue, even if physical classrooms reopen [48].

## 2.4 Social Presence and Learning Community

A major criticism of online learning and online components of blended learning is the anxiety, discomfort, and isolation students experience in their studies [38]. There could be a multitude of reasons why a student could be experiencing these troubles, including personality, perceived transactional distance in the online environment, or lack of trust in the community. The building of a social presence is vital to building relationships and success in online learning [49]. In turn, social presence cannot be built alone, as it must be facilitated through an effective group setting of an LC.

## 2.4.1 Social Presence

The Community of Inquiry (CoI) framework [50] proposed the three pillars of social presence, teaching presence, and cognitive presence as vital in online learning. A presence is defined as "a sense of being or identity created through interpersonal communication" [50], and a social presence is the "ability of learners to project themselves socially and emotionally, thereby being perceived as 'real people' in mediated communication" [51]. Empirically, it has been shown that social presence is related to higher student achievement [52]; positively impacts student motivation, participation, and retention; and enhances course satisfaction [53]. Whether synchronous or asynchronous, a high quality digital education is incomplete without prioritizing social presence [54].

Social presence is also tangibly related to the concept of student groups. Groups, when loosely defined, can mean anywhere between an entire class to three or four students. For a large online class of 30, requiring students to reply to each other's posts on discussion boards encourages students to make their social presence known [55]. In an assigned small group of three or four students, social presence depends on other group members' engagement in the discussion [55]. Social presence can also be developed through structured collaborative peer teams [56].

## 2.4.2 Learning Community

At the beginning of the twenty-first century, the emphasis of learning as an individual shifted to learning as a part of a community. This shift, facilitated by ICT, brought forth a growing interest in LCs [57]. Before LCs are defined, the associated concepts of "community" and "cohort" are examined.

There are some disagreements over how "community" should be defined, but it can be expressed as "a general sense of connection, belonging, and comfort that develops over time among members of a group who share purpose or commitment to a common goal" [58]. As a community grows, it will become intentional and sustainable, marked by an increased level of comfort, intimacy, self-reliance, and self-knowledge. For online learners, a community will "simulate the comforts of home, providing a safe climate, an atmosphere of trust and respect, an invitation for intellectual exchange, and a gathering place for like-minded individuals who are sharing a journey that includes similar activities, purpose, and goals" [58]. The community is social with tangible presence, and the bonds among the cohort that makes up the community is its strength.

In higher education, the word "cohort" may invoke images of professional schools, such as medicine, law, education, and engineering, where it was utilised to create peer learning environments for professional fields [59]. It has been argued that cohort creation must be intentional, as "a cohort structure does not in itself ensure a cohort" [60]. Two types of cohorts have been identified: a structural cohort and a communal cohort. A structural cohort is a small group of students enrolled in common courses together, while a communal cohort is a "tight-knit, reliable, common-purpose group" [61]. As both structural and communal components must function well for a cohort to succeed, the cohort should be integrated into a LC.

As an extension of the concept of community, a LC is "a common place where people learn through group activity to define problems affecting them, to decide upon a solution, and to act to achieve the solution" [62]. Scholars believe the origin of LCs dates back to 1927 as a two-year experiment at the University of Wisconsin [63]. In literature, LCs are also synonymous with "community of inquiry" or "community of practice" [58]. Characteristics of a LC can include shared values and vision, collective responsibility, reflective professional inquiry, collaboration, and promoting both group and individual learning [64].

In higher education, there are five major models of LCs: linked courses, learning clusters, freshman interest groups, federated learning communities, and coordinated studies [63]. Both linked courses and learning clusters are characterized by the amount of courses students share, where students will share more courses together if they are in learning clusters. The remaining models are characterized by themes, where freshman interest groups are first-year students who share courses together, federated learning communities are any students who shares courses with the same theme, and coordinated studies are full-time active learning based on an interdisciplinary theme. Outside of these five models, there are also Professional Learning Communities (PLCs), where the focus is on professional, instead of academic, achievements [64].

LCs are particularly important because knowledge construction is a social experience, even in an online environment, as knowledge is beneficial when socially distributed. Learning through interaction in an LC builds social capital, which in turn facilitates learning "by fostering trust, shared values, personal development, a sense of identity and access to the knowledge of others through networks that form a sound basis for sharing knowledge and skills" [57]. Because LCs increase the intellectual interaction of students, they promote coherence among students and create a common sense of purpose [63]. Hence, students in LCs show an increase in academic achievement, retention, motivation, satisfaction, intellectual development, gains in multiple areas of skill and competence, a willingness to form closer bonds, and a higher level of comfort with both students and instructors [63][65][66]. Faculty members have also reported the benefits of LCs, citing that the rich teaching experience fostered creativity and allowed them to increase their commitment to their academic institution, and the unity LCs bring improves overall campus climate [63].

Despite numerous benefits, scholars have also noted hesitancy around LCs. Large LCs are capable of forming cliques, which can cause further isolation and neglect of group members [65]. In addition, LCs can cause excessive socialization, misconduct, and disruptive and rebellious behaviours [67]. Scholars have also observed behaviours "that may hinder student learning, student development, and faculty-student relations", where LCs can cause "behavioural conformity", leading to "the groupthink phenomenon that produces mutually reinforced views and perspectives" about the students' assignments [68]. Much like "behavioural conformity", the familiarity of students within an LC can lead to a "comfort zone", where students are more concerned with maintaining a status quo than to challenge or grow intellectually [69].

As with any pedagogical decision, the design and operation of LCs must be intentional, as "learning communities, in and of themselves, do not cause a student to learn, make good grades or be retained" [70]. It has been posited that there are four characteristics that serves as the boundaries for defining a community, and by extension, a LC: access, relationship, vision, and function [71]. Access defines an LC by spatiality, the fact that students are in a class together [72]. As this is the easiest concept to grasp, access is often the boundary that groups student cohorts. LCs can also be defined by the relational or emotional ties unifying its members. Relational LCs are characterized by a sense of belonging, interdependence or reliance among the members, trust among the members, and faith or trust in the shared purpose of the community. Similarly, a LC built on a shared vision includes "individuals who share common purposes related to education" [73]. If a relational LC are those who feel like they are a member of a community, a vision LC are those who perceive themselves as a community. Finally, functional LCs are organized as a group aiming to achieve the same goal, exemplified by freshman learning communities where first year students take the same class and work on the same projects. Among these four types of boundaries, access and function can be observed externally, and relationship and vision can be observed internally.

Another design consideration for LCs is the size for each group. This is especially important, as some drawbacks of LCs are explicitly linked to the size of the group. The size of a group is critical in building a sense of community, and securing full and active participation [74]. In an online setting, scholars have found that an ideal online class size was twelve students [75], and the instructor/student ratio should be 1:30 [76]. The relationship between class size and student learning may also be curvilinear, and when class size exceeds the maximum ideal number, student learning will deteriorate [55]. Unlike the research around online class size, research on the ideal group size for engaging online discussion is lacking [55]. Scholars have noted that relatively small sizes for online groups means that non-active participants cannot become "invisible", and other group members can exert gentle "community pressure" to bring them into the discussion [74]. In addition, individuals in larger groups can exhibit social loafing, where "the larger the group, the higher likelihood of social loafing (sometimes called free riding) and the more effort it takes to keep members' activities coordinated. Small teams are more efficient – and far less frustrating" [77]. It has also been noted that PLCs should be kept small, but interconnected to other PLCs [78]. Despite the general consensus around LCs being "small", there has not been a quantitative definition of how many students constitutes "small", and how the size of an LC should be defined situationally.

## 2.5 Case Study: Electrical Engineering Integrated Learning Stream

A unifying case study for the three concepts discussed in this chapter, online learning, blended learning, and learning communities, can be found right in the Department of Electrical & Software Engineering at the University of Calgary. In the Winter term of 2019, the Integrated Learning Stream (ILS) project was piloted for 34 second-year electrical engineering students, where they were taught their course in an integrated format [79]. An integrated curriculum is the approach of teaching content from two or more domains "within an authentic context for the purpose of connecting these subjects to enhance student learning" [80]. In this case, five second-year electrical engineering courses were integrated, with the intention to "provide authentic learning experiences, with the ultimate goal of fostering deep learning in the students" [79]. The five courses were: ENEL 343 (circuits II), ENCM 369 (computer organization), ENEL 361 (electronic devices and materials), ENEL 327 (signals and transforms), and ENEL 300 (electrical and computer engineering professional skills). Students were grouped in LCs, provided team building activities, and ultimately worked on a real-world engineering project together, where they combined their knowledge from the five electrical engineering courses. In 2019, students were organized into three to four people teams to work on developing an audio player. Between the five courses, five instructors delivered their course content using a variety of approaches, from traditional lectures to the flipped classroom approach [79], which is a form of blended learning. Throughout the semester, in addition to regularly assessing students for grades, they were asked to perform a self-assessment, reflecting on the strategies which were successful and unsuccessful in helping them learn.

Thematic analysis was performed on the 2019 students' final reflection and logbooks for the themes of competence, relatedness, and autonomy. Through the use of LCs, students reported an increased sense of relatedness, exemplified by a higher sense of belonging, and a closer connection to peers and instructors [79].

Following the success of the 2019 pilot program, the ILS project expanded to 71 students for the 2020 Winter term. In the beginning, the scheduling, content delivery format, project, and assessment for the second iteration of ILS largely remained the same as the previous year. Project teams remained three to four students, but LCs have expanded to include two project teams, including upwards of eight students per LC. Students were also given the opportunity to name their LC, and develop their own cohort identity. In March 2020, at the onset of COVID-19 and the University of Calgary's decision to close all in-person instruction, the ILS project also shifted all content delivery from traditional and blended learning to entirely online learning. At that point, since most course assessments were completed, this largely impacted only the deliverable of the final audio player project, preventing most students from developing and demonstrating a physical prototype.

In the 2021 Winter term, the ILS project officially expanded to all 147 second year electrical engineering students. The goal, project, and assessment for ILS largely remained the same as previous years, but the scheduling and content delivery format significantly changed to adapt to the entirely online environment. Most course instructors opted for an asynchronous online content delivery, but some courses, including ENEL 300, included a synchronous component for assessment and discussion. The format of LCs remained similar to previous year, with each of the 18 LCs encompassing two project teams. Unlike previous years where the LCs can occupy a physical space, LCs this year have been able to entirely communicate online, using communication platforms such as Zoom or Discord. The reflection and logbook data from both the 2020 and 2021 ILS terms have not been thematically analyzed or published.

The case of the ILS project and its evolution from 2019 – 2021 demonstrated the viability of blended learning, and later online learning, as a practical content delivery format for engineering courses at the University of Calgary. The integration of LCs into this blended and online delivery format have proven to be largely beneficial for students, based on the thematic analysis from the 2019 data, and anecdotal evidence from students and instructors in the 2020 and 2021 sessions.

**2.6 Summary**

The goal of this chapter is to contextualize the non-F2F types of course content delivery, and the associated collaborative learning approach. The definition, benefits, and drawbacks were discussed for online learning and its two delivery modes, synchronous and asynchronous learning, were presented. Blended learning, as a mix of both F2F and online learning, was discussed. The concepts definition, benefits, and drawbacks around social presence and LCs were presented, in addition to discussion of how to construct and operationalize LCs. A case study from the University of Calgary was also presented to demonstrate the applicability of LCs and the viability of these content delivery models.

# Chapter 3 : Background in VLSI Algorithms for Clustering and Partitioning

## 3.1 Introduction

The research presented in this thesis aims to provide an optimized method to create student cohorts. In doing so, the methodology proposed draws upon algorithms first implemented for Integrated Circuit (IC) design. The goal of this chapter is to present an explanation for the Very Large-Scale Integration (VLSI) clustering and partitioning algorithms used in the creation and refinement of student cohorts.

Parts of this chapter is published as ***Optimized Cohort Creation for Hybrid Online Design Learning During COVID-19*** in the *Proceedings for the 2021 American Society for Engineering Education (ASEE) Virtual Conference [81]*.

The remainder of this chapter is organized as follows: in Section 3.2, the basics of ICs, VLSI, and Physical Design steps are outlined. Details surrounding partitioning, a major step in Physical Design, is summarized in Section 3.3. Clustering algorithms, including implementation details on the Hyperedge Coarsening, Modified Hyperedge Coarsening, and Best Choice algorithms, is explained in Section 3.4. Section 3.5 outlines Simulated Annealing, a global optimization process used in partitioning. Section 3.6 highlights the process of Multilevel Partitioning. Finally, the chapter is summarized in Section 3.7.

## 3.2 VLSI and Physical Design

An IC is defined as a circuit in which all or some of the circuit elements are inseparably associated and electronically interconnected so that it is considered to be indivisible for the purpose of construction and commerce [82]. An IC, which can also be called a "chip" or "microchip", is a set of electronic circuits placed on a flat semiconductor material, such as silicon. Early concepts of ICs, dated back to 1949, consists of five transistors arranged as a three-stage amplifier [83]. In 1959, the patent for the first IC prototype was filed by Jack Kilby who later won the Nobel Prize [84]. In 1965, Gordon Moore predicted that the number of transistors in an IC will double every year, thus coining Moore's Law [85]. Following the Moore's Law, the modern IC has grown simultaneously to be smaller and more powerful, consisting of millions of components.

In the following table, a summary of the IC components that are most relevant to this thesis is given [86].

*Table 1: Integrated Circuits Components*

| Term | Definition |
|---|---|
| Transistor | A semiconductor device for amplifying, controlling, and generating electrical signals. They are the active components of ICs. [87] |
| Cell | A logical or functional unit built from different components. Can represent Boolean logic such as AND, OR, XOR, NAND, etc. |
| Pin | An electrical terminal connecting a component to its external environment. |
| Net | A set of pins that must be connected, typically with wires. |
| Netlist | The collection of all nets and the components they are connected to. |

In order to represent and use IC components in algorithms, netlists are modelled as graphs. Derived from graph theory, first developed by Euler, a graph is defined as:

$$G = (V, E) \tag{3.1}$$

where V is the set of vertices (which are cells in an IC) and E is the set of edges (which are nets in an IC) connecting the vertices [88]. Due to the complexity of modern circuits, netlists closer resemble a hypergraph [89]. In a graph, each edge connects only two vertices. In a hypergraph, vertices are connected by a set of hyperedges, where each hyperedge is connected to two or more vertices. In the netlist of an IC, each net is typically connected to two or more cells. In Figure 1, an example of a hypergraph with nine vertices, connected by four hyperedges is showcased.

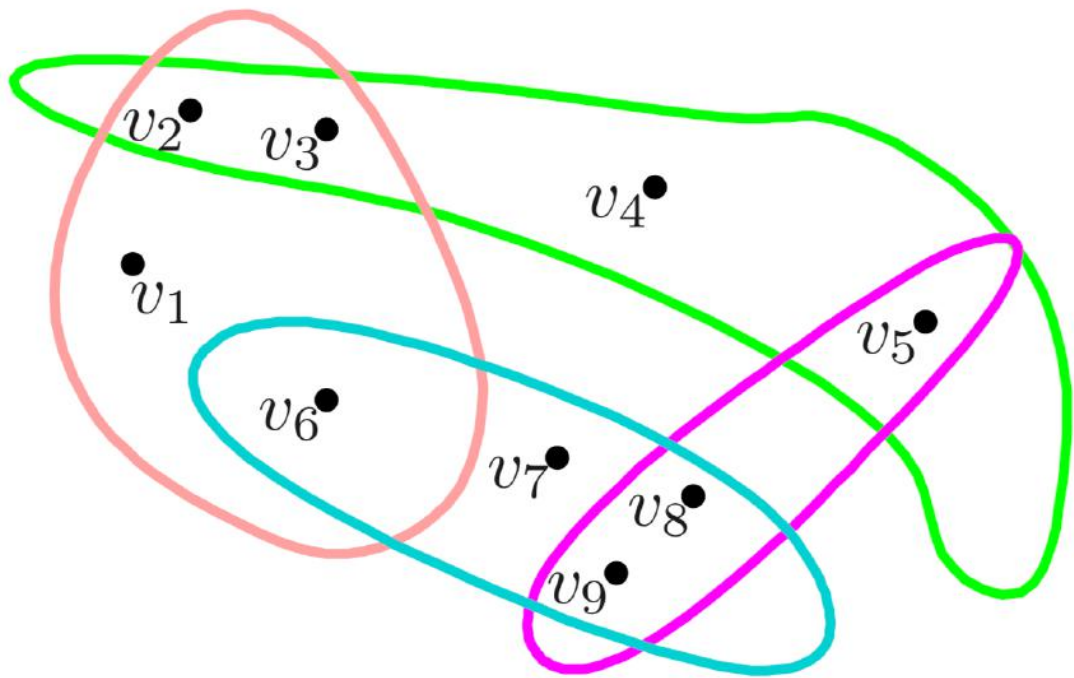

*Figure 1: Example of a Hypergraph [90]*

As ICs become more powerful, the process of creating ICs, known as VLSI, also becomes more complicated.

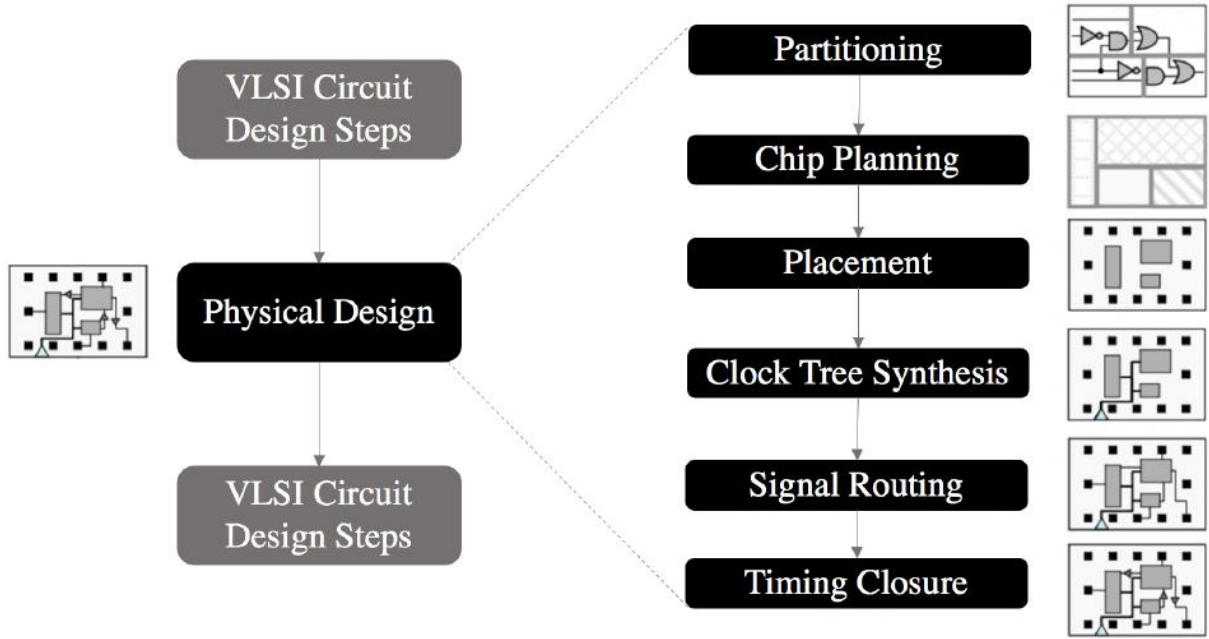

*Figure 2: Steps in the VLSI Circuit Design Flow [86]*

Of all the core steps in the steps in VLSI to design an IC, Physical Design, outlined in Figure 2, is the most relevant to this thesis. Physical Design translates abstract circuit requirements in a circuit design to a geometric chip layout [86]. All circuit design components are instantiated with their geometric representations, where circuit elements such as cells and nets (see Table 1), each with their fixed size and connections, are assigned a spatial location.

The Physical Design process is divided into several procedural steps to compartmentalize the complicated process of turning a circuit design into a 3D physical chip layout. The modern implementation of this process is aided by a class of software, hardware, and services called

Electronic Design Automation (EDA) [91]. The following Table 2 summarizes the major stages of Physical Design, as seen in Figure 2.

*Table 2: Physical Design Steps*

| Physical Design Step | Description |
|---|---|
| Partitioning | Breaking a large circuit into smaller subcircuits or modules, allowing each subcircuit to be designed and analyzed individually [86]. |
| Chip Planning | Determine the shape of the subcircuits [92]. |
| Placement | Placing cells within each subcircuit block [93]. |
| Clock Tree Synthesis | Determine buffering, gating, and routing of clock signals [94]. |
| Signal Routing | Finding the paths of the wires that connects the cells [86]. |
| Timing Closure | Ensure that the circuit timing constraints are satisfied [95]. |

## 3.3 Partitioning

The most consequential Physical Design process relevant to this thesis is the Partitioning stage. In order to tackle the design complexity of modern ICs, each circuit is commonly divided into smaller partitions, where cells in each partition have more connections with each other than cells outside of the partition.

The primary goal of any partitioning algorithm is to divide the circuit in such a way that minimizes the number of connections between subcircuits. In addition, partitioning algorithms must balance the complexity of the circuit with the solution quality and runtime, while meeting all design constraints. Due to this complexity, IC partitioning problems are categorized as NP-hard problems. First posed in 1971 [96], the P versus NP problem, or the polynomial time versus nondeterministic polynomial time problem, is one of the major unsolved questions in computational complexity theory. P versus NP asks whether a problem whose answer can be verified in polynomial time can also be solved in polynomial time. NP-hard problems are a class of problems at least as hard as any problem in NP [97]. In the context of IC partitioning, a partitioning problem size can grow linearly, but the time required to find an optimal solution grows faster than any polynomial function [86]. To address this, partitioning algorithms have been developed to find high-quality solutions while running in low-order polynomial time.

One of the most common partitioning algorithms are the Fiduccia-Mattheyses (FM) algorithm [98]. Of the common partitioning algorithms, the FM algorithm offers the best trade-off between solution quality and runtime, and is still the base of many partitioning algorithms used today.

Imagine students enrolled in a university to be equivalent to an IC. In that setting, students have already been partitioned into the smaller subcircuits of faculties and programs. Since the partitioning has already occurred, another class of algorithms should be explored in order further reduce the size of student groups.

## 3.4 Clustering

With the increasing complexity of modern ICs, partitioning algorithms were no longer able to achieve high quality results in reasonable times [99]. Clustering algorithms are introduced during the partitioning stage to group cells together and reduce the size of the netlist in a circuit. Clustering analysis is a method within statistical data analysis that groups objects that are similar together in a cluster. In context of VLSI, clustering is a process that exists within the partitioning stage, used to group tightly connected cells together, thus absorbing the connection between cells. Clustering algorithms takes the opposite approach from the aforementioned partitioning algorithms. While partitioning algorithms divides a netlist into smaller blocks, clustering algorithms complements this process by taking a bottom-up approach. Modern EDA software will use both partitioning and clustering to reduce runtime and preserve the structure of the original netlist [89].

Common clustering algorithms used for partitioning can in general be classified as hierarchical or agglomerative [89], where the difference lies in hierarchical clustering merges all cells into clusters at once, and agglomerative methods forms clusters one at a time based on the connectivity of the cells. In hierarchical clustering, vertices are matched into clusters [100][101][102][103][104], thus decreasing the size of the netlist. In contrast, clusters are formed one at a time using agglomerative methods, and the vertices already in clusters are removed from the netlist. Agglomerative clustering can be performed based on cell ordering [105], based on connectivity [106][107][108], or based on cell area [109].

In the rest of this section, the clustering algorithms used later in this thesis are described in detail.

### 3.4.1 Hyperedge and Modified Hyperedge Coarsening

Some of the first clustering algorithms developed for partitioning are: Hyperedge Coarsening (HC) [103] and Modified Hyperedge Coarsening (MHC) [103]. These algorithms aim to coarsen (or cluster) the netlist, and generate sets of roughly equal sized clusters. During the clustering phase, a series of smaller hypergraphs are constructed, reducing the size of the original circuit netlist into a simpler representation with fewer connections in low run times.

HC was developed based on Edge Coarsening (EC) [103], in order to remedy the problem in EC where during the production of smaller subcircuits, the number of nets in the circuit netlist does not decrease at the same rate. In Table 3, a generic HC algorithm is given.

*Table 3: Algorithm of Hyperedge Coarsening*

| **Input**: Flat Netlist<br><br>**Output**: Clustered Netlist |
|---|
| 1. Sort all hyperedges in increasing order<br>2. For each hyperedge visited in order:<br>    a. If the cells connected to the hyperedge has not been matched, they are matched together as a cluster and marked<br>    b. If any of the cells connected to the hyperedge has been marked, all of the cells connected to the hyperedge are skipped<br>        i. all of the unmarked cells become an individual cluster<br>3. The clusters of cells form the next level of coarser graphs. |

This algorithm gives preference to hyperedges with larger weights and small sizes, which may not always be applicable for every situation. While HC can reduce a significant amount of hyperedge weight to produce coarser graphs, especially compared to EC, it ultimately leaves a lot of unclustered cells. The MHC is an improved version of the HC, where it aims to address this issue with HC. MHC build upon the results from HC by revisiting the skipped cells during HC. The procedure for the MHC algorithm is as follows:

*Table 4: Algorithm of Modified Hyperedge Coarsening*

| |
|---|
| **Input**: Flat Netlist <br><br> **Output**: Clustered Netlist |
| 1. Sort all hyperedges in increasing order <br><br> 2. For each hyperedge visited in order: <br><br>      a. If the cells connected to the hyperedge has not been matched, they are matched together as a cluster and marked <br><br>      b. If any of the cells connected to the hyperedge has been marked, all of the remaining unmatched cells are matched together as a separate cluster and marked <br><br> 3. The clusters of cells form the next level of coarser graphs. |

Unlike HC which leaves many unclustered cells, MHC is capable of clustering more cells together, thus creating better clustering results.

**3.4.2 Best Choice Algorithm**

The Best Choice (BC) algorithm [109] is an agglomerative clustering technique that presents a bottom-up clustering technique that aims to solve large-scale placement problems This algorithm operates on the netlist and identifies the best pair of cells among all cell pairs to cluster. The algorithm cluster this pair, updates the ranking of the pair and clusters the next best pair together. This process is repeated until the circuit netlist size is reduced to the desired size.

A priority-queue data structure, also seen in [106] is adopted to rank and rapidly identify all best-pairs of cells. The priority-queue enables a global picture of clustering sequences, thus reducing suboptimal clustering choices. The high-level BC clustering algorithm is as follows:

*Table 5: High Level Algorithm for BC*

| **Input**: Flat Netlist <br> **Output**: Clustered Netlist |
|---|
| 1. Until *target object number* is reached: <br>   a. Find *closest pair* of cells <br>   b. Cluster them <br>   c. Update netlist |

The degree of clustering is controlled by *target clustering ratio* and *target object number*, where the *target object number* is the original number of cells divided by the *target clustering ratio*.

The detailed BC clustering procedure is as follows:

*Table 6: Detailed Algorithm for BC*

**Input**: Flat Netlist

**Output**: Clustered Netlist

Phase I. Priority-queue (PQ) Initialization:

1. For each cell *u*:

    a. Find closest cell *v*, and its associated clustering score *d*

    b. Insert tuple (*u*, *v*, *d*) into PQ with *d* as key

Phase II. Clustering:

1. While target object number is not reached and top tuple's score *d* > 0:

    a. Pick top tuple (*u*, *v*, *d*) of PQ

    b. Cluster *u* and *v* into new cell *u'*

    c. Update netlist

    d. Find closest cell *v'* to *u'* with its clustering score *d'*

    e. Insert tuple (*u'*, *v'*, *d'*) into PQ with *d'* as key

    f. Update clustering scores of all neighbours of *u'*

The clustering score function *d(u, v)* between cells *u* and *v* is defined as:

$$d(u, v) = \sum_{e} w_e / [a(u) + a(v)] \qquad (3.2)$$

where the weight $w_e$ of hyperedge e is defined as $\frac{1}{|e|}$, and $a(u)$ and $a(v)$ are areas of *u* and *v*, respectively. The clustering score is directly proportional to the total sum of edge weights between them, and inversely proportional to the sum of their areas.

The algorithm presents the clustering in two phases, with Phase I focusing on establishing the priority-queue. For each cell *u* in the netlist, its closest cell *v* and associated clustering score *d* are calculated. Using *d* as the key, the tuple *(u, v, d)* is inserted into the priority queue. Only one pair, cell *u* and its closest cell *v*, is inserted per cell *u*, thus allowing for a more efficient data structure management.

In Phase II, the top tuple *(u, v, d)* is chosen, and the pair *(u, v)* is clustered into a new cell *u'*. The netlist is also updated with the new cell *u'*. The closest cell *v'* and the associated clustering score *d'* are calculated for *u'*, and the new tuple *(u', v', d')* is inserted into the priority-queue. All clustering scores of neighbours of the new cell *u'* must be re-calculated, and the priority-queue re-adjusted accordingly.

BC clustering also faces some challenges, including:
- Using effective and efficient clustering score function, leading to higher quality solutions
- Accurately handling hyperedges
- Controlling cluster size for more balanced clustering

These challenges will require specific experimentation to determine what is the optimal solution for the clustering problem. Overall, this is a globally optimal clustering algorithm that operates directly on a netlist, and its priority-queue data structure ensures a faster implementation.

## 3.5 Simulated Annealing

Annealing describes a process in metallurgy, where a material is heated to a high temperature and cooled in a controlled manner to modify its material properties through its atomic configuration [110]. The process in which the atoms settle into a lattice is probabilistic in nature, and structural integrity of the lattice is affected by the rate at which the material is cooled. If the cooling rate is sufficiently and incrementally small, the atoms will settle into an optimum configuration at a higher probability.

Simulated Annealing (SA) [111] applies the annealing concept to solve difficult computational optimization problems, by utilizing a large search space to approximate a global optimum using a probabilistic approach. This is an iterative algorithm beginning at an arbitrary solution, where at each iteration, the local neighbourhood is searched to find a new solution. The new solution is formed by performing a small perturbation on the current solution within the neighbourhood. SA emulates the physical Annealing process by "cooling down" a high-cost (or high-energy) solution into a low-cost, structured solution. In VLSI, SA has been applied to many steps in the Physical Design process, including partitioning [112], where SA has been employed to solve particularly difficult partitioning problems.

The SA algorithm to find a global minimum is generally executed as follows:

*Table 7: Algorithm of SA*

| |
|---|
| **Input**: Initial Solution |
| **Output**: Optimized New Solution |

1. Initialize variables
    a. $T = T_0$
    b. Current Solution = Initial Solution
    c. Current Cost = **COST**(Current Solution)
2. **WHILE**($T > T_{min}$)
3.     $i = 0$
4.     **WHILE**($i < i_T$)
5.         $i = i + 1$
6.         $(a_i, b_i)$ = **SELECT_PAIR**(Current Solution)
7.         Trial Solution = **PERTURB**($a_i, b_i$)
8.         Trial Cost = **COST**(Trial Solution)
9.         $\Delta$Cost = Trial Cost – Current Cost
10.        **IF**($\Delta$Cost $< 0$)
11.            Current Cost = Trial Cost
12.            Current Solution = Trial Solution
13.        **ELSE**
14.            r = **RANDOM**(0, 1)
15.            **IF**($r < e^{\frac{-\Delta Cost}{T}}$)
16.                Current Cost = Trial Cost

| | |
|---|---|
| 17. | Current Solution = Trial Solution |
| 18. | $T = \alpha \cdot T$ |

Using an arbitrary initial solution, SA generates a new solution by perturbing, or moving, some aspects of the initial solution. After the new solution is computed, cost difference in changing the solution (ΔCost) is calculated and compared. To find a global minimum, ΔCost < 0, whereas for a global maximum, ΔCost > 0. If the criteria of ΔCost < 0 or ΔCost > 0 was not met, the change could still be probabilistically accepted. The most common method used in SA for probabilistically accepting a move is the Boltzmann acceptance criterion:

$$r < e^{\frac{-\Delta Cost}{T}} \qquad (3.3)$$

where $r$ is a random number generated between 0 and 1 using a uniform distribution. This acceptance also depends on the temperature parameter $T$, where a lower temperature will result in a lower chance of accepting the move. The temperature parameter begins with an initial temperature, and is slowly decreased by the cooling parameter $\alpha$ as the algorithm runs. The rate at which the temperature decreases is very important, as it should allow both a sufficiently high temperature initially, and a sufficiently slow decrease rate to allow the solution to settle, especially in the cooler temperatures.

SA is a stochastic algorithm, where two runs using identical parameters will usually not yield the same results. The difference between the runs is a result of random perturbations used to generate moves and probabilistically accepting these moves.

A benefit of SA is that it is not greedy, meaning that it does not always reject potential solutions with a higher cost, especially in the beginning. A greedy algorithm would have difficulty distinguishing between a local optimum and a global optimum, as it only ascends or descends in one direction from the initial solution. An algorithm capable of accepting a non-improving move has the potential to move away from the local optimum into the global optimum.

## 3.6 Multilevel Partitioning

Multilevel Partitioning is a technique designed to improve the scalability of netlist partitioning. In Multilevel Partitioning, there are four major phases that involves both clustering and partitioning algorithms. Initially, the original netlist is clustered hierarchically during the clustering phase. After which, a partitioning algorithm is used to divide this clustered circuit netlist into smaller subcircuits. The netlist is then unclustered and refined to reduce the number of connections between the clusters [86]. In Figure 3, the general steps of Multilevel Partitioning are shown.

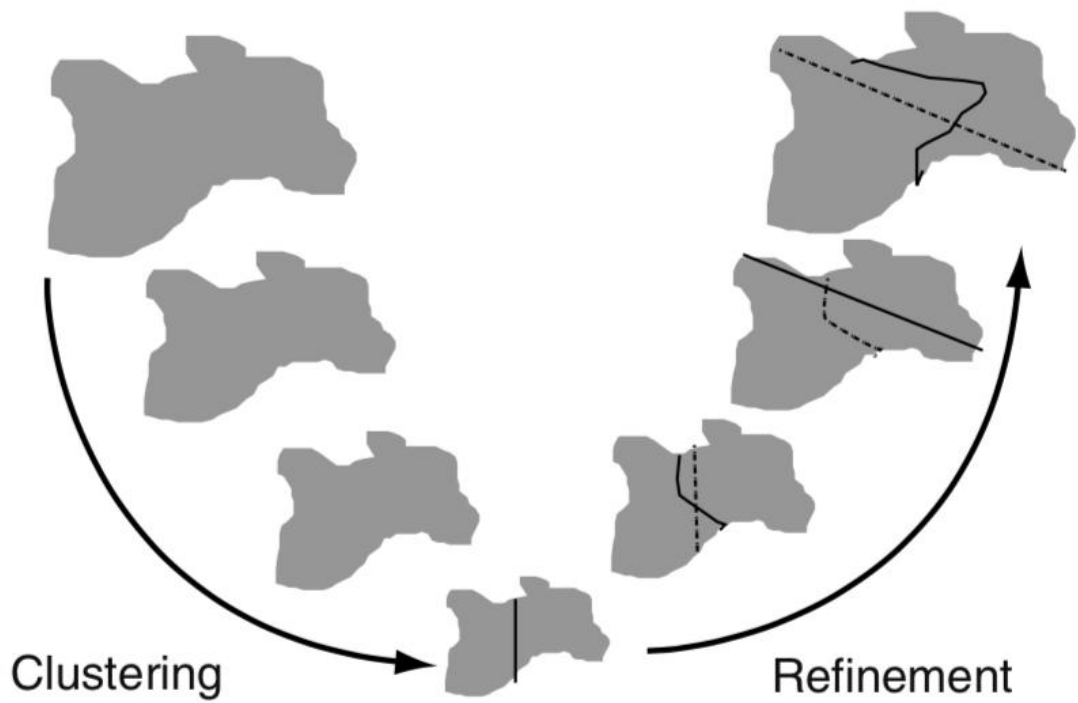

*Figure 3: Multilevel Partitioning[89]*

**3.7 Summary**

The goal of this chapter is to summarize the background technical information related to the implementation of this thesis. General concepts around ICs were introduced, including IC components and the concept of VLSI. Steps in the Physical Design process were introduced, with an emphasis on partitioning. Clustering and Partitioning algorithms were further emphasized, including details on the HC, MHC, BC, and SA algorithms. Finally, Multilevel Partitioning is briefly outlined.

# Chapter 4 : Learning Community Creation and Refinement Framework

## 4.1 Introduction

Community-based learning is gaining significance in post-secondary education, especially for uses in online and blended learning models. In order to leverage learning communities (LCs), they must be designed and refined with optimization in mind. The goal of this chapter is to explore how to best identify and refine optimal student LC configurations. The proposed models are tested using student enrollment data in 3$^{rd}$ year Electrical Engineering (ENEL) Fall 2020 courses. The enrollment data can be represented as students in a network, with individual students connected by the classes they mutually share. Using methodological inspiration from very large-scale integration (VLSI) clustering algorithms, student LCs can be abstracted and clustered based on the enrollment data. The resulting LCs can then be used as a starting point to be refined into higher-quality LCs with Simulated Annealing (SA), a partitioning algorithm that performs a global optimum search. Since a circuit netlist is a form of a network, the techniques used in optimizing integrated circuits (ICs) can easily be adapted to address student LC assignment problem. Ultimately, the goal of both the clustering and refinement algorithms are to increase internal LC contact and reduce unnecessary contacts between LCs by grouping students that have the most amount of contact time in courses together.

Parts of this chapter is published as ***Optimized Cohort Creation for Hybrid Online Design Learning During COVID-19*** in the *Proceedings for the 2021 American Society for Engineering Education (ASEE) Virtual Conference [81]*.

The remainder of this chapter is organized as follows: Section 4.2 outlines the proposed methodology for optimal LC creation, and the experiments used to determine the optimal clustering parameters is described in Section 4.3. Section 4.4 outlines the proposed methodology for stochastic LC refinement, and the experiments needed to fine-tune the algorithm is described in Section 4.5. Section 4.6 introduces the LC Creation and Refinement Framework. Finally, Section 4.7 summarizes the contributions of this chapter.

**4.2 Proposed Methodology for Optimal Learning Community Creation**

The rationale behind using VLSI clustering algorithms to create student LCs is because students, when enrolled in classes, form a network which is similar to a netlist in a circuit. Several clustering algorithms, such as Hyperedge Coarsening (HC) [102], Modified Hyperedge Coarsening (MHC) [102], and the Best Choice (BC) [109] algorithms have been developed to cluster the highly interconnected components of a circuit netlist. These algorithms can be modified from their intended design of creating coarsened, or clustered, hypergraphs, to create student LCs. These algorithms are relatively computationally inexpensive, and easily translate to student LC creation with some modifications. Optimal LC creation must take into account of the difference in student's scheduling, and cannot be achieved through simple course registration into a LC. Therefore, having an automated method to create optimized student LCs is necessary. The boundary of these LCs will be defined based on their access and function [71]. In the following sections, the details of the proposed algorithms and how the clustering algorithms have been modified to generate student LCs will be discussed.

**4.2.1 Representing Student Enrollment as a Network**

Student enrollment are data maintained by the Office of the Registrar, including information associated with individual students' IDs and the courses they are enrolled in each semester. The attributes of the student enrollment dataset include:

- Enrolled academic program of the student
- Primary academic plan of the student
- Projected academic year
- Term ID
- Class number
- Subject of the enrolled class
- Catalog number of the enrolled class
- Class section
- Class start time
- Class end time
- Class schedule (from Monday to Friday)
- Standard meeting pattern of the class

Using these data attributes, this thesis proposes that students enrolled in any program can be represented as a network so that clustering algorithms can be used to make LCs. This network will show the connectivity of students. In the circuit-like network of students, a student is a cell in the network, and a common course between two students is a net connecting the two cells. If two students share twelve courses together, they have twelve nets between them. The following figure represents a sample student enrollment dataset as a network.

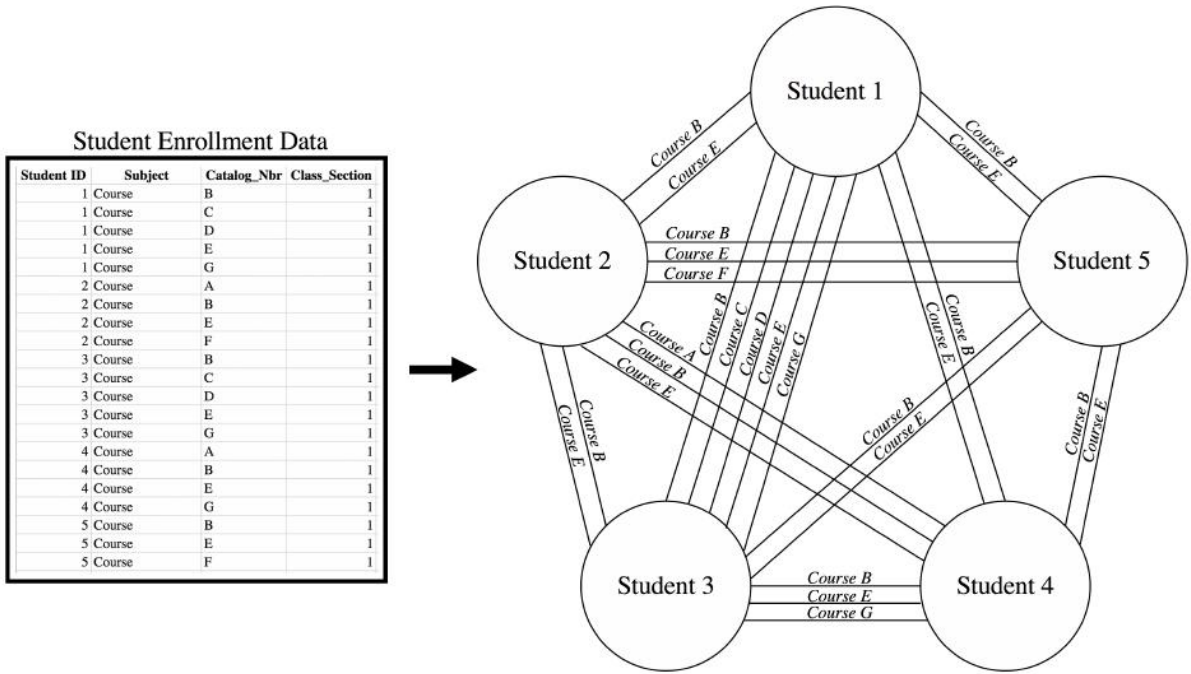

*Figure 4: Student Enrollment Data (left) and Student Enrollment Network (right) of Sample Dataset*

Figure 4 demonstrates how to transform student enrollment data into a network. Five unique students are represented as cells in the network, and they are connected by the seven courses they mutually share, represented by nets in the network. A network that includes all course enrollments, similar to Figure 4, would inevitably become very dense if the student enrollment dataset becomes very large. This thesis proposes calling this type of student enrollment network a Fully Dense Network. Another method of representing the student enrollment network is by constructing the network for students who are connected via mutually enrolled non-lecture courses, such as tutorial or laboratory sections. The rationale behind removing lectures is to reduce the density of the connections in the network. There would be more opportunities for group-based collaborations in tutorial and laboratory sections, as opposed to during lectures. Hence, effective LCs could still be formed by disregarding lecture sections. This type of network is proposed to be called a Sparse Network. Both types of networks are viable methods of representing student enrollment data.

In addition to student enrollment being represented as networks, it can also be represented as two matrices, the Adjacency and the Connectivity matrices. The Adjacency matrix, $A$, is a sparse matrix representing the relationship between all students and all courses in a given dataset. The matrix elements would populate with the value of 1 if the student (indicated by the row of the matrix) is enrolled in a course (indicated by the column of the matrix), and 0 otherwise. The value of matrix element $A_{ij}$ of Adjacency matrix $A$ is populated by the following equation:

$$A_{ij} = 1 \text{ if student } i \text{ is enrolled in course } j$$
$$A_{ij} = 0 \text{ if student } i \text{ is not enrolled in course } j$$
(4.1)

The Connectivity matrix, $C$, is a symmetric matrix showing the connectivity between students with each other. This connectivity is based on the Adjacency matrix, and can be calculated by:

$$C = AA^T \tag{4.2}$$

In the Connectivity matrix, the number of connections between two students is indicated by the value populating the cell with the row and column matching the values of the two students' assigned IDs. In Figure 5, both matrices are visualized.

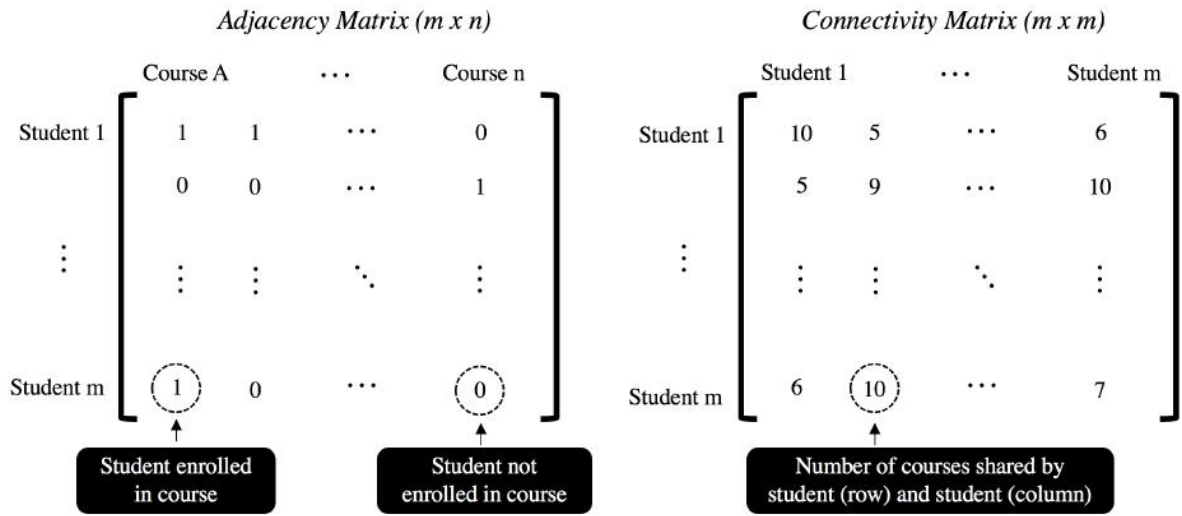

*Figure 5: Adjacency Matrix, A (left) and Connectivity Matrix, C (right)*

The preceding Figure 5 demonstrates what an Adjacency matrix and a Connectivity matrix could look like. In the Adjacency matrix, the elements are valued at 1 or 0, as indicated by equation (4.1). In the Connectivity matrix, each element of the matrix $C_{ij}$ is valued at the number of connections between student $i$ and student $j$. Both the Adjacency and the Connectivity matrices are built using the student enrollment network.

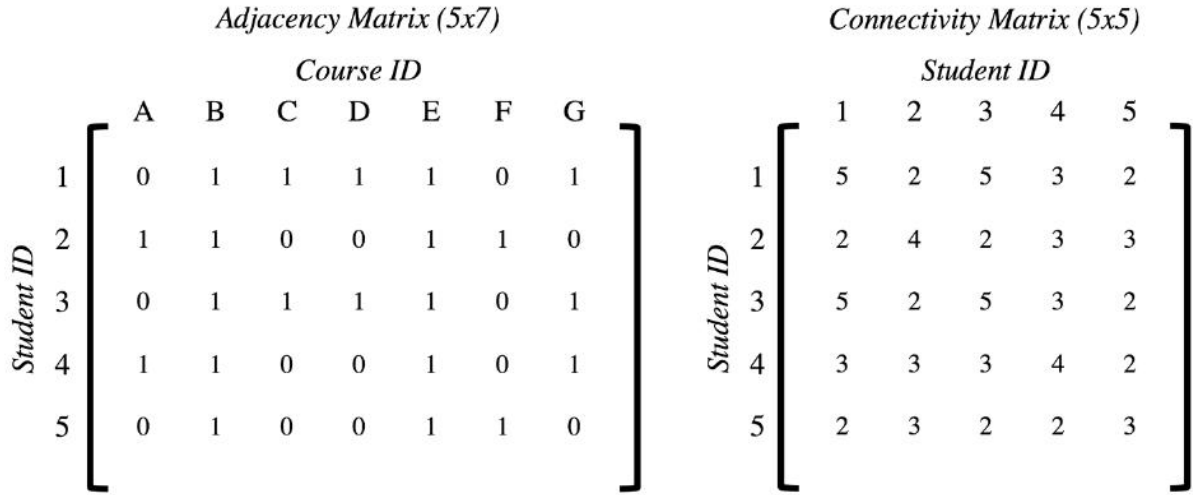

*Figure 6: Adjacency Matrix, A (left) and Connectivity Matrix, C (right) of Sample Dataset*

The preceding Figure 6 showcases the Adjacency and Connectivity matrices for the student enrollment network in Figure 4. The rows of the Adjacency matrix reflect the five unique students, and the columns of the matrix reflects the seven enrolled courses. The Connectivity matrix is populated with values indicating the number of connections between any two students, which reflects the number of nets between the same two students in the student enrollment network in Figure 4.

**4.2.2 Implementing Hyperedge and Modified Hyperedge Coarsening**

Once the Adjacency and Connectivity matrices are made for each group of students, the HC, MHC, and BC algorithms can be used to make student LCs. HC (Table 3) and MHC (Table 4) can be directly implemented for both the Fully Dense and the Sparse student enrollment networks, requiring minimal changes to the algorithm in order to accommodate the new context of student LC creation. The output of HC and MHC would be a group of clusters based on the student enrollment network, where each final cluster is equivalent to a student LC.

### 4.2.3 Implementing Best Choice Algorithm

Unlike HC and MHC, the BC clustering algorithm (Table 5 and Table 6) required significant modifications to the original algorithm in order to accommodate the changes induced by using student enrollment data. Using the two types of student enrollment networks, the Fully Dense Network and the Sparse Network, the Adjacency matrices and the Connectivity matrices for each are constructed.

In the process of applying the BC algorithm to the network, the clustering score function $d$ must be defined for this particular case. This thesis proposes that the closeness of any two clusters $u$ and $v$ in the network can be defined as the following Linear Clustering Score Function:

$$d_l = \frac{connectivity}{size(u+v)} \qquad (4.3)$$

where connectivity is the number of connections between clusters $u$ and $v$, and $size(u+v)$ is the number of students in both clusters. The larger the $d$ score, the closer the two clusters are, where each cluster is a student LC.

The BC algorithm has a general tendency of selecting larger clusters to merge with each other, resulting in an uneven distribution of cluster sizes. In order to minimize this tendency, this thesis also proposes an alternative definition for cluster scoring using the Nonlinear Clustering Score Function:

$$d_n = \frac{connectivity}{size(u+v)^2} \qquad (4.4)$$

The clustering score definition in equation (4.4) penalizes the selection and merger of two larger clusters, allowing for smaller clusters to be selected. Since equation (4.4) prioritizes pairing smaller clusters, the final cluster output may lead to more even cluster size distribution. However, that does not necessitate that use of equation (4.4) will lead to better cluster quality than using equation (4.3). Therefore, both definitions are implemented and compared to see which clustering score yields a better result.

Another proposed modification to the BC algorithm is the addition of random selection. During the run of the algorithm, there are often more than one cluster that form the closest pair with the target cluster. In the original implementation of the BC algorithm, the first cluster in the list all potential clusters that can form the closest pair with the target is chosen. For this modification, a random cluster among all the potential clusters will be chosen instead. Injecting random selection will allow the algorithm to search for other possible combinations of clusters that would yield better results. This modification will render BC as a stochastic algorithm, where the outcomes of each run will vary, despite the same input. Due to the numerous combinatorics outcomes of the clustering algorithm, the additional stochasticity to the algorithm will ensure all clustering combinations has an opportunity to be explored. Without the random selection, only the first cluster in the list of all potential closest clusters will form a pair. However, this addition also brings the possibility of producing poor cluster results due to the inherent randomness. Therefore, the thesis also proposes introducing a Monte Carlo process to the algorithm to search for the best possible outcome.

The Monte Carlo method is a mathematical technique used to estimate the possible outcomes of an uncertain event [113]. This method can also be used to evaluate stochastic algorithms, where given a fixed input, a Monte Carlo simulation can be run to examine the distribution of the variable results. The proposed BC algorithm will run for a specified number of times as a Monte Carlo simulation, where the best cluster network will be stored. Each time a better cluster network is found, the result will be stored and updated.

**4.2.4 Cluster Quality Measurement**

Each proposed modified clustering algorithm pre-sets a maximum LC size, and will eventually yield a group of clusters representing student LCs. In order to evaluate the quality of these LCs, this thesis proposes a single cluster score as a quantitative measure of LC quality, based on a series of calculated scores. In the construction of this measure, two values must be determined, the internal connection existing within each LC and the external connections existing between LCs. The internal connection, representing the number of connections between students within a single LC, is calculated by counting the number of shared courses between the students in the LC. The external connections, representing the number of connections between students in different LC groups, is calculated by counting the number of shared courses between students not in the same LC group. The Total Cluster Quality Score is calculated by the sum of all internal connections subtracted by the sum of all external connections, and calculated as follows:

$$S_T = \sum_{j=1}^{n} (S_{I_j} - S_{E_j}) \quad (4.5)$$

where $S_T$ is the Total Cluster Quality Score, $S_{I_j}$ is the total number of internal cluster connections for any given cluster $j$, for $n$ clusters in a set, and $S_{E_j}$ is the total number of external cluster connections between any given cluster $j$ and every other cluster in the set. With the goal of maximizing the internal cluster connection and minimizing external connections, the larger the Total Cluster Quality Score, the more preferable the clustering result is.

## 4.3 Experimental Design for Optimal Learning Community Creation

The HC and MHC algorithms can be executed with the proposed changes directly on student enrollment networks to produce clustering results, and does not require additional experimentation. The BC algorithm can also be executed for student enrollment networks, with the following proposed adjustments. In order to evaluate the clustering results by using the BC algorithm, several clustering parameters must be finely tuned through experimentation. Among which, the ideal size of a student LC, and the number of runs in a Monte Carlo simulation needed to discover the best clustering results must be determined. All experiments for clustering will be done on the 3$^{rd}$ year ENEL Fall 2020 dataset.

### 4.3.1 Optimal Maximum Cluster Size for Best Choice Algorithm

One of the major tuning parameters of the BC algorithm is the maximum size of the cluster generated. In context, this parameter determines the size of each student LC. For operational reasons of having student LCs, the algorithm will be prompted to create as many equal-sized clusters as possible. The goal of this experiment is to determine what is the ideal LC size that can yield an optimal set of clusters. This experiment is executed by generating 200 runs per Monte Carlo simulation, the BC algorithm is used to create student LC sets with a maximum cluster size of six to twelve, at increments of one.

### 4.3.2 Optimal Number of Monte Carlo Simulation Runs for Best Choice Algorithm

The other tuning parameter required is the number of runs in a Monte Carlo simulation required to yield the best possible result. The BC algorithm was modified to accept a predetermined number of runs to capitalized on the random selection process introduced in the formation of the closest pair. The goal of this experiment is to determine a threshold of how many Monte Carlo simulation runs are needed to consistently produce the best possible clustering results. This experiment is executed by using a maximum cluster size of nine, and generating clustering results using 100 to 500 runs per simulation, at the increment of 100 runs.

**4.4 Proposed Methodology for Stochastic Learning Community Refinement**

Much like the BC clustering algorithm, Simulated Annealing (SA) also requires significant modification to the original partitioning algorithm in order to accommodate the changes brough forth by using student enrollment data. Specifically, SA must be modified to be able to search for and find the optimal combination of students in order to form LCs. The modified SA algorithm should also accept LC results from the previous clustering methods as a starting point, and refine the LCs by improving the configuration of students in each LC. Therefore, the SA algorithm was adapted to accept student LCs created from HC, MHC, or BC clustering algorithms, and would further improve upon the resulting quality of the LCs. Ultimately, the SA algorithm should complement the clustering algorithm in the creation of student LCs.

**4.4.1 Initial Input Cluster**

Due to the nature of SA as a global optimization algorithm, the quality of the starting point of the search is vital. A good starting solution, one that is already closer to the desired global optimum, will yield that optimum faster. In theory, any cluster configuration could be used as a starting point for SA, but clusters with poor quality, where they are not close to the optimum, will take much longer to be refined. Using the student enrollment for the 3$^{rd}$ year ENEL Fall 2020 dataset, the optimal LC configuration determined by the BC algorithm will be used as a starting point for SA.

Only the LCs produced by the BC algorithm will be used, because BC algorithm produces the highest quality LCs, as seen in Section 5.3.3.

**4.4.2 Implementing Simulated Annealing**

Defining the initial input cluster is merely the first step in modifying the SA algorithm (Table 7). As is the case with the BC clustering algorithm, various functional details of SA must be defined with respect to the goal of refining student LCs. Although several parameters need to be determined through experimentation, the cost and the perturbation functions can be directly defined in this use case.

The cost function serves as a measure for the quality of the cluster, where better clusters, or LCs, should yield lower cost values. For this application of SA, the cost function is defined as:

$$Cost = \sum_{j=1}^{n} (S_{E_j} - S_{I_j}) \tag{4.6}$$

where $Cost$ is the cost value, $S_{E_i}$ is the total number of external cluster connections between any given cluster $j$ and every other cluster in the set, and $S_{I_j}$ is the total number of internal cluster connections for any given cluster $j$, for $n$ clusters in a set. The goal of the cluster refinement remains the same: minimizing the number of external cluster connections while maximizing the number of internal cluster connections. Therefore, the cost function must also be minimized, where the lower the cost value, the higher the quality of the clusters. The cost function is related to the Total Cluster Quality function in equation (4.5) as follows:

$$Cost = -S_T \tag{4.7}$$

where $S_T$ is the total cluster quality score. The difference between the two functions resides in the fact that cost must be minimized for the SA algorithm to work, whereas the Total Cluster Quality function was designed to be maximized for ease of understanding. It is intuitively easier to associate a larger positive number for a cluster to the cluster being of a higher quality.

The change in cost is calculated as:

$$\Delta Cost = Cost_{trial} - Cost_{current} \qquad (4.8)$$

where $Cost_{trial}$ is the calculated cost of the trial solution, and $Cost_{current}$ is the cost of the current solution. A value of $\Delta Cost < 0$ indicates an increase in the quality of the clusters, and therefore the trial solution is considered better than the current solution, as it is closer to the global optimum. For example, the following table outlines the change in cost for a SA system:

*Table 8: Cost Change Example*

| $Cost_{current}$ | $Cost_{trial}$ | $\Delta Cost$ |
|---|---|---|
| -178 | -170 | 8 |
| -170 | -181 | -11 |

Table 8 provides two examples of a SA system's current and trial cost, as well as the change in cost. The first row's example yielded a $\Delta Cost$ of 8, therefore $\Delta Cost > 0$, and classifies the trial solution as worse than the current solution. While not preferable, a key characteristic for SA is that it is not a greedy algorithm. It would be willing to accept worse solutions, especially in the beginning, in order to find better solutions later on. As seen in the example in second row, the worse trial solution and cost from the first example was accepted, and now acts as the current cost of the system. The trial solution of the second example is better than the current solution, as $\Delta Cost < 0$. The acceptance of the trial solution relies on the value of $\Delta Cost$, but is stochastically determined by the Boltzmann acceptance criterion in equation (3.3).

During the perturbation phase, there are several methods that could be implemented to create a new clustering solution. The two primary techniques are insertion and swapping, as shown in Figure 7 below.

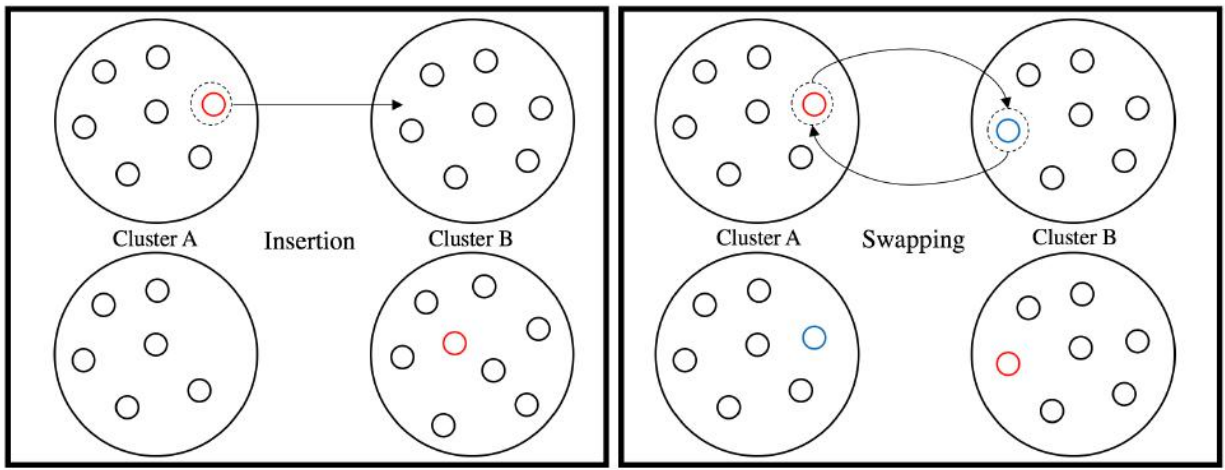

*Figure 7: Insertion (left) and Swapping (right) as Cluster Perturbation Techniques*

In this implementation, swapping was used to as the perturbation method to find better clusters, or better student LCs. Swapping was chosen to preserve the maximum size of the clusters, to improve upon the solution presented by the BC algorithm. The perturbation function implemented is a fully random swapping between any two students from any two LCs during any point of the SA run. The only limiting condition to swapping was that no two students from the same LC should be selected.

The insertion method would create an unbalanced cluster size within the set. If cluster size was factored into the cost function and optimized as a variable, insertion could be used as the perturbation function. However, the design of the clustering algorithm is to create similarly equal sized LCs, and the goal of the refining algorithm is to preserve this feature, therefore employing swapping as a perturbation technique, as to not exceed the set maximum cluster size. Keeping LCs at relatively similar sizes, and not exceeding a maximum cluster size, would prevent the creation of overly large LCs, as they can be detrimental to students by causing clique formation and isolation [65].

In addition to determining the cost and perturbation functions, there are five major parameters that must be set prior to an SA run, that would have a significant impact on the result. These five parameters are: number of initial perturbations (N), acceptance probability ($AP$), cooling rate ($\alpha$), minimum temperature ($T_{min}$), and maximum number of iterations at each temperature ($i_T$).

$AP$ is the probability at which an initial bad move would be accepted. A larger $AP$ means the SA algorithm will initially accept almost any bad move, allowing for a larger search window while looking for the global optimum. $AP$ is also the value critical in determining the initial temperature ($T_0$) of the SA run, since:

$$AP \leq e^{\frac{-\overline{\Delta Cost}}{T_0}} \qquad (4.9)$$

The preceding equation is equivalent to:

$$T_0 \leq \frac{-\overline{\Delta Cost}}{\ln(AP)} \qquad (4.10)$$

where $AP$ is the acceptance probability and $0 < AP < 1$, $T_0$ is the initial temperature of the system. In this case, $\overline{\Delta Cost}$ is the initial cost change. The value of $-\overline{\Delta Cost}$ can be computed as follows:

$$-\overline{\Delta Cost} = Cost_{initial} - \overline{Cost_{trial}} \tag{4.11}$$

where $-\overline{\Delta Cost}$ is the difference between the cost value of the starting point to the system ($Cost_{initial}$) and the average cost of the system after the first perturbation ($\overline{Cost_{trial}}$). Latter of the value can be calculated by:

$$\overline{Cost_{trial}} = \frac{\Sigma Cost_{trial}}{N} \tag{4.12}$$

where the average cost of the first perturbation ($\overline{Cost_{trial}}$) is the sum of costs to an initial perturbation ($Cost_{trial}$), averaged by the number of times an initial perturbation was made ($N$). In order to set the $T_0$ for a SA system, both $AP$ and $N$ must first be determined.

The temperature of a SA run is constantly changing, starting from $T_0$, and gradually cooled by the parameter $\alpha$. The temperature cooling function for SA is as follows:

$$T_{next} = \alpha \cdot T_{current} \tag{4.13}$$

where the next temperature of the system ($T_{next}$) is cooled from the current system temperature ($T_{current}$) at the rate $\alpha$, where $0 < \alpha < 1$. The value of $\alpha$ impacts the speed at which the temperature ($T$) cools, thus the speed at which the SA algorithm reaches termination. The smaller the $\alpha$ value, the faster the speed at which $T$ cools.

Eventually, the system will be cooled to a point where almost no new moves will be accepted, and thus the SA run must be terminated. The SA algorithm implemented is designed to be terminated once the system temperature reaches below a set $T_{min}$. The value of $T_{min}$ would impact the quality of the final solution, where if $T_{min}$ is too high, it could reduce the probability of better solutions being accepted. However, if $T_{min}$ is too low, it could add unnecessary runtime to the system, since new moves are hardly accepted at very low temperatures.

During the SA run, at each $T$, a set number of perturbations are made, set by $i_T$. The number of total moves in a single SA run is determined by $i_T$ multiplied by the number of times $T$ is reduced prior to reaching $T_{min}$. Therefore, the value of $i_T$ can impact the solution quality of SA the same way $T_{min}$ does. If $i_T$ is too small, it could cause an insufficient number of moves made in the search for the global optimum. However, if $i_T$ is too large, it could drastically increase system runtime, making the SA algorithm run very slow.

### 4.5 Experimental Design for Stochastic Learning Community Refinement

All five of the aforementioned parameters, being $N$, $AP$, $\alpha$, $T_{min}$, and $i_T$ must be experimentally determined in order to optimize the final output of the algorithm. The following experiments will be conducted, and the performance of impact of changing these parameters will be assessed using the Total Cluster Quality Score function (4.5), used to benchmark LC quality in Section 4.4. All experiments will be done using LCs produced by BC algorithm as a starting point, in conjunction with the 3rd year ENEL Fall 2020 dataset.

#### 4.5.1 Number of Initial Perturbations for Average Initial Cost

The average initial cost of the system is calculated by equation (4.12). In order to compute for $\overline{C_{trial}}$, the number $N$ initial swaps must be determined in order to consistently approximate some average initial cost value. Considering that SA is a stochastic algorithm, the cost value of the initial perturbation would differ depending which two random students are being swapped. Therefore, $N$ must be determined as the threshold at which a consistent $\overline{C_{trial}}$ can be computed for $N$ and every value larger than $N$.

In this experiment, for every $N$ value between 1 and 1000:

1. Performing a random swap on the starting cluster configuration $N$ times
2. Calculate the individual $C_{trial}$ value from each new cluster configuration
3. Average all $C_{trial}$ values over the N number of initial random swaps

The result of this experiment will yield the threshold $N$ value, in which a consistent $\overline{C_{trial}}$ value can therefore be approximated.

### 4.5.2 Initial Acceptance Probability

The relationship between $AP$ and the $T_0$ of the SA system is highlighted by equation (4.10). In order to compute for $T_0$, both $AP$ and $-\overline{\Delta Cost}$ must be determined. The previous experiment for $N$ will be used in the computation of $-\overline{\Delta Cost}$, but $AP$ must be experimentally determined.

The following experiment was conducted to examine the effect of a changing $AP$ value on the quality of the final LCs. Using $N$ determined from the previous experiment to calculate $\overline{C_{trial}}$ and $-\overline{\Delta Cost}$, the following parameters are set to explore the impact of the parameter $AP$:

- $\alpha = 0.95$
- $T_{min} = 0.0001$
- $i_T = 50$

A Monte Carlo simulation with 200 SA runs is executed for each experimental value of $AP$, and these simulations will be performed for the following AP values:

- $AP = 0.5$
- $AP = 0.6$
- $AP = 0.7$
- $AP = 0.8$
- $AP = 0.9$
- $AP = 0.95$
- $AP = 0.99$

The result of this experiment will examine the final quality of LCs refined by the SA algorithm, as impacted by the $AP$ value.

### 4.5.3 Cooling Rate

The relationship between $\alpha$ and the $T$ of a SA system is determined by equation (4.13). $\alpha$ controls the speed at which the SA system is cooled, as each new $T$ of the system is computed using $\alpha$. Using $N$ determined by the experiment in Section 4.5.1, the impact of $\alpha$ is experimentally examined using these following parameters:

- $AP = 0.95$
- $T_{min} = 0.0001$
- $i_T = 50$

A Monte Carlo simulation of 200 SA runs is executed for each value of $\alpha$, and the simulations will be performed for the following $\alpha$ values:

- $\alpha = 0.5$
- $\alpha = 0.6$
- $\alpha = 0.7$

- $\alpha = 0.8$
- $\alpha = 0.9$
- $\alpha = 0.95$
- $\alpha = 0.99$

The results of this experiment will determine an optimal $\alpha$ range, where the temperature is cooled at a rate that allows for the search of the optimal LC configuration without overextending the duration of the search.

### 4.5.4 Minimum Temperature

$T_{min}$ is the parameter responsible for terminating the SA run, where if $T$ of the system is reduced to below $T_{min}$, the system will stop, and the final LCs will be produced. Due to the design of the SA algorithm and how $T$ is cooled, $T$ would approach, but never reach, 0°C. Therefore, $T_{min}$ would also never be equal to or below 0°C, being $T_{min} > 0$. Determining the value of $T_{min}$ is a balance between prolonging the search for the global optimum without unnecessarily extending the runtime of the system. The impact of $T_{min}$ is experimentally examined using the following parameters:

- $AP = 0.95$
- $\alpha = 0.95$
- $i_T = 50$

The $N$ value determined by the experiment in Section 4.5.1 will be used. A Monte Carlo simulation with 200 SA runs is executed for each experimental value of $T_{min}$, and these simulations will be performed for the following $T_{min}$ values:

- $T_{min} = 10$

- $T_{min} = 5$
- $T_{min} = 1$
- $T_{min} = 0.1$
- $T_{min} = 0.01$
- $T_{min} = 0.001$
- $T_{min} = 0.00001$

The result of this experiment will determine the value of $T_{min}$ necessary to prolong the global optimum search without overextending the search duration.

### 4.5.5 Maximum Number of Iterations at Each Temperature

$i_T$ determines the number of moves made at each $T$ of a SA run. Therefore, the value of $i_T$, together with the number of times $T$ is reduced before $T_{min}$, influences the total number of moves made in a SA run. In theory, the more moves imply more chances for SA to find its global optimum. However, the acceptance of a new move is tied to only $T$. Once the system is cooled, the system will become very unlikely to accept new moves, no matter the $i_T$ value at that stage. Therefore, the following experiment is conducted to determine the impact of $i_T$ on the final cluster solution, using the following parameters:

- $AP = 0.95$
- $\alpha = 0.95$
- $T_{min} = 0.0001$

The $N$ value determined by the experiment in Section 4.5.1 will be used. A Monte Carlo simulation of 200 SA runs will be executed for each $i_T$ value, and these simulations will be performed for the following $i_T$ values:

- $i_T = 10$
- $i_T = 20$
- $i_T = 30$
- $i_T = 40$
- $i_T = 50$
- $i_T = 60$
- $i_T = 70$
- $i_T = 80$
- $i_T = 90$
- $i_T = 100$

The result of this experiment will determine the value of $i_T$ necessary to prolong the global optimum search without overextending the search duration.

## 4.6 Proposed Methodology for Learning Community Creation and Refinement Framework

This chapter discussed the LC creation process, both the enrollment network modeling and LC creation in section 4.2, as well as LC refinement in section 4.4. The LC creation process takes inspiration from VLSI clustering algorithms, whereas the LC refinement process uses a VLSI partitioning algorithm. Therefore, inspired by multilevel partitioning in Figure 3, all steps introduced thus far can be combined to create a multilevel framework to generate LCs from student enrollment data. The following Figure 8 proposes the complete enrollment data to LC processing framework.

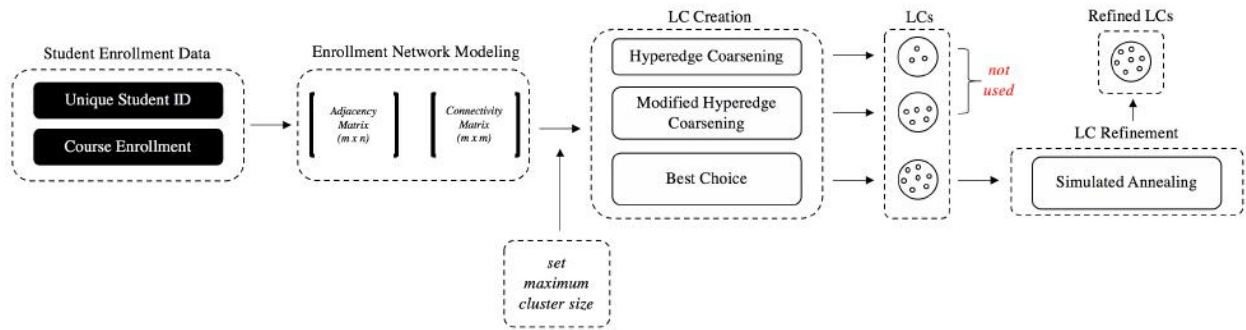

*Figure 8: Learning Community Creation and Refinement Framework*

The preceding Figure 8 outlines the entire LC creation and refinement process, from the raw student enrollment data obtained from the University Registrar to the refined LCs in the end. This framework can serve as the backbone for large-scale student LC creation used at the administration level of the university. Since the modified VLSI clustering algorithms prioritizes students' course enrollment as the clustering parameter, the algorithm should be able to perform clustering without distinguishing the students' faculty or department affiliations. Furthermore, SA will perform refinement on top of the clustering results as a mean of additional reassurance. Therefore, this framework can be implemented directly for large-scale LC creation without the need for additional data processing on the student enrollment data.

The capabilities of this framework's application on large enrollment datasets will be demonstrated using a Fall 2020 and Winter 2021 enrollment dataset released from the Office of the Registrar, containing 3939 students across five different faculties.

**4.7 Summary**

The goal of this chapter is to outline the proposed framework for VLSI-inspired LC creation and refinement, from raw student enrollment data to finalized LCs. This methodology is split into three stages: enrollment network modeling, optimized LC creation, and stochastic LC refinement. In the enrollment network modeling stage, a student enrollment network is constructed from the raw enrollment data. The network is clustered using the HC, MHC, and BC algorithms. The clustering results from the algorithms will be compared, and additional parameter-tuning experiments were designed for the BC algorithm. Resulting LCs can then be refined by SA, where the SA parameters will also be experimentally examined. Finally, the entire framework will be tested on a large student enrollment dataset. The experimental results will be presented in Chapter 5.

# Chapter 5 : Learning Community Creation and Refinement Results

## 5.1 Introduction

Chapter 4 introduced the modified VLSI-based methodology for creating and refining learning communities (LCs). It also introduced the experiments that must be performed for the modified algorithms in order to fine tune its various parameters. The goal of this chapter is to demonstrate and discuss the results from the experimentations highlighted in Chapter 4. The dataset used during the experimentation will also be visualized. In addition, the entire proposed framework will be tested on a large Fall 2020 and Winter 2021 student enrollment dataset. Ultimately, the results and discussion highlighted in this chapter will explore the viability of using VLSI-based algorithms to create and refine student LCs.

Parts of this chapter is published as ***Optimized Cohort Creation for Hybrid Online Design Learning During COVID-19*** in the *Proceedings for the 2021 American Society for Engineering Education (ASEE) Virtual Conference [81]*.

The remainder of this chapter is organized as follows: Section 5.2 visualizes the enrollment dataset used for the following experiments. Section 5.3 provides the results for the optimal LC creation experiments, as well as visualization of created LCs. Section 5.4 showcases the experimental results from the stochastic LC refinement stage, where the refined LCs are visualized. In Section 5.5, the LC Creation and Refinement Framework is tested on a large student enrollment network. The implications of the results in this chapter are discussed in Section 5.6. Finally, Section 5.7 summarizes the findings and contributions of this chapter.

## 5.2 Third Year Electrical Engineering Fall 2020 Enrollment Network

The dataset used in the following sections for the purpose of obtaining created and refined student LCs was a 3rd year electrical engineering (ENEL) student course enrollment dataset for Fall 2020. This official student enrollment data was released by the Registrar's Office. The dataset has been fully anonymized by the office, with the uniquely identifiable student IDs removed, and replaced with IDs for each student, increasing sequentially from one. From this dataset, there are 52 unique course sections registered between 81 students. The 3rd year ENEL Fall 2020 student enrollment dataset can be visualized through its Adjacency and Connectivity matrices.

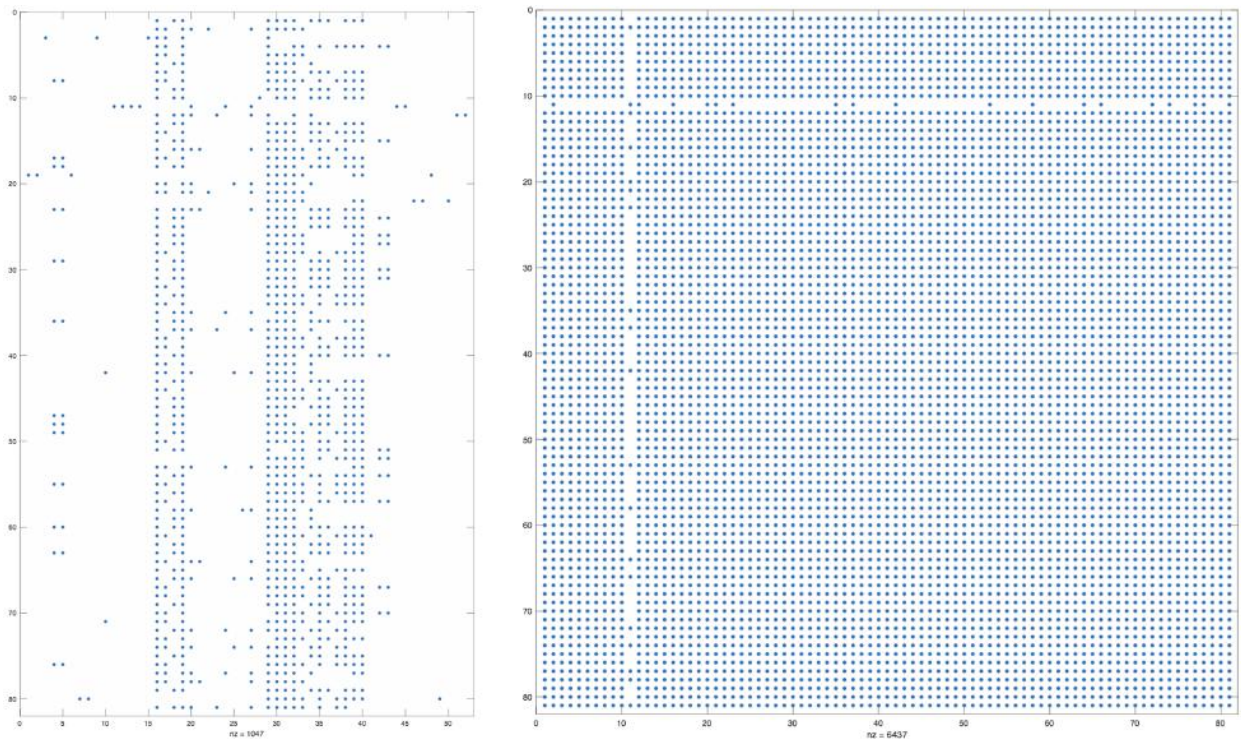

*Figure 9: Adjacency Matrix (left) and Connectivity Matrix (right) of 3rd Year ENEL Fall 2020 Student Enrollment (Fully Dense Network)*

Figure 9 shows the dense connectivity of this student enrollment network, where a dot is present if a student is enrolled in a course (left) or connected to another student (right). All students except for one (student ID #11) are connected to all other students. Student ID #11 shares common courses with only a few students, thus loosely connected to the network. In Figure 10, the connectivity network of the dataset is visualized.

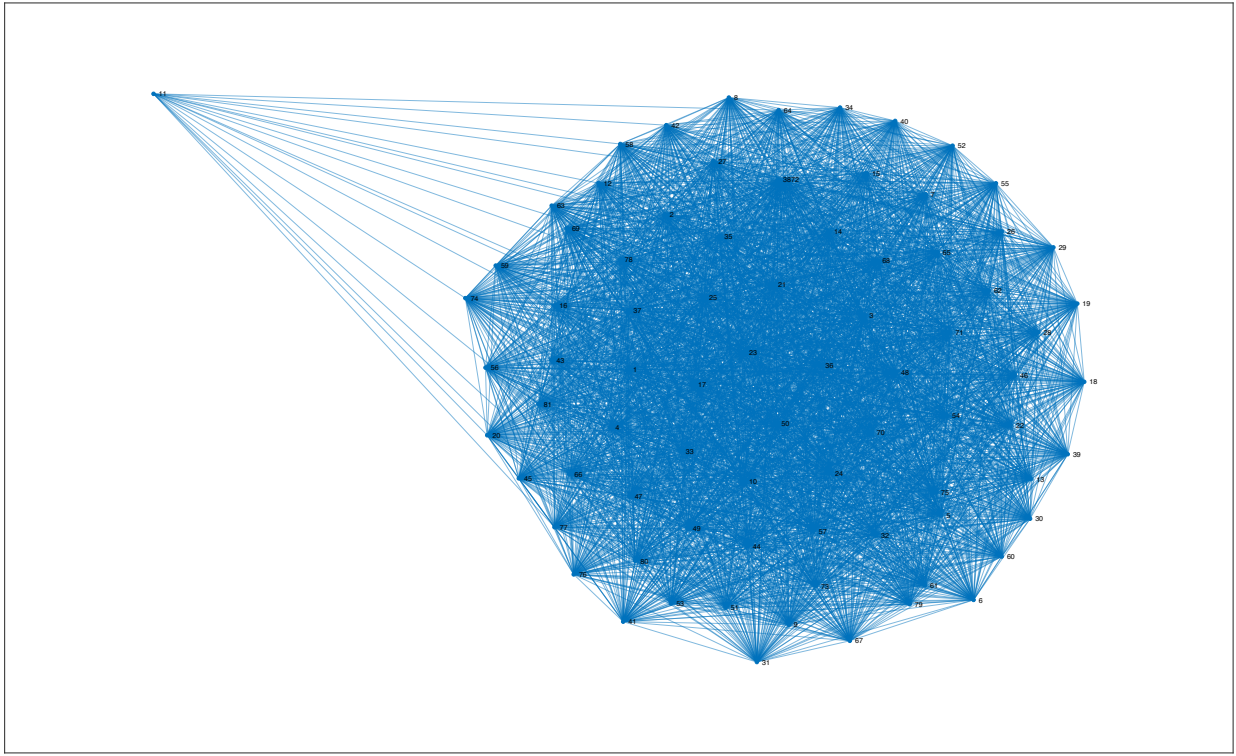

*Figure 10: Connectivity Graph of 3rd Year ENEL Fall 2020 Student Enrollment (Fully Dense Network)*

Figure 10 further illustrated the dense connectivity of this student network, where it can be classified as a Fully Dense Network. The density of the network poses significant challenges for any clustering algorithm. Good clustering results should maximize the number of connections within each cluster, while minimizing the number of connections between clusters. In this case, each cluster is a student LC. It would be difficult to minimize connections between clusters for a densely connected network, so it would be worthwhile to explore reducing the density of this network.

A Sparse Network can be formed from this student enrollment dataset, since a predominant number of connections between students were formed due to common large lectures, containing more than 30 student enrollments. With the large lectures removed as a connection between students, the Sparse Network can be visualized only through connections represented by tutorial and laboratory sections. The number 30 was chosen as it was the maximum number of students allowed in any room at the time of the Fall 2020 semester.

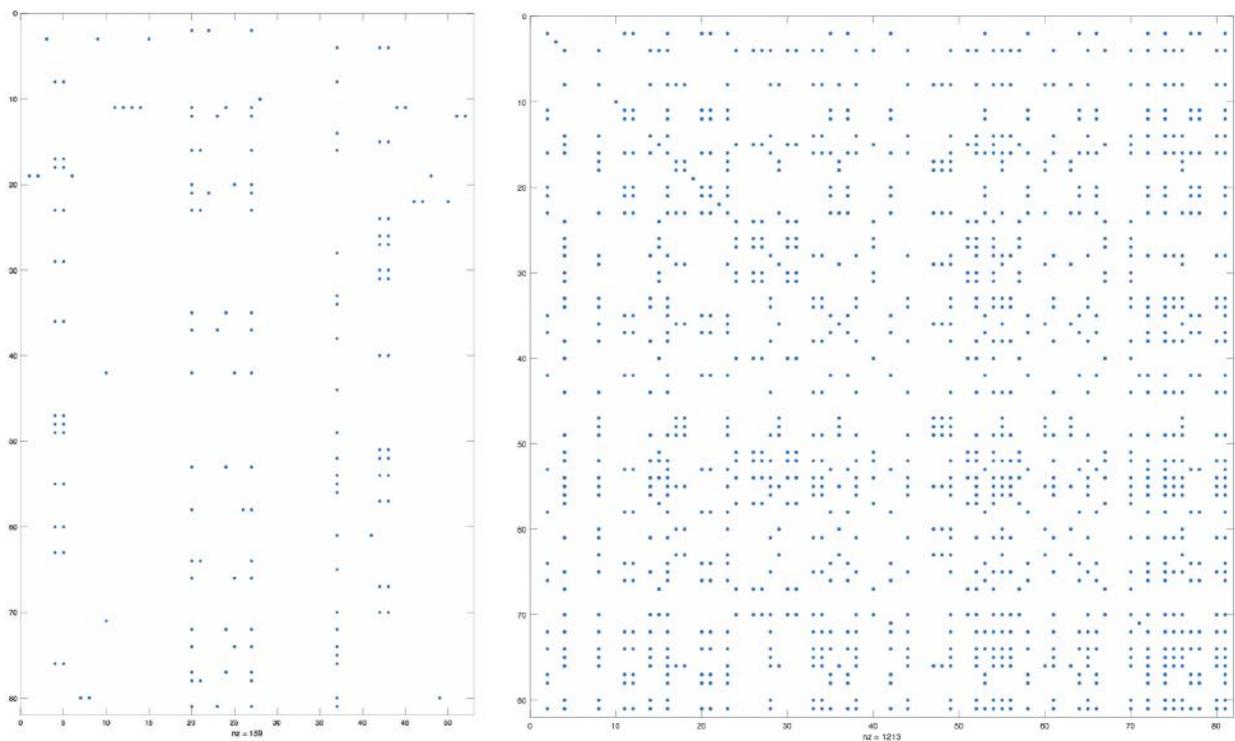

*Figure 11: Adjacency Matrix (left) and Connectivity Matrix (right) of 3rd Year ENEL Fall 2020 Student Enrollment (Sparse Network)*

Compared to Figure 9, Figure 11 showcases a much sparser Adjacency and Connectivity matrices. The student enrollment network became significantly less dense once the large lectures have been removed. This is further demonstrated in the following figure.

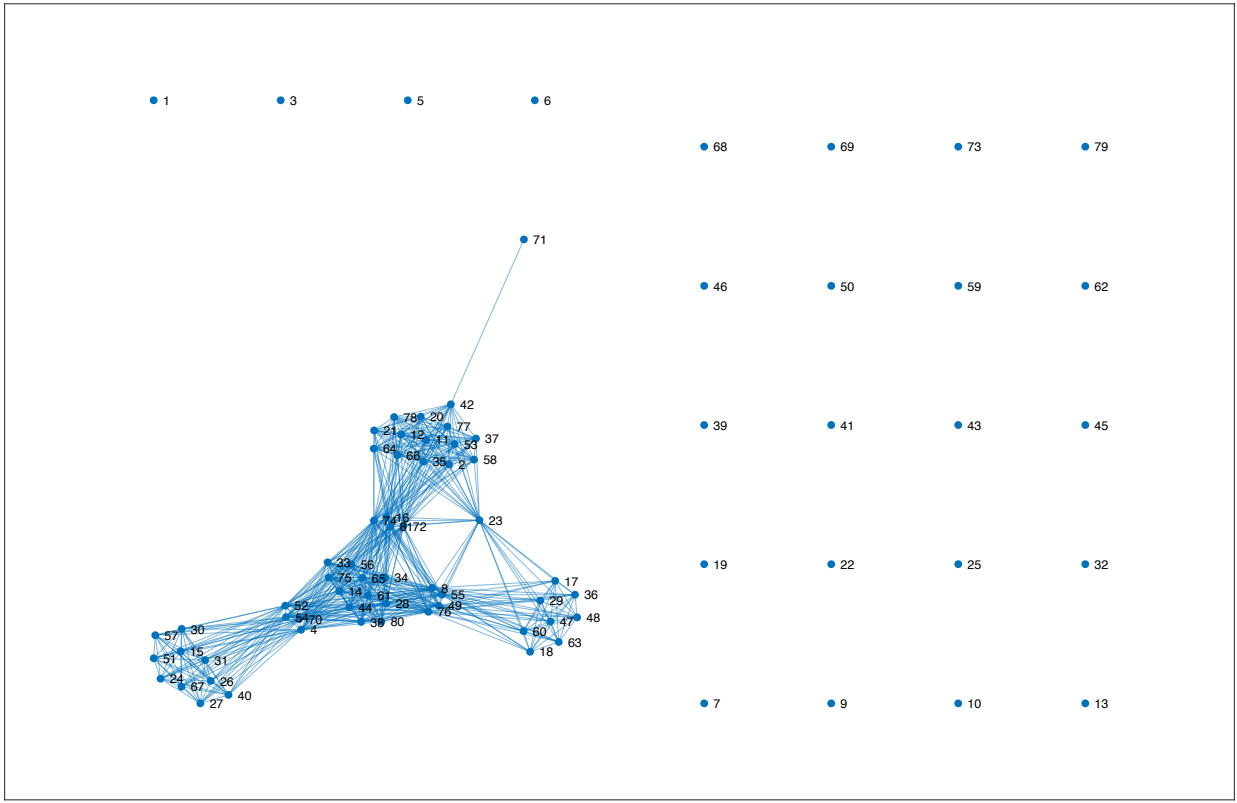

*Figure 12: Connectivity Graph of 3rd Year ENEL Fall 2020 Student Enrollment (Sparse Network)*

Figure 12 highlighted that 24 students are no longer connected within the network once large lectures are no longer used as connections. A Sparse Network could potentially allow the clustering algorithm to form more meaningful LCs, as the LCs are centered around their connections in smaller tutorial and laboratory settings, where they would have more opportunities for collaboration.

Two types of student enrollment networks can be produced based on this dataset, the Fully Dense Network with students connected by all course enrollments, and the Sparse Network with the large lectures removed. Both of these networks will be used by the following clustering algorithms.

## 5.3 Optimal Learning Community Creation Results

Following the network modeling of converting 3rd year ENEL Fall 2020 student course enrollment dataset into networks, they are clustered using Hyperedge Coarsening (HC), Modified Hyperedge Coarsening (MHC), and Best Choice (BC) algorithms. The implementation behind these algorithms is described in Sections 3.4 and 4.2. The following sections describes the clustering results using each of the three algorithms, as well as the experimental results for fine-tuning BC clustering parameters. The quality of the final clusters is evaluated using the Total Cluster Quality Score ($S_T$) Function in equation (4.5). Ultimately, the goal is to create high quality clusters that can be used as student LCs.

### 5.3.1 Hyperedge Coarsening Results

HC was implemented on the 3rd year ENEL student connectivity network, for both the Fully Dense Network and the Sparse Network. The clustering results for both type of networks was completely identical, meaning the density of the network connectivity has no impact on the usage and results of HC.

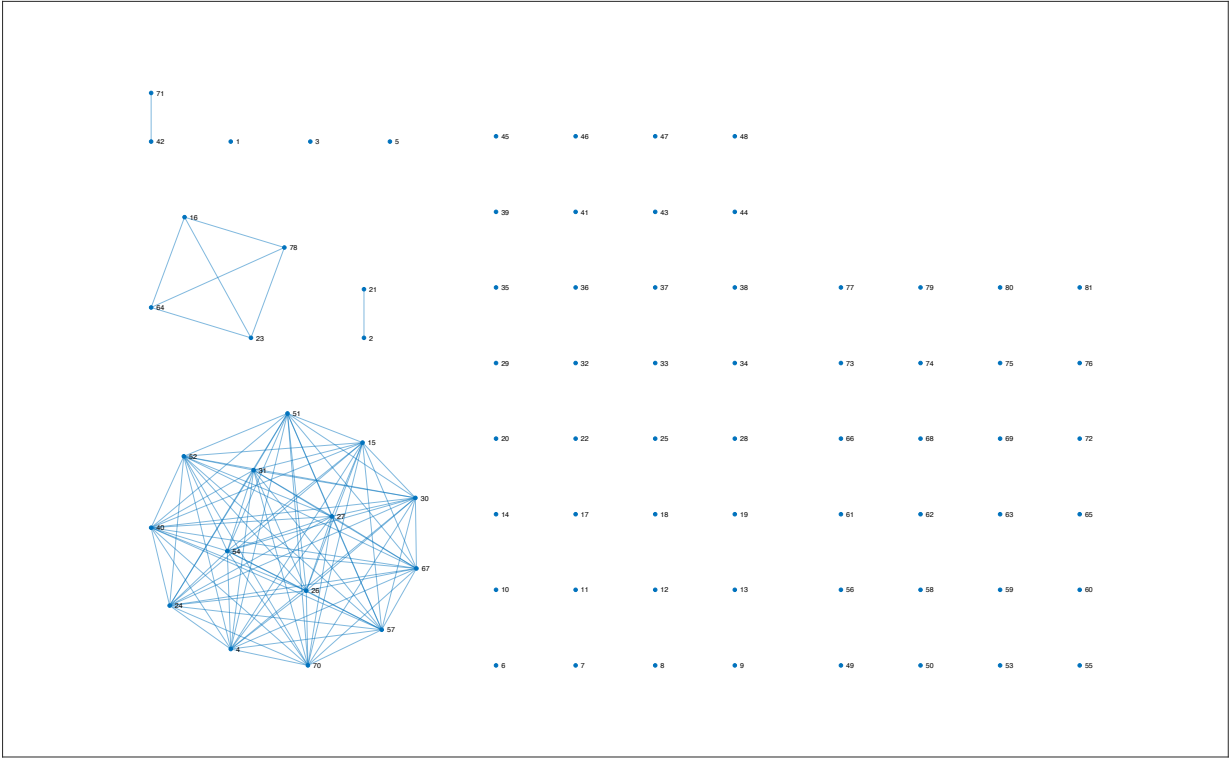

*Figure 13: Hyperedge Coarsening Learning Community Creation Results*

Figure 13 illustrates the clustering result using HC. HC produced very poor clustering results, where the algorithm was unable to form more than one major LC of 14 students. The remaining 67 students were not clustered, or clustered as pairs or a group of four. The resulting LCs using HC yielded $S_T$ = -34122. Since the quality of a cluster is measured by maximizing $S_T$, HC failed to produce high quality clusters for student LC creation. The following Table 9 tabulates the full clustering results using the HC algorithm.

*Table 9: Hyperedge Coarsening Learning Community Creation Results*

| LC ID | Learning Community Membership |||||||||||||
|---|---|---|---|---|---|---|---|---|---|---|---|---|
| 1 | 19 |||||||||||||
| 2 | 80 |||||||||||||
| 3 | 11 |||||||||||||
| 4 | 58 |||||||||||||
| 5 | 10 |||||||||||||
| 6 | 22 |||||||||||||
| 7 | 12 |||||||||||||
| 8 | 42 |||||| 71 ||||||
| 9 | 2 |||||| 21 ||||||
| 10 | 16 ||| 23 ||| 64 ||| 78 |||
| 11 | 4 | 15 | 24 | 26 | 27 | 30 | 31 | 40 | 51 | 52 | 54 | 57 | 67 |
| 12 | 1 |||||||||||||
| 13 | 3 |||||||||||||
| 14 | 5 |||||||||||||
| 15 | 6 |||||||||||||
| 16 | 7 |||||||||||||
| 17 | 8 |||||||||||||
| 18 | 9 |||||||||||||
| 19 | 13 |||||||||||||
| 20 | 14 |||||||||||||

| | |
|---|---|
| 21 | 17 |
| 22 | 18 |
| 23 | 20 |
| 24 | 25 |
| 25 | 28 |
| 26 | 29 |
| 27 | 32 |
| 28 | 33 |
| 29 | 34 |
| 30 | 35 |
| 31 | 36 |
| 32 | 37 |
| 33 | 38 |
| 34 | 39 |
| 35 | 41 |
| 36 | 43 |
| 37 | 44 |
| 38 | 45 |
| 39 | 46 |
| 40 | 47 |
| 41 | 48 |
| 42 | 49 |
| 43 | 50 |
| 44 | 53 |
| 45 | 55 |
| 46 | 56 |
| 47 | 59 |
| 48 | 60 |
| 49 | 61 |
| 50 | 62 |
| 51 | 63 |
| 52 | 65 |
| 53 | 66 |
| 54 | 68 |
| 55 | 69 |
| 56 | 72 |
| 57 | 73 |
| 58 | 74 |
| 59 | 75 |
| 60 | 76 |
| 61 | 77 |
| 62 | 79 |
| 63 | 81 |

The poor clustering result using HC is as expected, due to the nature of the algorithm. All cells in this student enrollment network are connected via hyperedges, and given the denseness of the network, cells will be connected by multiple hyperedges. In this case, cells are students, and hyperedges are the common courses students are enrolled in. Given HC's tendency to skip all cells if one has been previously marked, it is expected to be unable to cluster many cells. Therefore, it is difficult to use the resulting clusters as LCs for students, since most of the students are not grouped together. Overall, HC is not an ideal clustering method for this type of densely connected network.

**5.3.2 Modified Hyperedge Coarsening Results**

MHC was also implemented for both the Fully Dense Network and the Sparse Network for the ENEL Fall 2020 dataset. Once again, the clustering results for both type of networks was completely identical. MHC was also not affected by the density of network connectivity.

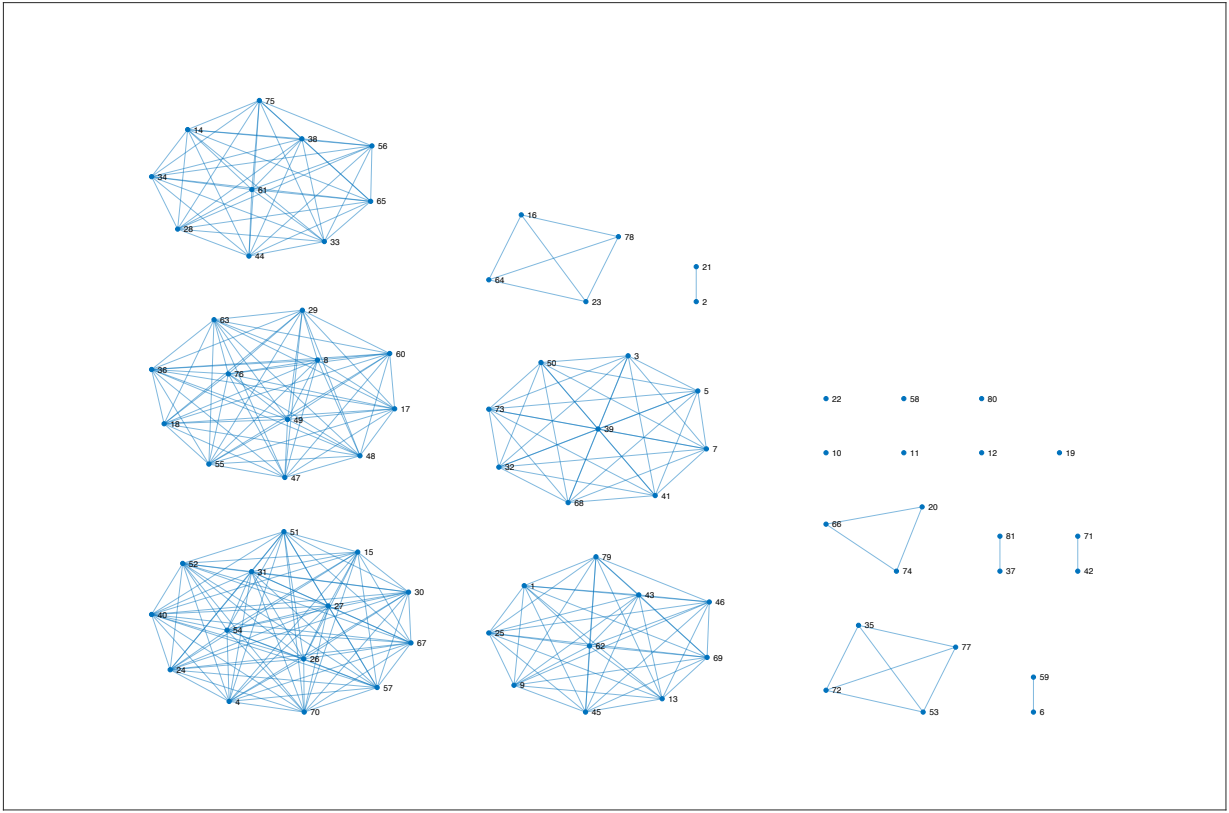

*Figure 14: Modified Hyperedge Coarsening Learning Community Creation Results*

The preceding Figure 14 demonstrated the MHC clustering results, using the Sparse Network with large lectures removed. The MHC algorithm produced five major LCs with varying sizes, and left 26 students not clustered, or paired in groups of two, three, and four. The score for MHC's resulting clusters is $S_T = -1906$. The score also proves that MHC produced better clusters than HC. The full clustering result using MHC is tabulated in the following Table 10:

*Table 10: Modified Hyperedge Coarsening Learning Community Creation Results*

| LC ID | Learning Community Membership |
|---|---|
| 1 | 19 |
| 2 | 80 |
| 3 | 11 |
| 4 | 58 |
| 5 | 10 |
| 6 | 22 |
| 7 | 12 |
| 8 | 42, 71 |
| 9 | 2, 21 |
| 10 | 81, 37 |
| 11 | 16, 64, 78, 23 |
| 12 | 74, 20, 66 |
| 13 | 72, 53, 35, 77, 48, 17, 18, 49, 55, 60, 29, 63 |
| 14 | 36, 8, 76, 47, 15, 51, 52, 54, 24, 57, 26, 27, 30, 31 |
| 15 | 67, 4, 70, 40, 75, 44, 14, 56, 28, 61 |
| 16 | 33, 34, 65, 38, 39, 7, 41, 73, 50 |
| 17 | 32, 3, 68, 5, 13, 45, 46, 79, 25, 62 |
| 18 | 1, 69 |
| 19 | 59, 6 |

MHC produced better clustering results than HC, by yielding five sizable clusters instead of one, and the results yielded a much larger $S_T$ value when compared to HC. This method worked better than HC because MHC can revisit the hyperedges skipped by HC and cluster the unmarked cells accordingly. This resulted in more LCs being formed compared to the results of HC. While MHC appears to work better than MHC, especially for this densely connected network, the resulting LCs are uneven in size, thus lower in quality. Therefore, it is still difficult to use the clustering results produced by MHC as student LCs.

### 5.3.3 Best Choice Algorithm Results

BC was implemented for the same student connectivity network as HC and MHC in the previous sections. As referenced in Section 4.3, two clustering parameters must be fine-tuned. The following sections will explore the experimental results of the optimal maximum cluster size and number of Monte Carlo simulation runs needed for BC. The quality of the clusters is evaluated using the Total Cluster Quality Score in equation (4.5), where high quality clusters can be used as student LCs.

*5.3.3.1 Optimal Maximum Cluster Size for Best Choice Algorithm Results*

This experiment aims to discover the relationship between the maximum cluster size and the quality of the resulting clusters. In this case, each resulting set of clusters can possibly act as a student LC configuration. A 200-run Monte Carlo simulation was used in these experiments, as it was assumed to be a reasonable number of runs to allow the best $S_T$ to be found. Both the Linear ($d_l$) and the Nonlinear ($d_n$) Clustering Score Functions in equations (4.3) and (4.4) were implemented and evaluated based on the quality of the resulting clusters.

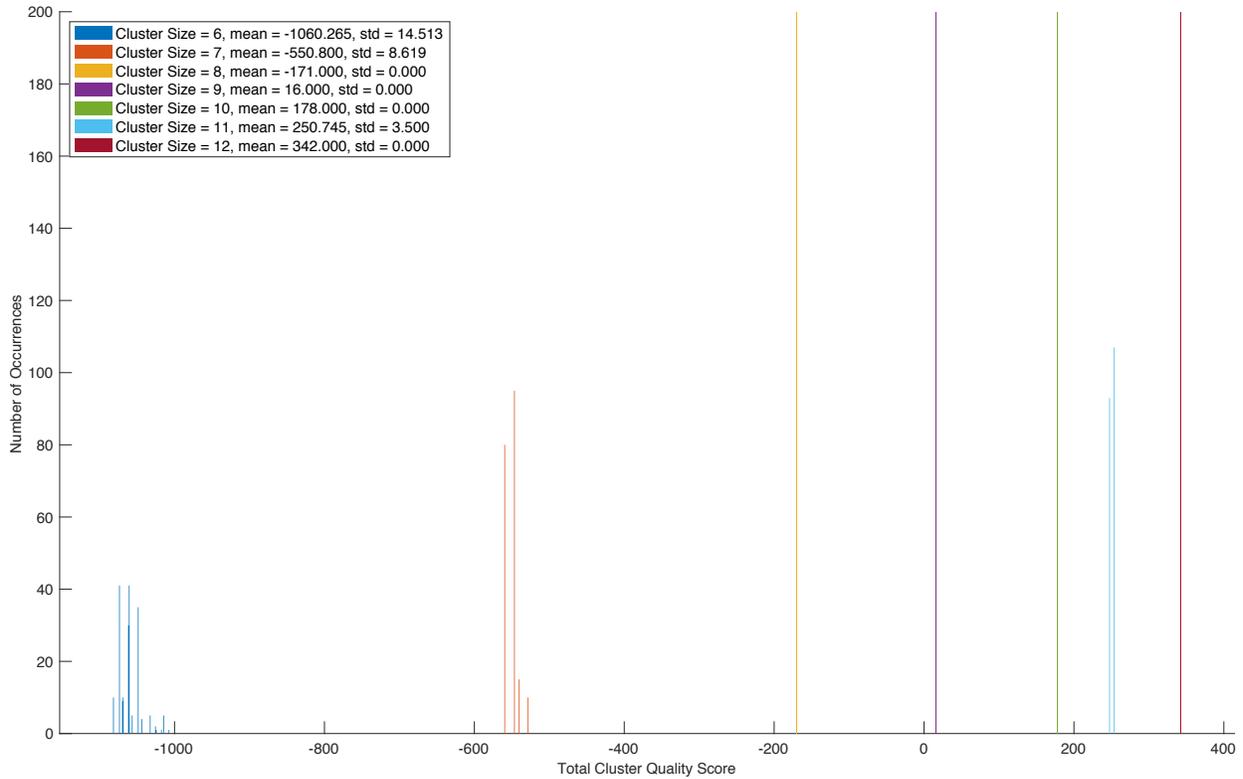

*Figure 15: Histogram of Maximum Cluster Size Results, using Linear Clustering Score Function (Fully Dense Network)*

Figure 15 captures the results for the maximum cluster size experiment on the Fully Dense Network using $d_l$ as a combined histogram, where a 200-run Monte Carlo simulation was performed at each maximum cluster size. The mean and standard deviation (std) of each simulation was also visualized. Using $d_l$, cluster quality visibly grows with the maximum cluster size, whereas the maximum cluster size increases, the mean $S_T$ also grows. In addition, the standard deviation for each simulation shows that for many maximum cluster sizes, there is only one cluster solution.

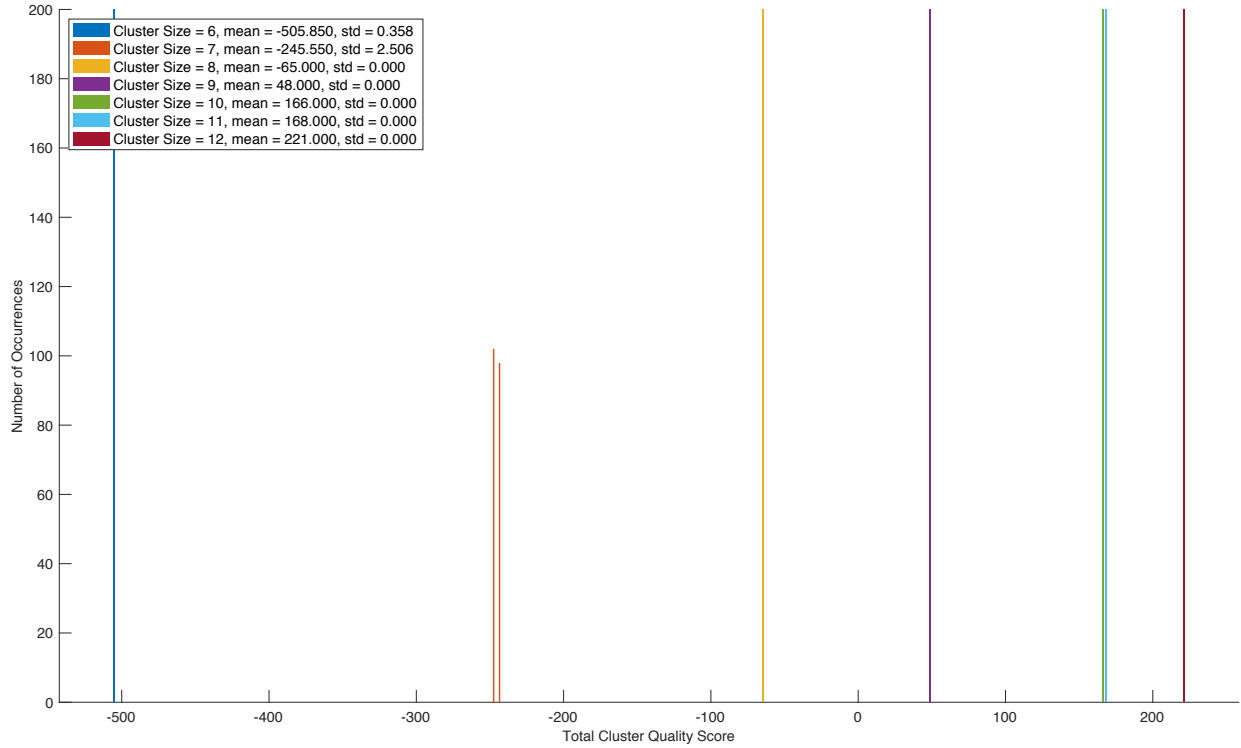

*Figure 16: Histogram of Maximum Cluster Size Results, using Linear Clustering Score Function (Sparse Network)*

Figure 16 captures the results for the maximum cluster size experiment on the Sparse Network using $d_l$. The results are represented as a combined histogram of 200-run Monte Carlo simulations performed for each maximum cluster size. Much like the similar figure in Figure 15, maximum cluster size is correlated with $S_T$, where cluster quality increases when maximum cluster size increases. Also, at many maximum cluster sizes, there is only one solution. In fact, when compared to the Fully Dense Network, the use of the Sparse Network yielded more instances where only one clustering solution is present at each maximum cluster size.

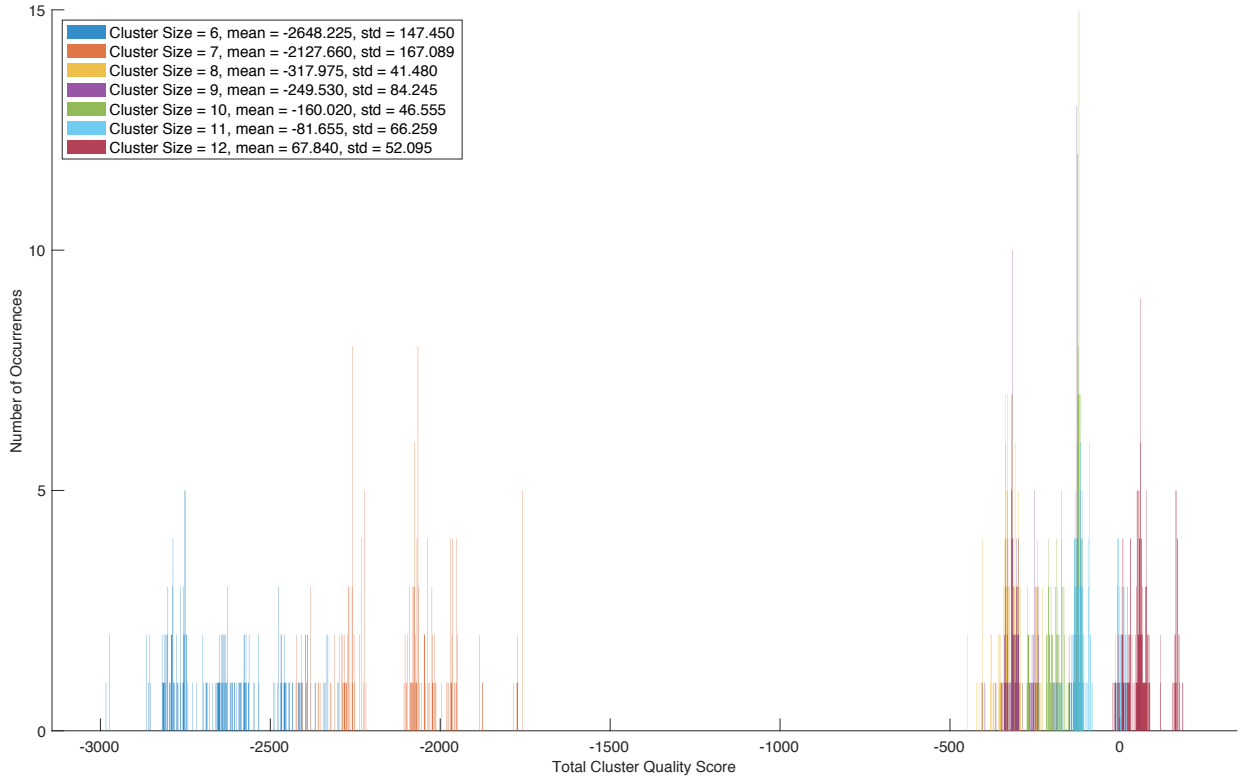

*Figure 17: Histogram of Maximum Cluster Size Results, using Nonlinear Clustering Score Function (Fully Dense Network)*

In a similar histogram, Figure 17 captures the results for the maximum cluster size experiment on the Fully Dense Network using $d_n$. Following the same trend as the previous figures, $S_T$ grows as maximum cluster size increases, indicating that cluster quality is improving as the clusters gets larger. However, unlike the previous figures which used $d_l$, the use of $d_n$ yielded varied results at each maximum cluster size, demonstrated by the standard deviation. Therefore, there are less guarantee when it comes to the quality of the clustering results when using $d_n$, especially when compared to using $d_l$.

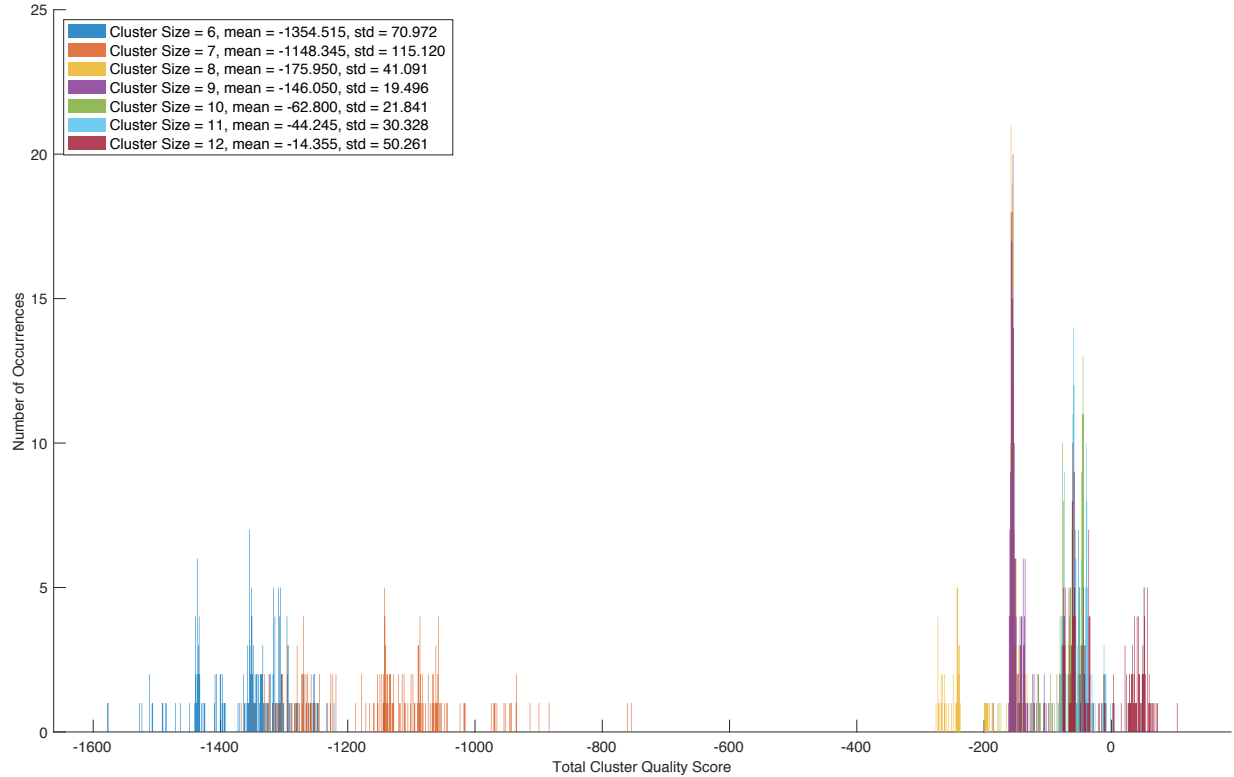

*Figure 18: Histogram of Maximum Cluster Size Results, using Nonlinear Clustering Score Function (Sparse Network)*

Figure 18 captures the results of the maximum cluster size experiment on the Sparse Network using $d_n$. Much like Figure 17 before it, $S_T$ is positively correlated to maximum cluster size, where the quality of the cluster increases as size increases. Also similar to the previous figure, the cluster quality varies in each Monte Carlo simulation, as seen by the standard deviation value. However, when compared to the previous figure using the Fully Dense Network, this set of results using the Sparse Network yielded results with smaller standard deviations. For $d_n$, the use of the Sparse Network produces more consistent results at each maximum cluster size. This is likely due to the fact that there are overall less connections in the Sparse Network, thus removing some alternative options for clusters to be formed. If there are less options for cluster formation, there would be more consistently formed clusters.

The following Table 11 and Table 12 tabulates the results of the previous four figures. The results presented in these tables are visualized in the figures to follow.

*Table 11: Best Choice Maximum Cluster Size Results, using Linear Clustering Score Function*

| Network Type | Maximum Cluster Size | Mean $S_T$ | Best $S_T$ | Standard Deviation of $S_T$ |
|---|---|---|---|---|
| Fully Dense | 6 | -1060.265 | -1007 | 14.513 |
| Sparse | 6 | -505.85 | -505 | 0.358 |
| Fully Dense | 7 | -550.8 | -528 | 8.619 |
| Sparse | 7 | -245.55 | -243 | 2.56 |
| Fully Dense | 8 | -171 | -171 | 0 |
| Sparse | 8 | -65 | -65 | 0 |
| Fully Dense | 9 | 16 | 16 | 0 |
| Sparse | 9 | 48 | 48 | 0 |
| Fully Dense | 10 | 178 | 178 | 0 |
| Sparse | 10 | 166 | 166 | 0 |
| Fully Dense | 11 | 250.745 | 254 | 3.5 |
| Sparse | 11 | 168 | 168 | 0 |
| Fully Dense | 12 | 342 | 342 | 0 |
| Sparse | 12 | 221 | 221 | 0 |

*Table 12: Best Choice Maximum Cluster Size Results, using Nonlinear Clustering Score Function*

| Network Type | Maximum Cluster Size | Mean $S_T$ | Best $S_T$ | Standard Deviation of $S_T$ |
|---|---|---|---|---|
| Fully Dense | 6 | -2648.225 | -2299 | 147.45 |
| Sparse | 6 | -1354.515 | -1142 | 70.972 |
| Fully Dense | 7 | -2127.66 | -1757 | 167.089 |
| Sparse | 7 | -1148.345 | -754 | 115.12 |
| Fully Dense | 8 | -317.975 | -228 | 41.48 |
| Sparse | 8 | -175.95 | -132 | 41.091 |
| Fully Dense | 9 | -249.53 | -116 | 84.245 |
| Sparse | 9 | -146.05 | -48 | 19.496 |
| Fully Dense | 10 | -160.02 | -110 | 46.555 |
| Sparse | 10 | -62.8 | -12 | 21.841 |
| Fully Dense | 11 | -81.655 | 59 | 66.259 |
| Sparse | 11 | -44.245 | 52 | 30.328 |
| Fully Dense | 12 | -67.84 | 186 | 52.095 |
| Sparse | 12 | -14.355 | 104 | 50.261 |

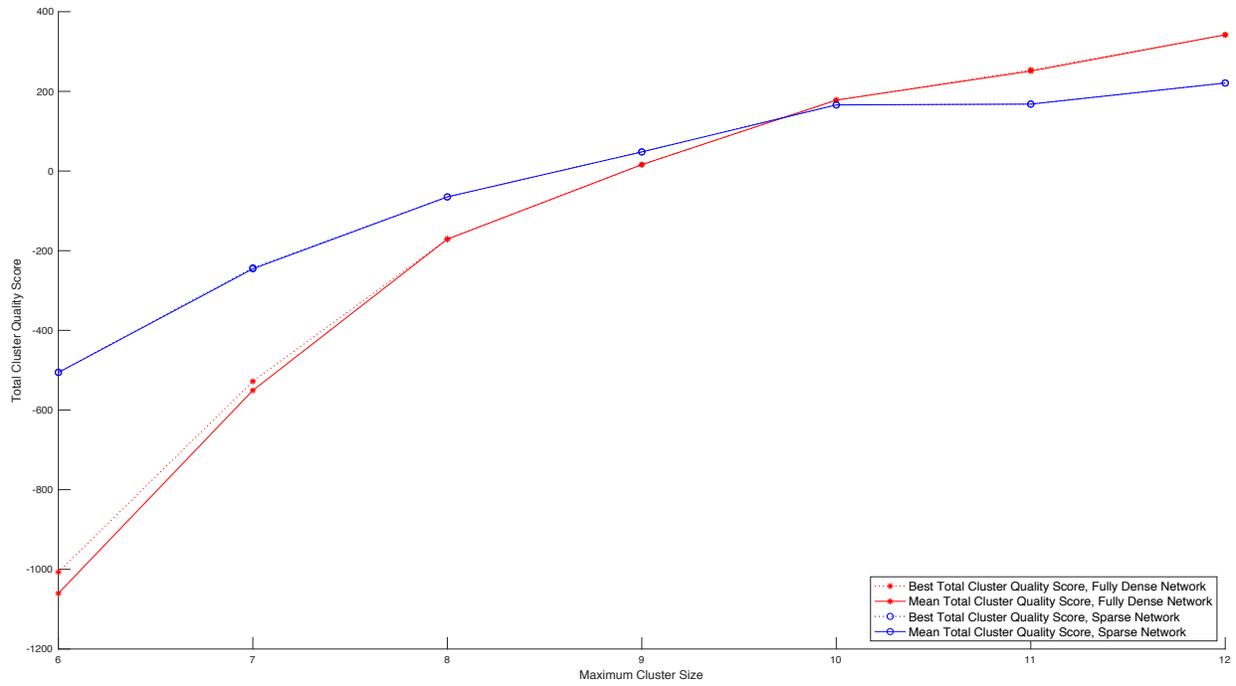

*Figure 19: Maximum Cluster Size Results, using Linear Clustering Score Function*

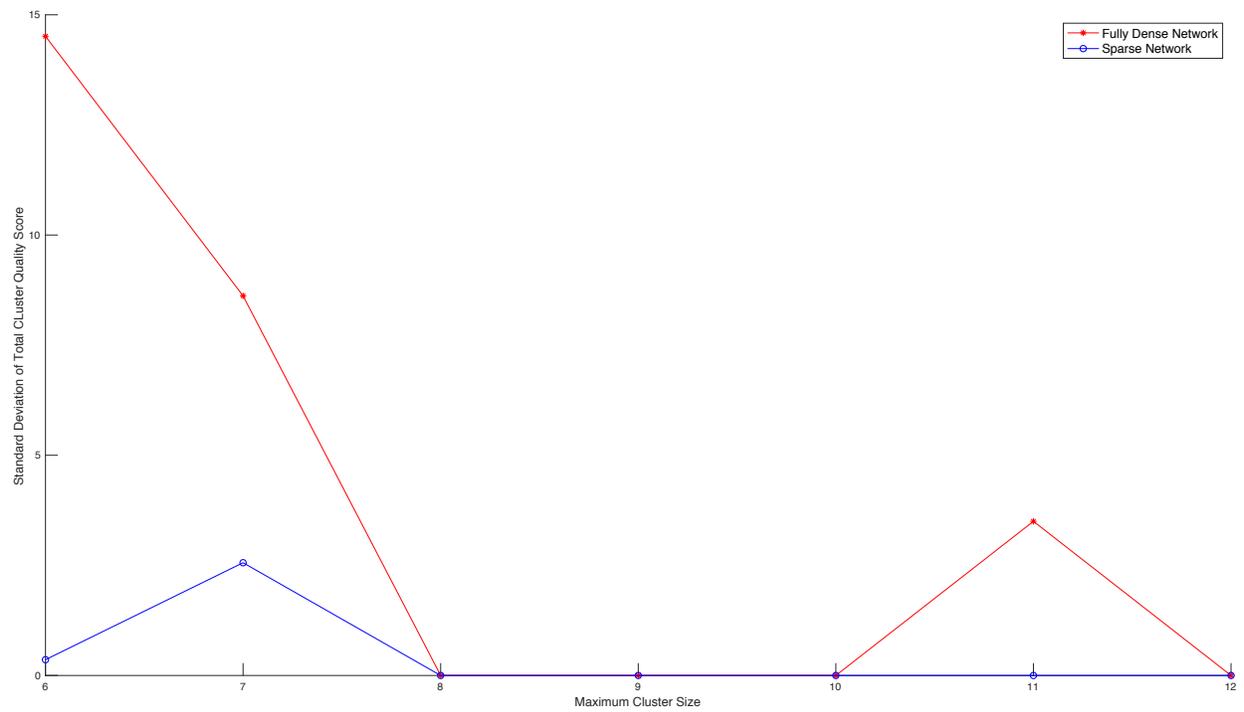

*Figure 20: Standard Deviation of Maximum Cluster Size Results, using Linear Clustering Score Function*

Figure 19 compares the maximum cluster size of the clustering solution with its respective mean $S_T$ and best $S_T$ obtained in the Monte Carlo simulation. Figure 20 compares the same maximum cluster size with the standard deviation of $S_T$ obtained from the simulation. The clusters were defined using $d_l$, for maximum cluster sizes from six to twelve. The results for using the Fully Dense Network and the Sparse Network are represented separately in the figures as red and blue respectively. From Figure 19, it can be observed that $S_T$ increases as the maximum cluster size increases. Furthermore, there are very little differences between the mean $S_T$ and best $S_T$ for each Monte Carlo simulation. Figure 20 demonstrated that the standard deviation of the Monte Carlo simulations ran on the Fully Dense Network was larger than those ran on the Sparse Network. In addition, as cluster size increases, standard deviation for $S_T$ tends to decrease, but most of the Monte Carlo simulations yielded a standard deviation of 0. These observations reflect those from the histograms in Figure 15 and Figure 16.

While the $S_T$ sees rapid improvement for smaller cluster sizes, as the maximum cluster size reaches nine and larger, the improvement of the $S_T$ becomes less significant. Also, the algorithm appears to perform much better on the Sparse Network for smaller clusters. Once the maximum cluster size reaches beyond nine, the algorithm performs slightly better on the Fully Dense Network. While larger cluster sizes would yield less contact between clusters, the trade-off becomes that a very large cluster represents a very large student LC, which would not be beneficial for their learning.

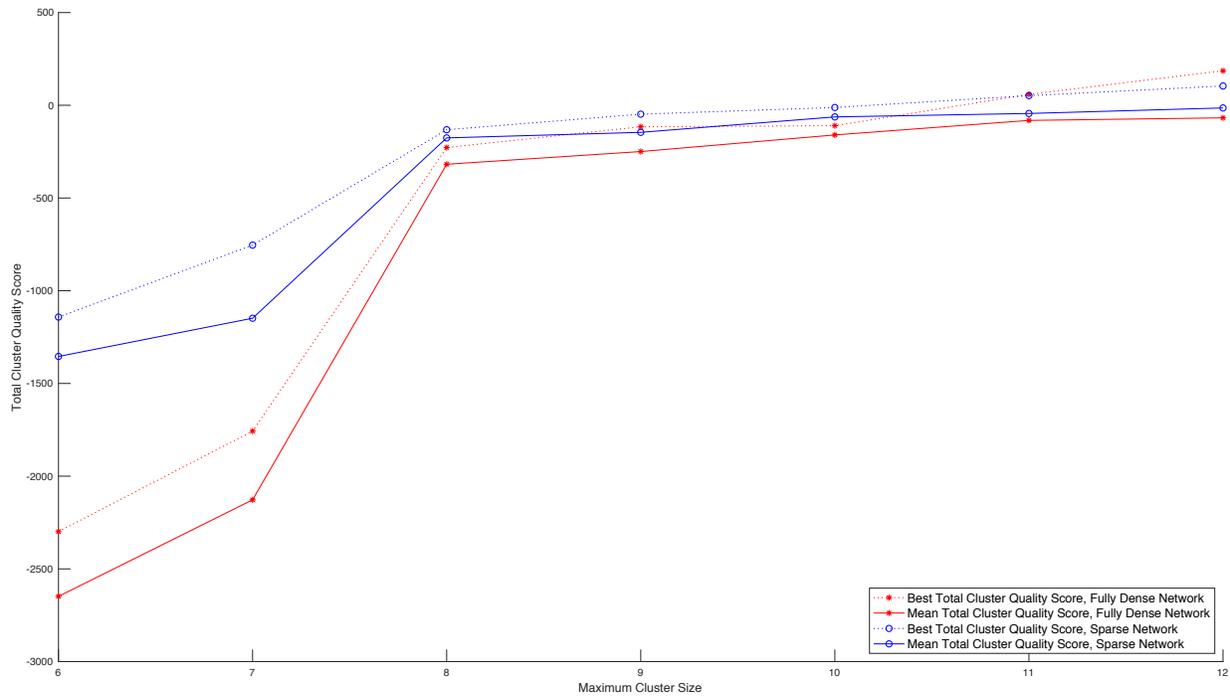

*Figure 21: Maximum Cluster Size Results, using Nonlinear Clustering Score Function*

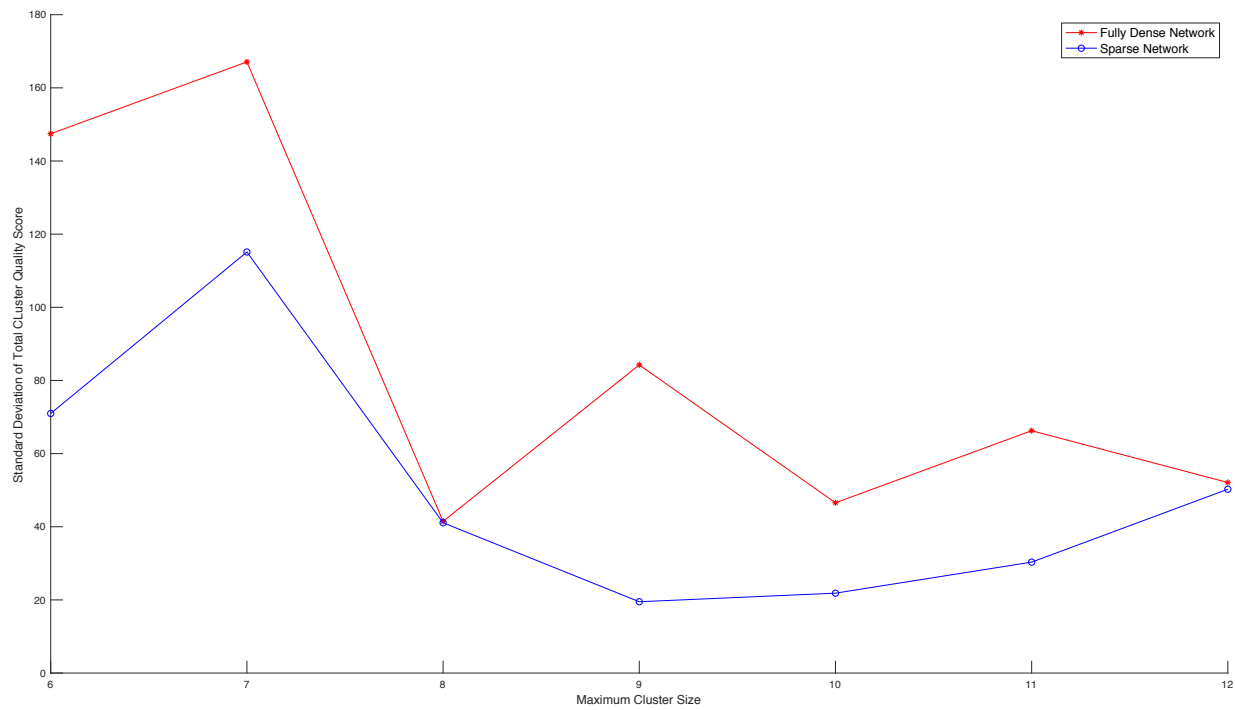

*Figure 22: Standard Deviation of Maximum Cluster Size Results, using Nonlinear Clustering Score Function*

In Figure 21, the maximum cluster size with the mean $S_T$ and the best $S_T$ of each Monte Carlo simulation is compared, where the results were generated using $d_n$. Figure 22 compares the standard deviation of $S_T$ for each Monte Carlo simulation at different maximum cluster sizes. Overall, the results of Figure 21 generated using $d_n$ demonstrated the same trends as Figure 19, where the results were generated using $d_l$. As the maximum cluster sizes increase, the quality of the generated clusters also increases, as indicated by both the mean and the best $S_T$ in the figure. However, compared to results using $d_l$, small clusters created using $d_n$ demonstrated poorer quality, especially when applied to the Fully Dense Network. In addition, the mean $S_T$ and best $S_T$ hardly overlaps for this set of results. This fact is equally reflected in Figure 22, where the standard deviation of $S_T$ are significant at each maximum cluster size. The large standard deviation leads to a large gap between the mean $S_T$ and the best $S_T$ for each set of results. It should be noted that much like Figure 20, standard deviation of $S_T$ when Fully Dense Network was used is larger than when the Sparse Network was used. These results reflect the findings in the histograms in Figure 17 and Figure 18.

As the cluster size increased, the quality of the clusters improved dramatically, especially when cluster size changed between seven to eight. Overall, the algorithm performed better for the Sparse Network when using $d_n$. The performance difference was more significant when the maximum cluster sizes are small, but it became less apparent once the size reached past the threshold of eight. When using $d_n$, the size of the cluster matters significantly, as it intrinsically links to the quality of the final result.

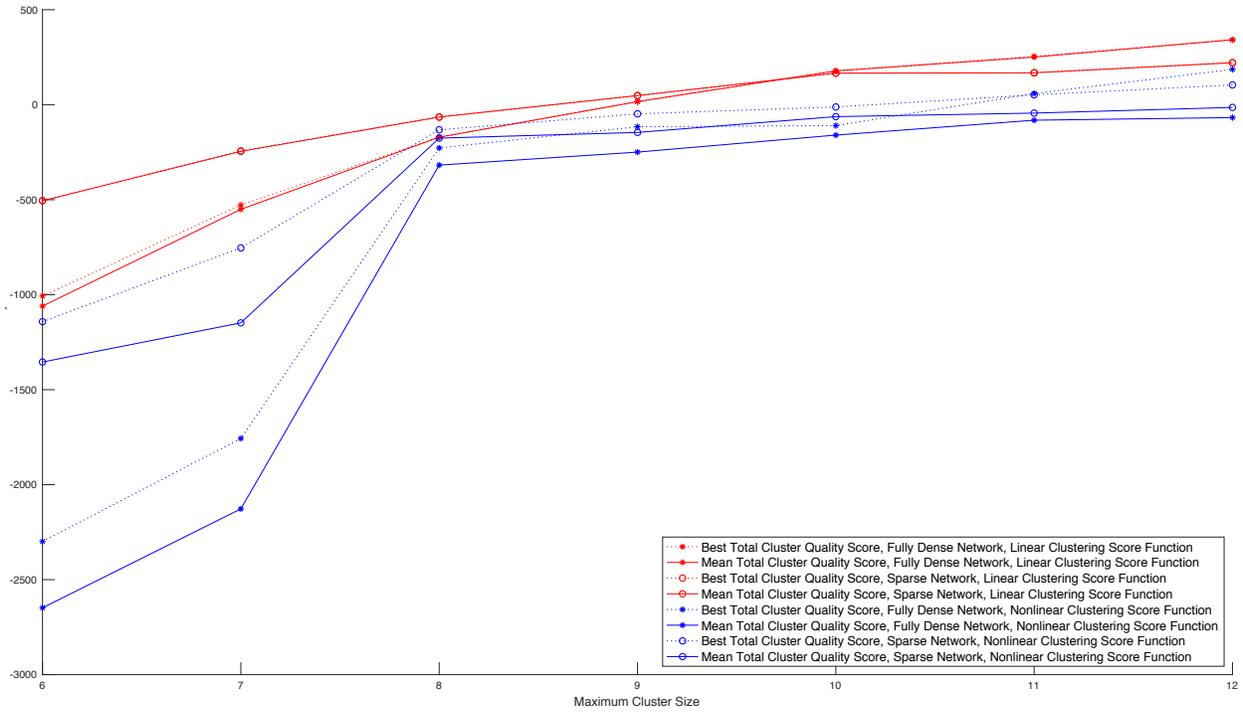

*Figure 23: Maximum Cluster Size Results*

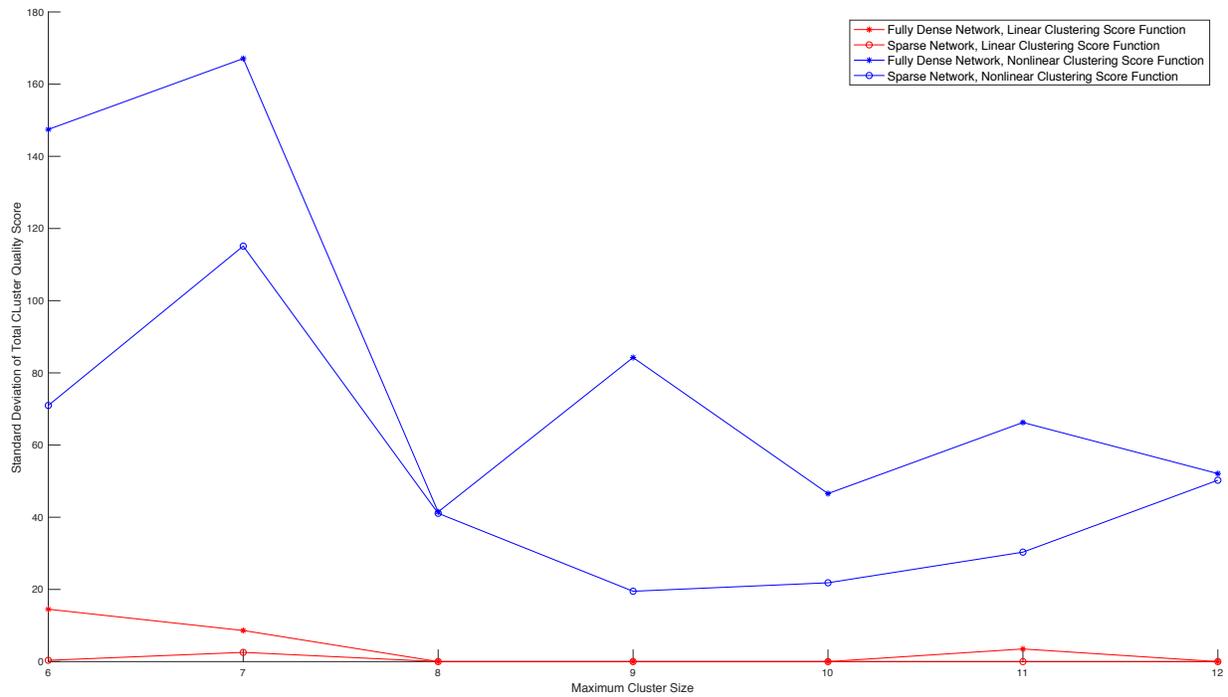

*Figure 24: Standard Deviation of Maximum Cluster Size Results*

The preceding Figure 23 combines the results in the previous Figure 19 and Figure 21, while Figure 24 combines the results from Figure 20 and Figure 22. The results generated using $d_l$ are in red, while the results generated using $d_n$ are in blue. Overall, $d_l$ tends to yield better clustering results than $d_n$, especially for the smaller cluster sizes. For cluster sizes eight and above, the performance becomes more similar. In addition, the results generated using $d_l$ shows much smaller standard deviation for each set of simulations, when compared to $d_n$. Therefore, results generated with $d_l$ will be more consistent in quality.

Considering these results, a maximum cluster size of nine or ten would be ideal for this network. The clusters, thus student LCs, should be generated using $d_l$. For $d_l$, the type of network does not matter, as both the Fully Dense and the Sparse networks can produce results with comparable quality at a maximum cluster size of nine or ten. If $d_n$ must be used, the same recommendation of a maximum cluster size of nine or ten would also suffice. However, $d_n$ should only be used on the Sparse Network, as it consistently produces clusters with better quality when compared to the Fully Dense Network. Cluster sizes of nine or ten is an ideal compromise of where it is large enough to generate reasonable quality LCs, without being too large and defeating the purpose of forming these student LCs.

*5.3.3.2 Optimal Number of Monte Carlo Simulation Runs for Best Choice Algorithm Results*

The aim of this experiment is to determine the number of runs in a Monte Carlo simulation required to yield the best clusters, or student LCs, produced by the BC algorithm, as determined by the number of runs at which the best $S_T$ was reached. A maximum cluster size of nine was chosen as a medium-sized benchmark to run the experiments, whereas in theory, any reasonable cluster size would suffice. This experiment hopes to answer whether having more runs in a Monte Carlo simulation for the BC algorithm would lead to better student LC creation.

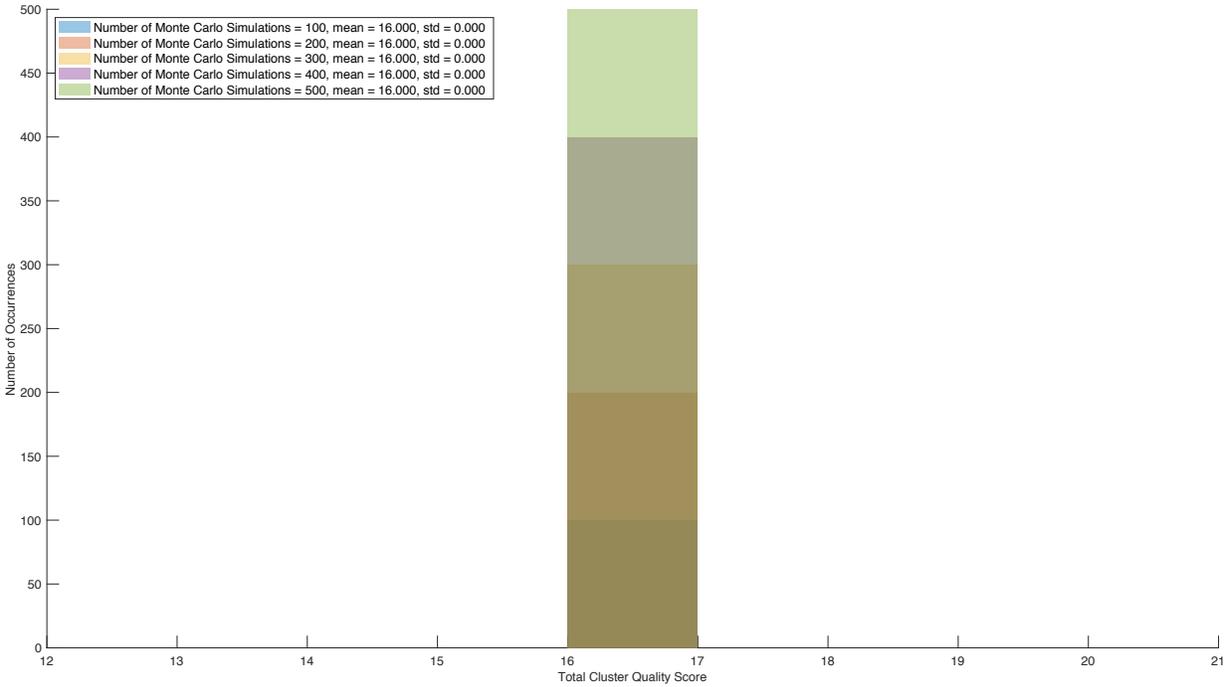

*Figure 25: Histogram of Number of Monte Carlo Runs Results, using Linear Clustering Score Function (Fully Dense Network)*

Figure 25 compares the results of differently sized Monte Carlo simulations ran using $d_l$, for a maximum cluster size of nine, for the Fully Dense Network. Regardless of how many runs are in the Monte Carlo simulation, the mean $S_T$ is the same, where mean $S_T = 16$. The standard deviation of $S_T$ for each set of simulation is also 0. For this case, no matter how large the Monte Carlo simulation is, there will only be one set of clusters produced for using $d_l$ and a maximum cluster size of nine when the Fully Dense Network is used.

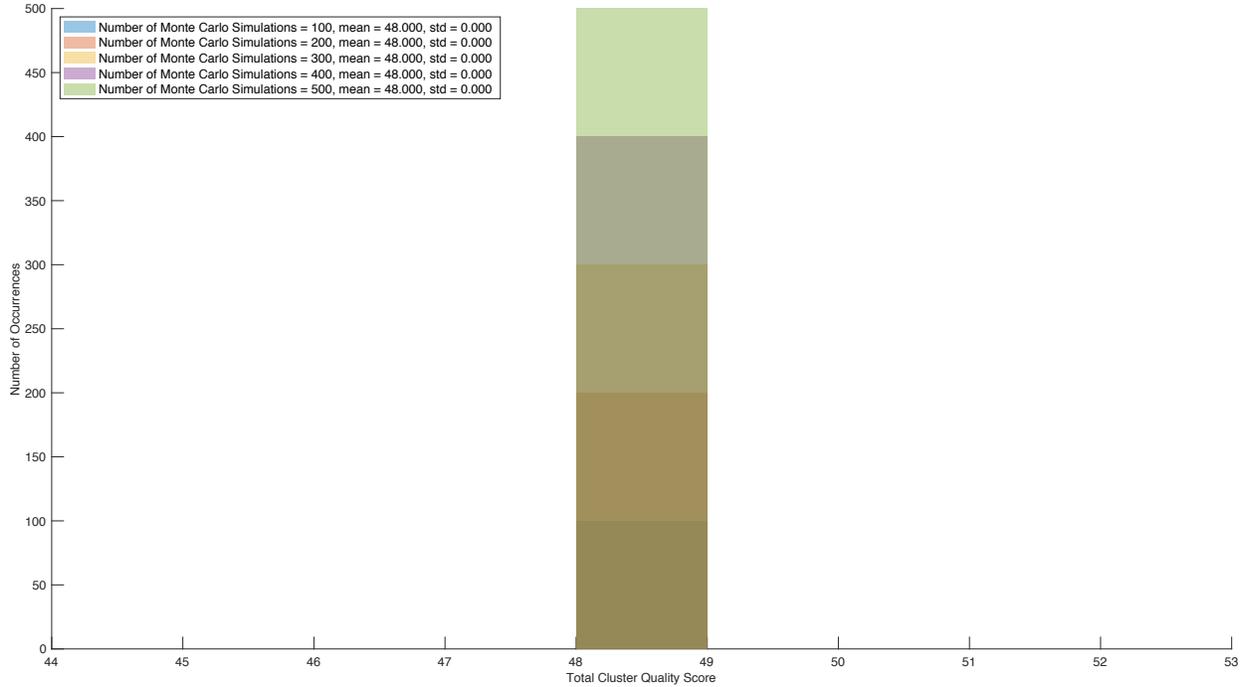

*Figure 26: Histogram of Number of Monte Carlo Runs Results, using Linear Clustering Score Function (Sparse Network)*

Figure 26 compares the results of differently sized Monte Carlo simulations ran using $d_l$, for a maximum cluster size of nine, for the Sparse Network. Much like Figure 25 previously, the mean $S_T$ stays the same as the number of runs in the Monte Carlo simulation grows. In this case, the mean $S_T = 48$. The standard deviation of $S_T$ is also 0. Once again, despite the size of the Monte Carlo simulation, there will only be one set of clustering results produced for $d_l$ and a maximum cluster size of nine when the Sparse Network is used.

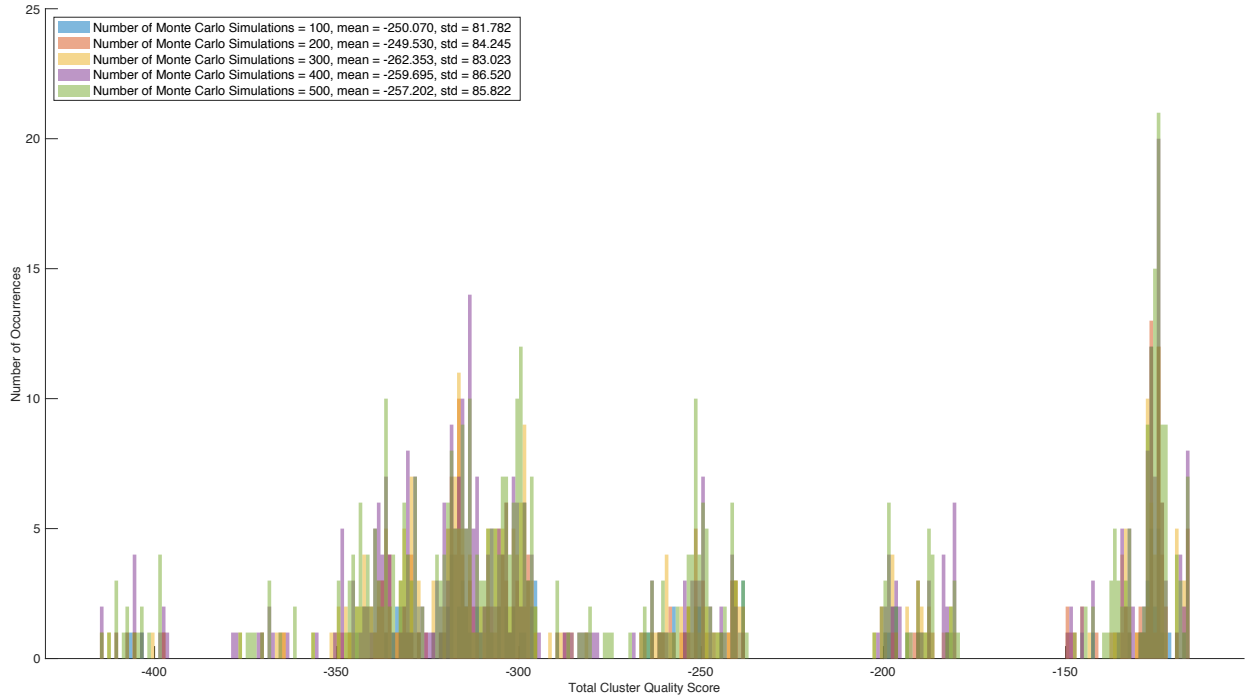

*Figure 27: Histogram of Number of Monte Carlo Runs Results, using Nonlinear Clustering Score Function (Fully Dense Network)*

Figure 27 compares the results of differently sized Monte Carlo simulations ran using $d_n$, for a maximum cluster size of nine, for the Fully Dense Network. Unlike the previous results generated using $d_l$, this set of results show a varied mean $S_T$ and standard deviation of $S_T$ as the number of runs in the Monte Carlo simulation changes. There is not an observable trend between the size of the Monte Carlo simulation and the overall quality of the resulting clusters. From the figure, the mean $S_T$ appears to fluctuate between values within a certain range, whereas the standard deviation of $S_T$ seems to increase slightly as the number of runs in the Monte Carlo simulation increased. While not neatly, the results from each Monte Carlo simulation seem to generally overlap in the histogram.

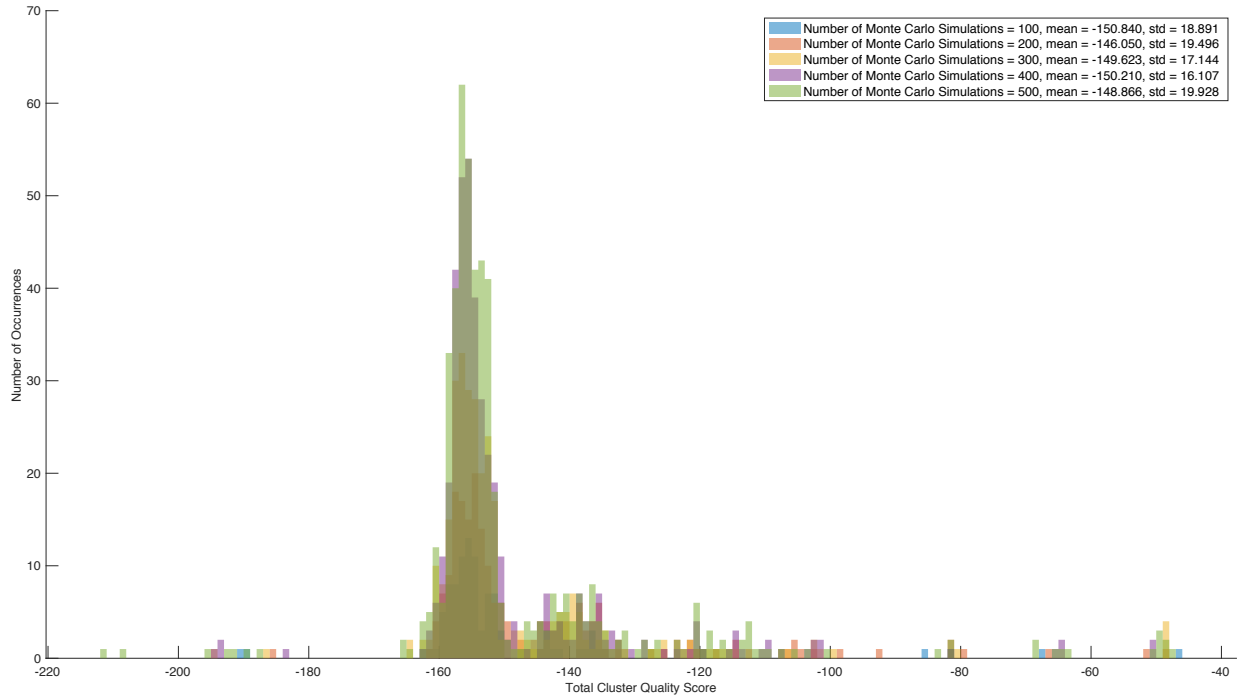

*Figure 28: Histogram of Number of Monte Carlo Runs Results, using Nonlinear Clustering Score Function (Sparse Network)*

Figure 28 compares the results of differently sized Monte Carlo simulations ran using $d_n$, for a maximum cluster size of nine, for the Sparse Network. This set of results also showed a varied mean $S_T$ and standard deviation of $S_T$. However, when compared to the previous figure, the clusters generated are more similar, as reflected by the smaller standard deviation in this figure. The value of mean $S_T$ also seem to fluctuate less, when compared to Figure 27. Once again, there is not necessarily an observable trend between the number of Monte Carlo simulation runs and the overall quality of the resulting clusters. From the histogram, the results from each set of Monte Carlo simulation does appear to generally overlap.

The following Table 13 and Table 14 tabulates the results of the previous four figures. The results presented in these tables are visualized in figures to follow.

*Table 13: Best Choice Number of Monte Carlo Runs Results, using Linear Clustering Score*

*Function*

| Network Type | Number of Monte Carlo Simulation Runs | Mean $S_T$ | Best $S_T$ | Standard Deviation of $S_T$ |
|---|---|---|---|---|
| Fully Dense | 100 | 16 | 16 | 0 |
| Sparse | 100 | 48 | 48 | 0 |
| Fully Dense | 200 | 16 | 16 | 0 |
| Sparse | 200 | 48 | 48 | 0 |
| Fully Dense | 300 | 16 | 16 | 0 |
| Sparse | 300 | 48 | 48 | 0 |
| Fully Dense | 400 | 16 | 16 | 0 |
| Sparse | 400 | 48 | 48 | 0 |
| Fully Dense | 500 | 16 | 16 | 0 |
| Sparse | 500 | 48 | 48 | 0 |

*Table 14: Best Choice Number of Monte Carlo Runs Results, using Nonlinear Clustering Score Function*

| Network Type | Number of Monte Carlo Simulation Runs | Mean $S_T$ | Best $S_T$ | Standard Deviation of $S_T$ |
|---|---|---|---|---|
| Fully Dense | 100 | -250.070 | -117 | 81.782 |
| Sparse | 100 | -150.840 | -46 | 18.891 |
| Fully Dense | 200 | -249.530 | -116 | 84.245 |
| Sparse | 200 | -146.050 | -48 | 19.496 |
| Fully Dense | 300 | -262.353 | -117 | 83.023 |
| Sparse | 300 | -149.623 | -48 | 17.144 |
| Fully Dense | 400 | -259.695 | -116 | 86.520 |
| Sparse | 400 | -150.210 | -50 | 16.107 |
| Fully Dense | 500 | -257.202 | -116 | 85.822 |
| Sparse | 500 | -148.866 | -47 | 19.928 |

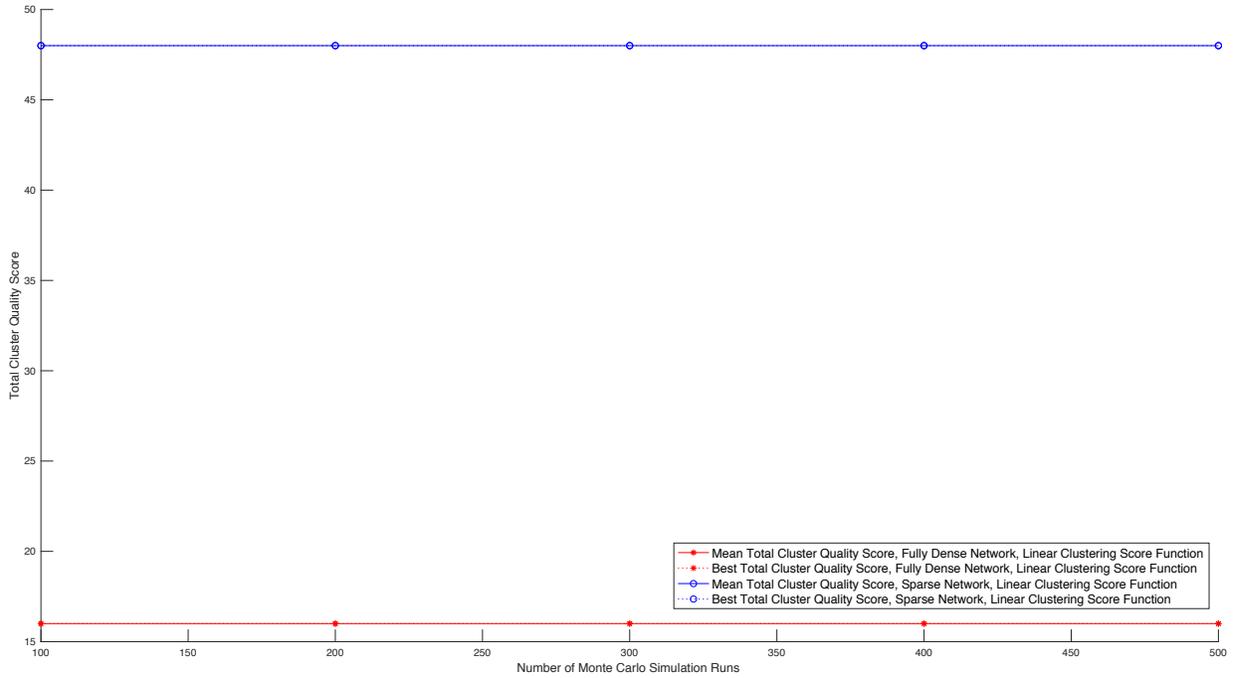

*Figure 29: Number of Monte Carlo Runs Results, using Linear Clustering Score Function*

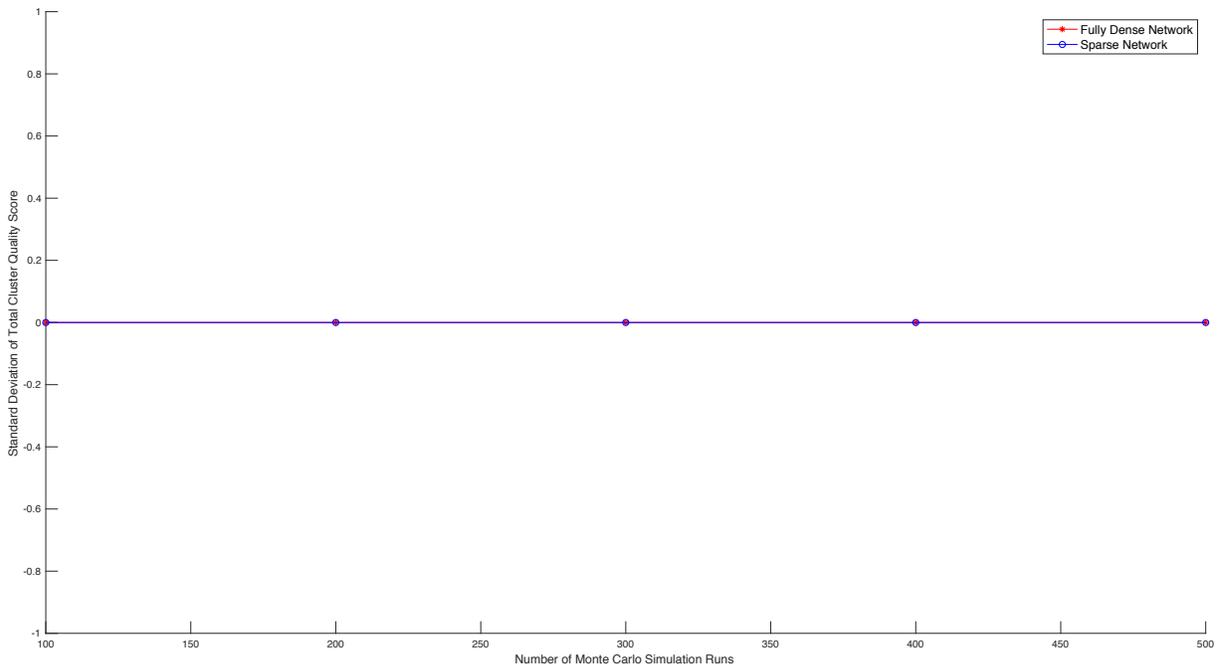

*Figure 30: Standard Deviation of Number of Monte Carlo Runs Results, using Linear Clustering Score Function*

Figure 29 compares the size of the Monte Carlo simulations with the mean $S_T$ and best $S_T$ of each simulation, produced using $d_l$. Figure 30 compares the size of the Monte Carlo simulations with the standard deviation of $S_T$ produced in each simulation. For both figures, Fully Dense Network results are displayed in red, whereas the Sparse Network results are displayed in blue. These figures compare the results represented in Figure 25 and Figure 26. From Figure 29, it can be seen that the mean $S_T$ overlaps with the best $S_T$ for both the Fully Dense and the Sparse networks. The Sparse Network produced better results than the Fully Dense Network, as the $S_T$ value for the former is larger. However, the value of $S_T$ did not change as the size of the simulations changed. For this case of using $d_l$ to produce clusters with the maximum size of nine, the number of runs in the Monte Carlo simulation does not matter, because only one set of solution is every produced under these conditions. This fact is further corroborated by Figure 30, demonstrating that none of the simulations had a standard deviation value larger than 0.

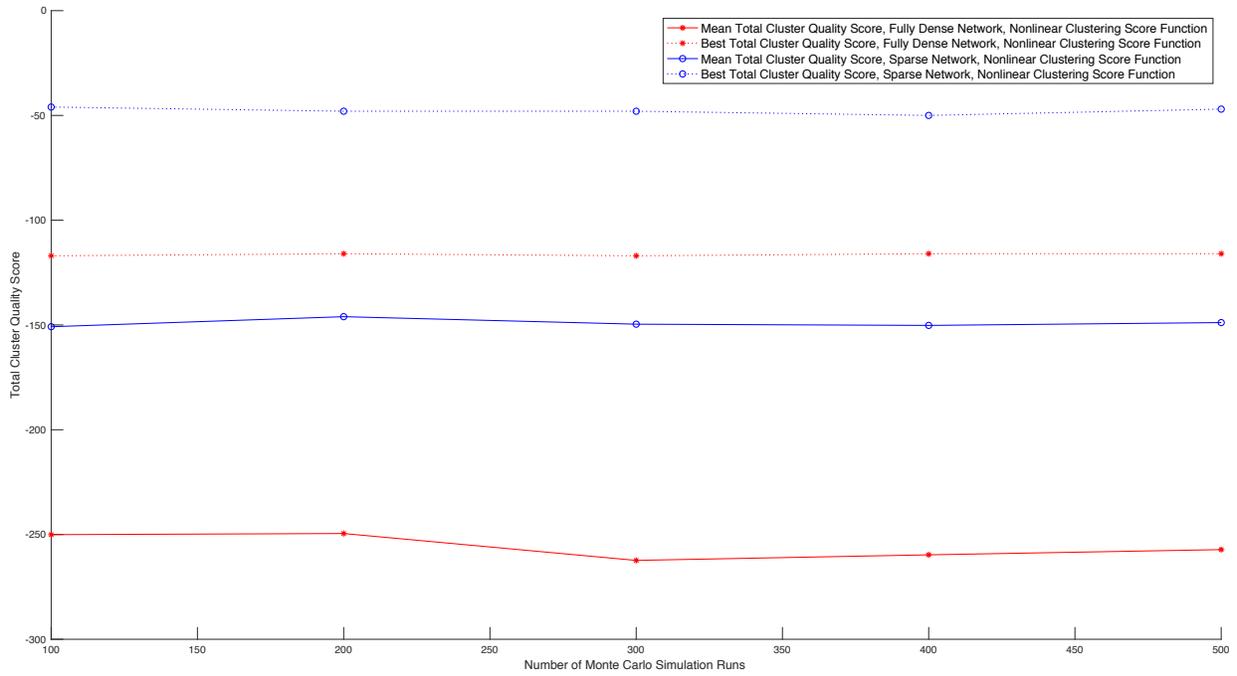

*Figure 31: Number of Monte Carlo Runs Results, using Nonlinear Clustering Score Function*

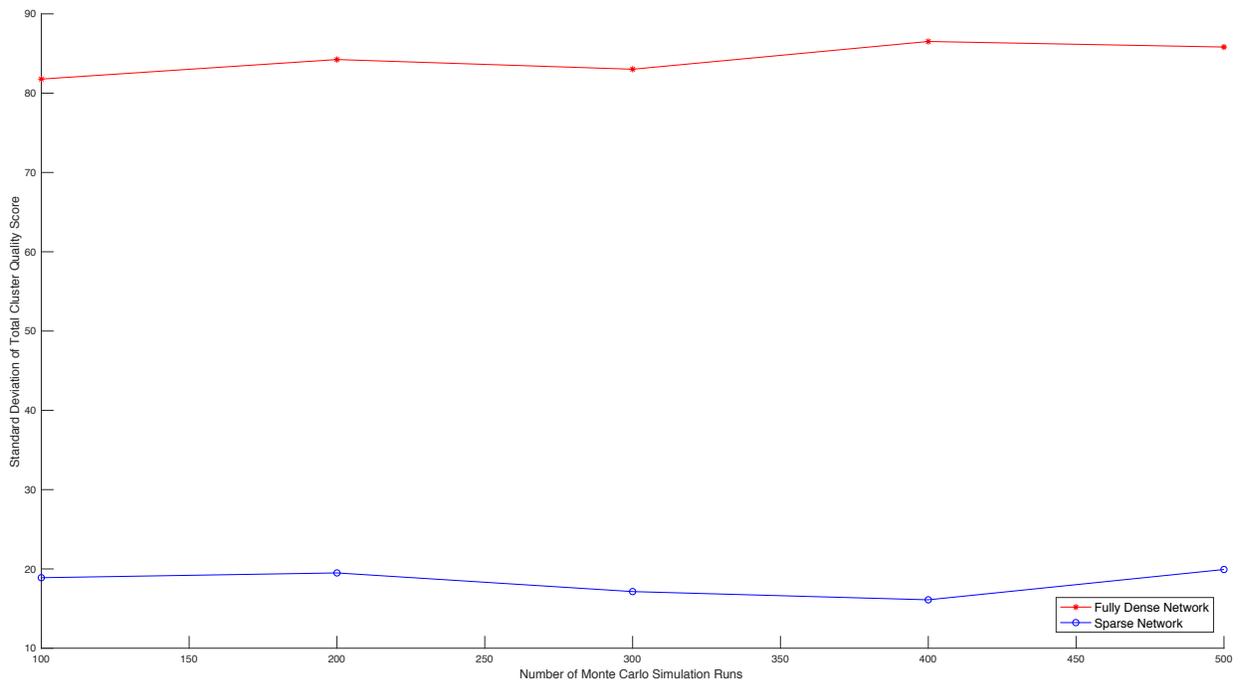

*Figure 32: Standard Deviation of Number of Monte Carlo Runs Results, using Nonlinear Clustering Score Function*

Figure 31 compares the size of the Monte Carlo simulation with the mean $S_T$ and best $S_T$ of each simulation, produced using $d_n$. Figure 32 makes a similar comparison, between the size of the Monte Carlo simulation with the standard deviation of $S_T$ for each simulation. For both figures, the Fully Dense Network is displayed in red, while the Sparse Network is displayed in blue. These figures compare the results captured in the histograms in Figure 27 and Figure 28. Unlike the previous set of results, the results produced using $d_n$ showed a clear difference between the mean $S_T$ and the best $S_T$ values. The difference in the Fully Dense Network is much larger than the difference in the Sparse Network. This fact is corroborated by Figure 32, where the standard deviation of $S_T$ for the Fully Dense Network is larger than the standard deviation of $S_T$ for the Sparse Network. There are some small observed changes in the mean $S_T$, best $S_T$, and the standard deviation of $S_T$ as the number of runs in the Monte Carlo simulation changes. When $d_n$ is used, the Sparse Network produces clusters with higher quality and less variation between runs. However, the number of runs in the Monte Carlo simulation has very little impact on both the quality and the variability of the results, for both the Fully Dense and the Sparse networks.

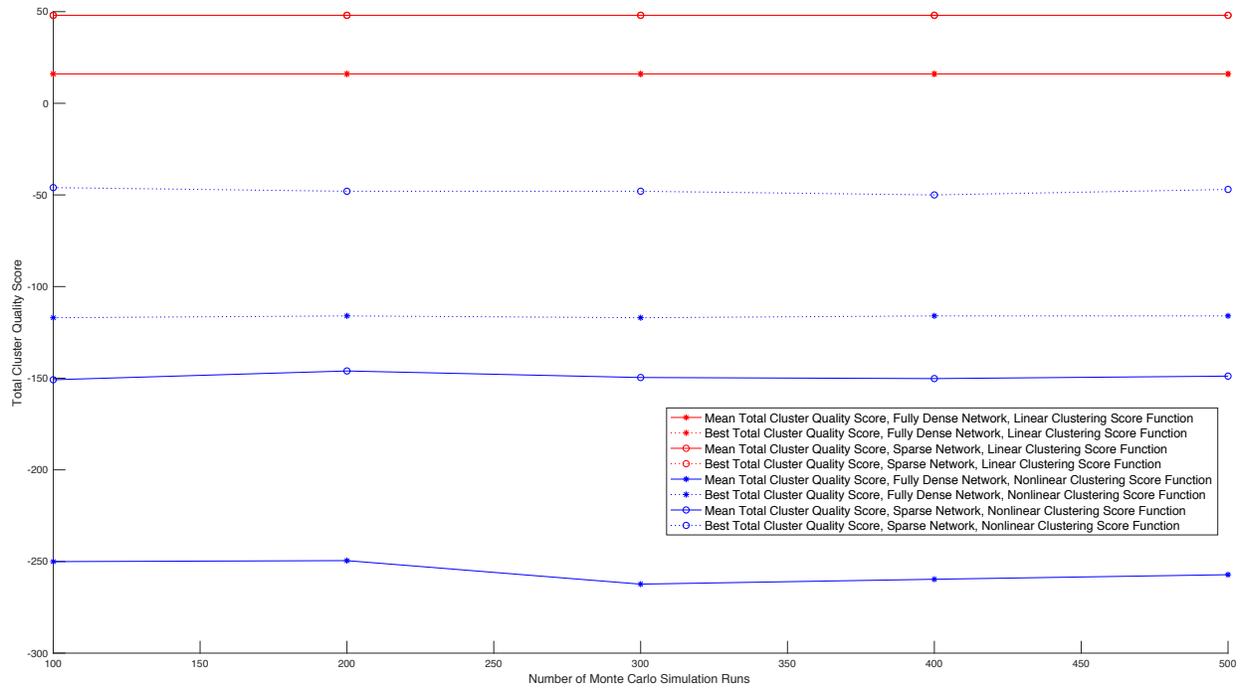

*Figure 33: Number of Monte Carlo Runs Results*

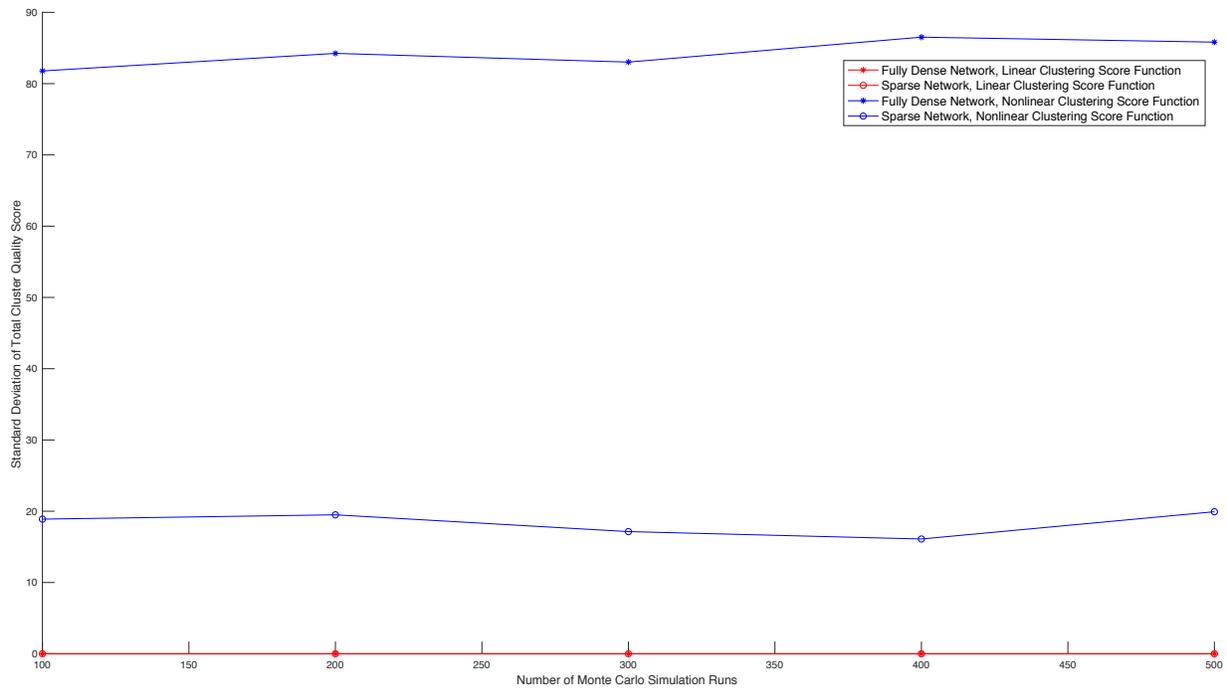

*Figure 34: Standard Deviation of Number of Monte Carlo Runs Results*

Figure 33 combines the results from Figure 29 and Figure 31, showing the changes in mean $S_T$ and best $S_T$ as the size of the Monte Carlo simulation changes, using both $d_l$ and $d_n$, and both the Fully Dense and the Sparse networks. Figure 34 combines the results from Figure 30 and Figure 32, showing the changes in standard deviation for $S_T$ as the size of the Monte Carlo simulation changes. For both figures, results generated using $d_l$ are shown in red, whereas the results generated using $d_n$ are shown in blue. From both Figure 33 and Figure 34, $d_l$ produced better clusters with less variability than $d_n$. Furthermore, $d_l$ does not need a Monte Carlo simulation for this student enrollment network, because it was able to consistently find the same cluster solution, regardless of the number of runs. Monte Carlo simulation does have an impact when $d_n$ is used, as the results generated vary more (when using the Fully Dense Network) or less (when using the Sparse Network). Therefore, for this set of student enrollment data, $d_l$ should be used with no additional need for Monte Carlo simulation.

If $d_n$ must be used, a moderately sized Monte Carlo simulation should suffice. The size of the simulation has little impact on the quality of results produced by $d_n$, so the simulation does not need to be large. However, since the results when using $d_n$ varies, there should still be a number of runs conducted to gather a decent sample. From this experiment, 100 to 200 runs for a Monte Carlo simulation should suffice when $d_n$ is used.

*5.3.3.3 Best Choice Learning Community Creation Results*

Based on the previous experiments, this Fully Dense student course enrollment network could best be clustered into LCs of maximum cluster size ten using $d_l$.

*Table 15: Best Choice Learning Community Creation Results*

| LC ID | Learning Community Membership | | | | | | | | | |
|---|---|---|---|---|---|---|---|---|---|---|
| 1 | 8 | 17 | 28 | 34 | 38 | 44 | 49 | 55 | 61 | 76 |
| 2 | 1 | 9 | 13 | 15 | 18 | 24 | 30 | 31 | 40 | 54 |
| 3 | 23 | 25 | 29 | 36 | 43 | 46 | 48 | 60 | 62 | 63 |
| 4 | 7 | 32 | 39 | 45 | 50 | 57 | 67 | 68 | 71 | 73 |
| 5 | 14 | 16 | 33 | 56 | 65 | 66 | 72 | 74 | 75 | 81 |
| 6 | 4 | 10 | 26 | 27 | 47 | 51 | 52 | 69 | 70 | 79 |
| 7 | 2 | 20 | 21 | 37 | 42 | 53 | 58 | 64 | 77 | 78 |
| 8 | 3 | 5 | 6 | 12 | 19 | 22 | 35 | 41 | 59 | 80 |
| 9 | 11 | | | | | | | | | |

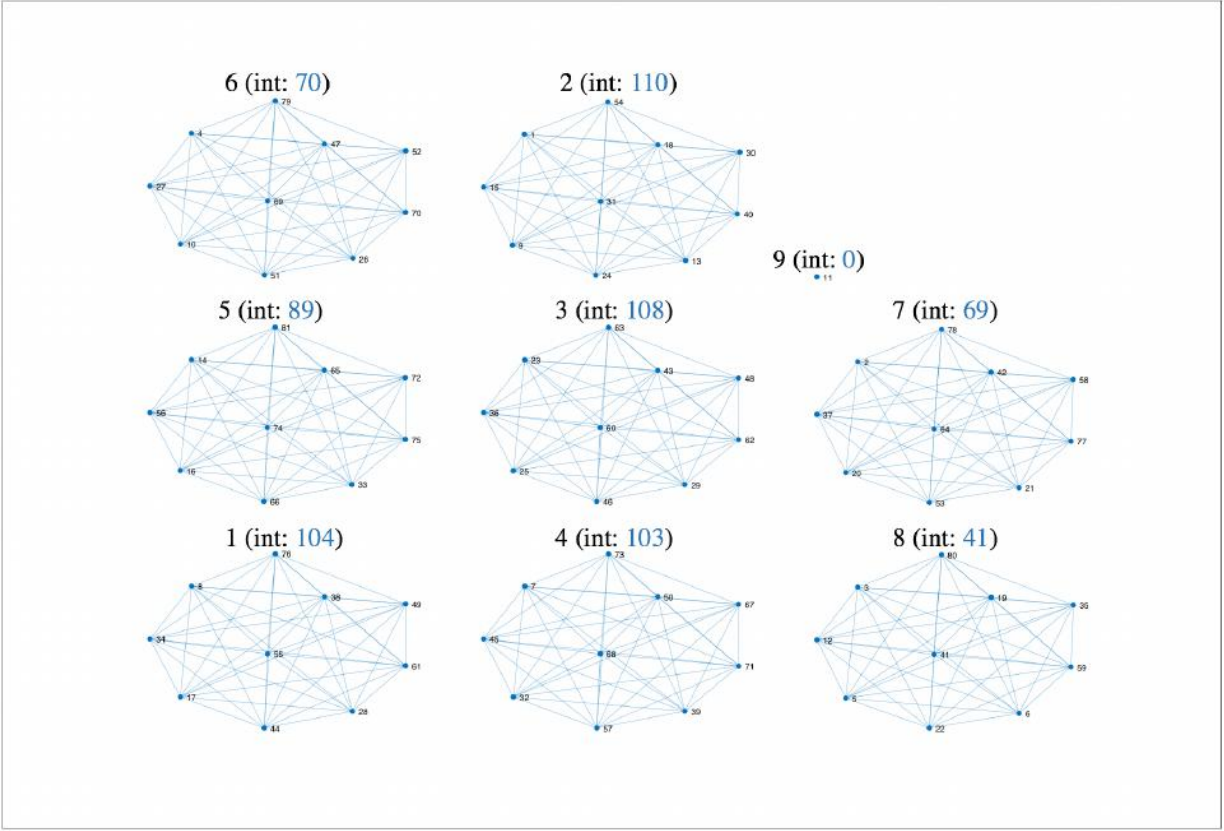

*Figure 35: Best Choice Learning Community Creation Results using Maximum Cluster Size of Ten*

The preceding Table 15 tabulates the final results of LC creation, and the preceding Figure 35 illustrates the final LCs, where each LC contains a maximum of ten students. There are eight full LCs and one individual student serving as a single LC, each with a LC ID indicated on top of the LC. The number of internal connections (int) is also indicated for each LC in blue. Overall, the LCs vary in quality. Some LCs, such as Cluster #2, demonstrates much stronger internal connections than other LCs, such as Cluster #8.

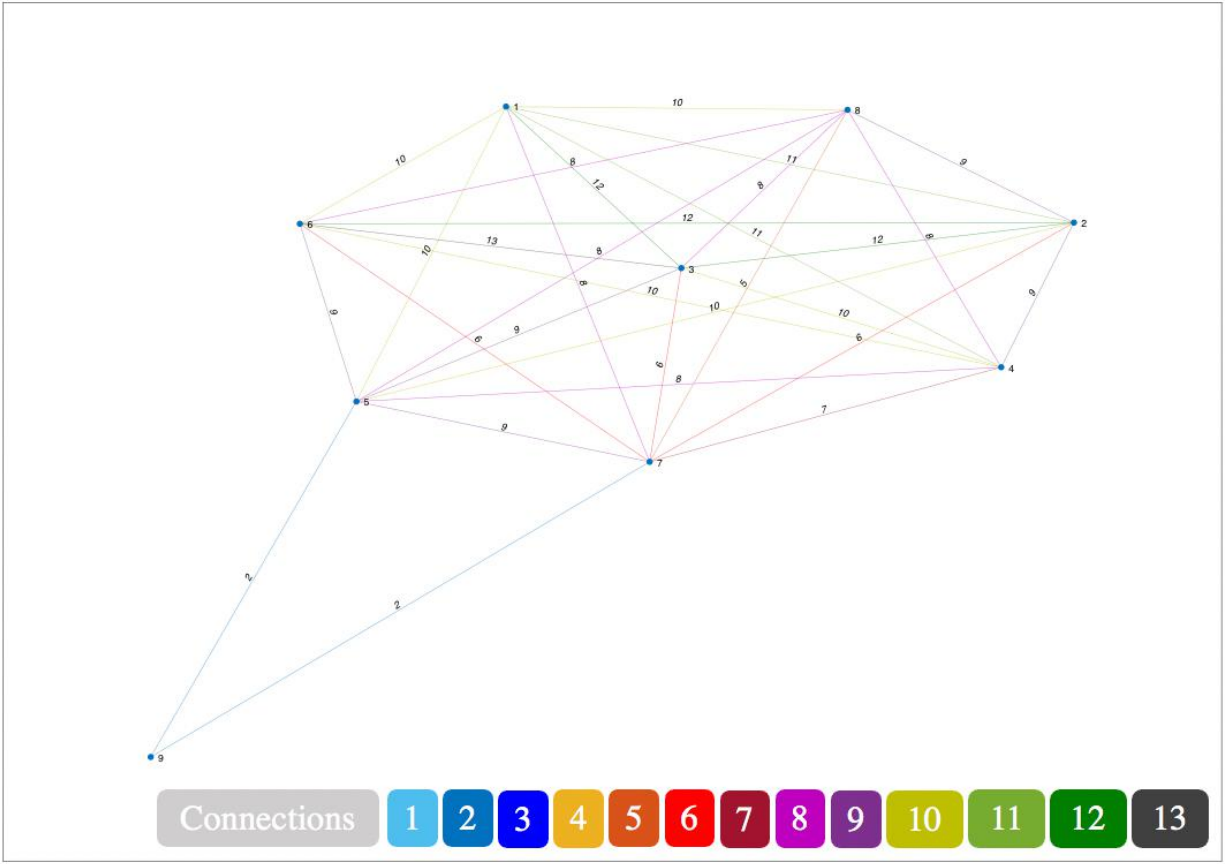

*Figure 36: External Connections Between Best Choice Learning Community Results*

The preceding Figure 36 illustrates the external connections remaining between LCs, where the different colours indicates the different number of connections between LCs. These external connections have been minimized as much as possible by the clustering algorithm, but external connections still remain, with as many as thirteen connections unsevered between LC #3 and #6. However, the overall connections between LCs are fairly minimal for a densely connected network. The resulting LCs is the best compromise for being able to produce decently isolated LCs without making the size of the LCs too large.

The resulting clusters are balanced, and the reduced cross-LC contact means that these LCs can be implemented in blended learning scenarios with a face-to-face component that still preserve health and safety of both students and educators. For example, if this method is enacted for creating LCs for Integrated Learning Stream (ILS) introduced in Section 2.5, it would create balanced student LCs that further facilitates their learning experience, especially during the semester-long design project.

## 5.4 Stochastic Learning Community Refinement Results

Following the creation of optimized LCs using the BC algorithm, the LCs can be refined using Simulated Annealing (SA), a global optimization function. The implementation behind SA is described in Section 4.4. The following sections describes the experimental results for fine-tuning SA parameters, as well as the LC refinement result using SA. As SA is meant to build on the clustering results of the previous section, the LCs in Table 15, visualized in Figure 35 and Figure 36, will be used as the starting point of all experiments in this section.

### 5.4.1 Number of Initial Perturbations for Average Initial Cost Results

This experiment was conducted to determine the $N$ amount of times the initial cost value ($C_{trial}$) should be averaged to compute for the average initial cost value ($\overline{C_{trial}}$). The details of the experiment are described in Section 4.5.1. A threshold must be determined at which a consistent $\overline{C_{trial}}$ can be computed for every $N$ and values larger than $N$.

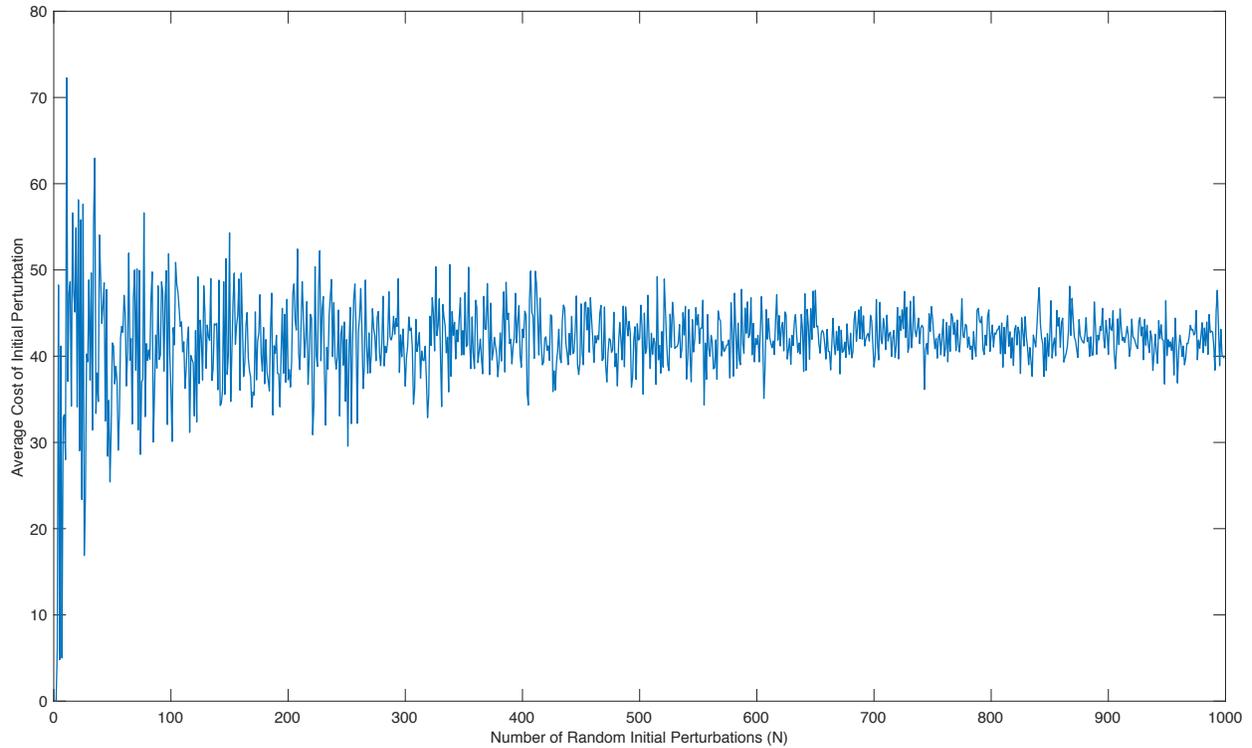

*Figure 37: Average Cost of Initial Perturbations per N Random Initial Perturbations*

Figure 37 compares the value of $\overline{C_{trial}}$ with the *N* number of times it has been averaged. $\overline{C_{trial}}$ fluctuates between 0 and 72 for small values of *N*. As *N* grows larger, $\overline{C_{trial}}$ is averaged over more values, and reaches a steadier state of fluctuation, between 30 to 50, and later between 35 to 45. For a cluster network of this size, there are too many initial perturbation possibilities due to the fully random perturbation selection, resulting in the wide fluctuations in $C_{trial}$, leading to the fluctuation of $\overline{C_{trial}}$ in the beginning. As more $C_{trial}$ are averaged, steady state is reached for the initial average cost value. At around *N = 1000*, the $C_{trial}$ value generally sits at around 40. From the figure, it can be seen that this steady state was reached at around *N = 600*. In the SA run for this clustering network, *N = 600* will be used to calculate the $\overline{C_{trial}}$ value.

### 5.4.2 Initial Acceptance Probability Results

This experiment will examine the relationship between acceptance probability (*AP*) and initial temperature ($T_0$), and the impact *AP* has on a SA run. The details of the experiment are described in Section 4.5.2. SA was performed for the following *AP* values, listed in Table 16. All other variables used in a SA run are kept the same.

*Table 16: Experimental Values of Initial Acceptance Probability*

| Parameter | Experimental Value | | | | | | |
|---|---|---|---|---|---|---|---|
| *AP* | 0.5 | 0.6 | 0.7 | 0.8 | 0.9 | 0.95 | 0.99 |

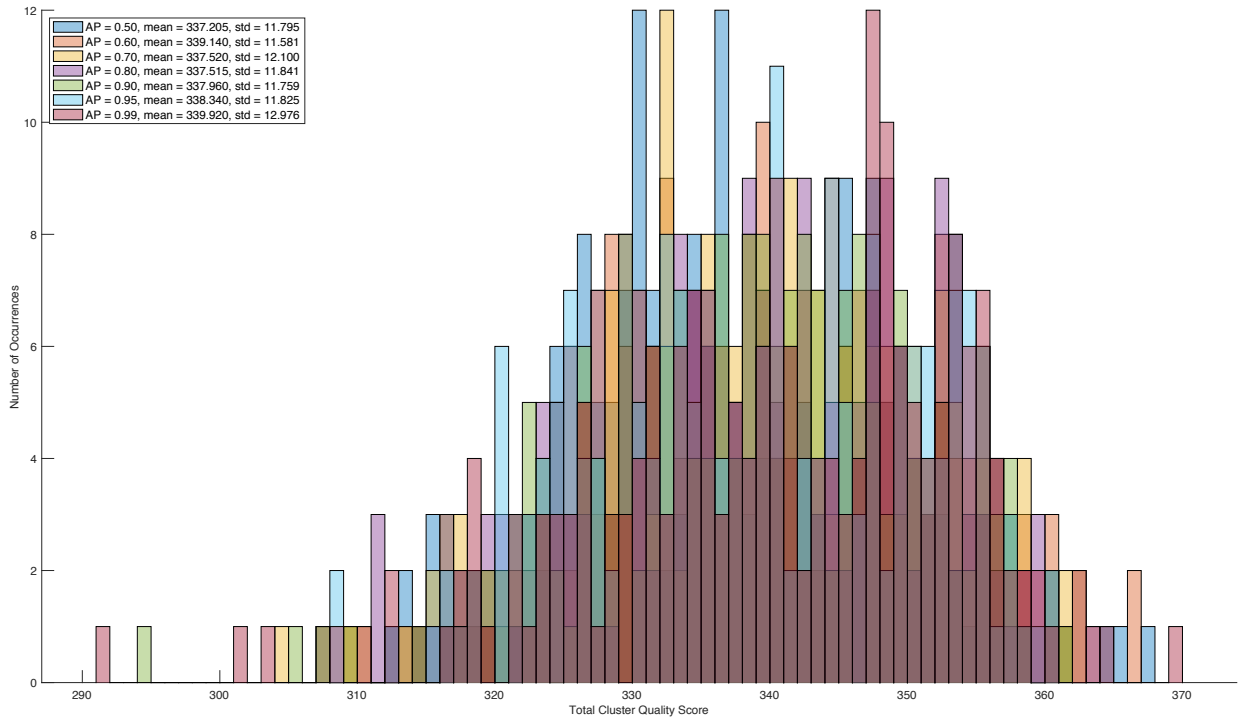

*Figure 38: Histogram of Cluster Refinement Results with Different Acceptance Probabilities*

Figure 38 captures the $S_T$ value of the cluster, or LC, refinement results using different *AP* values. For each *AP* value, a 200-run Monte Carlo simulation was executed. This figure shows the histogram of the collected results, alongside both the mean and the standard deviation of $S_T$ for each simulation. The results from each simulation largely overlapped, and it appears very difficult to determine if there is a correlative relationship between *AP* and the quality of refined LCs using the *AP* values.

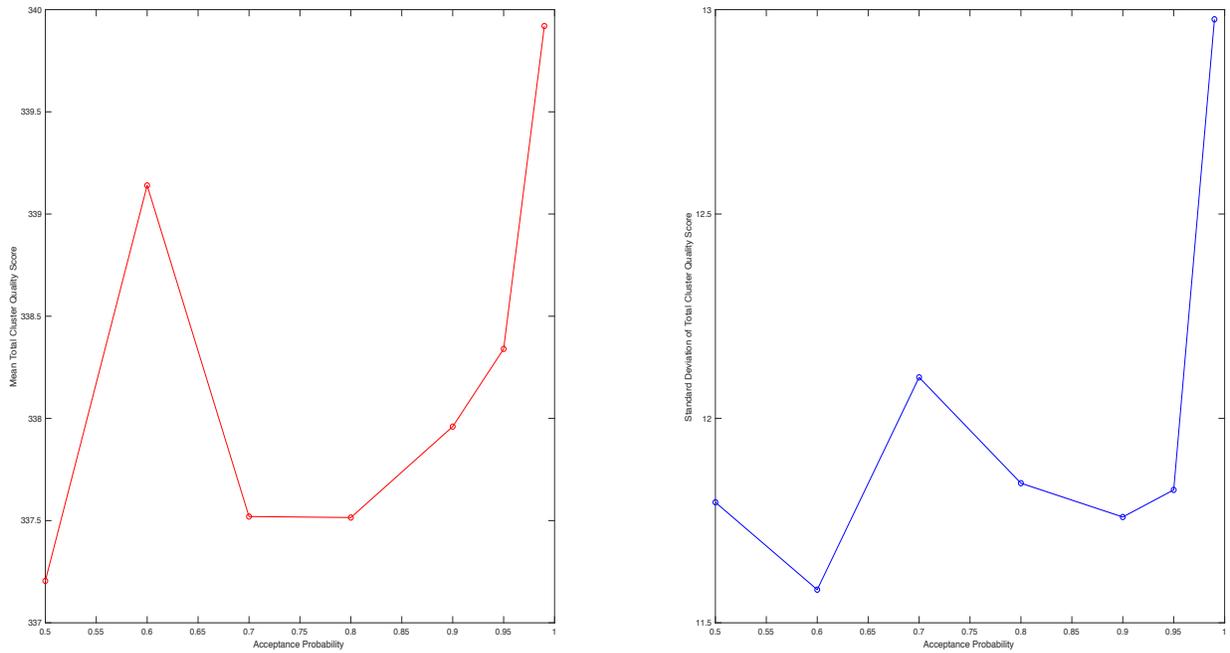

*Figure 39: Mean Total Cluster Quality Score (left) and Standard Deviation of Total Cluster Quality Score (right) with Different Acceptance Probabilities*

Figure 39 compares the *AP* values with the mean $S_T$ and standard deviation of $S_T$ for each simulation, visualizing the values on the histogram in Figure 38. The mean $S_T$ is graphed on the left in red, whereas the standard deviation of $S_T$ is graphed on the right in blue. From the left figure, it is difficult to determine a correlation between the mean $S_T$ of each Monte Carlo simulation and the *AP* values used for each simulation. The mean $S_T$ values fluctuated between 337 and 340, which is not a large difference in terms of cluster quality. Likewise, the standard deviation of $S_T$ also fluctuates between 11.5 and 13, meaning for each simulation, the clusters produced are somewhat similar in terms of the distribution of the quality. It appears that the strictness of the *AP* value has no bearing on the quality of clusters refined by the SA algorithm.

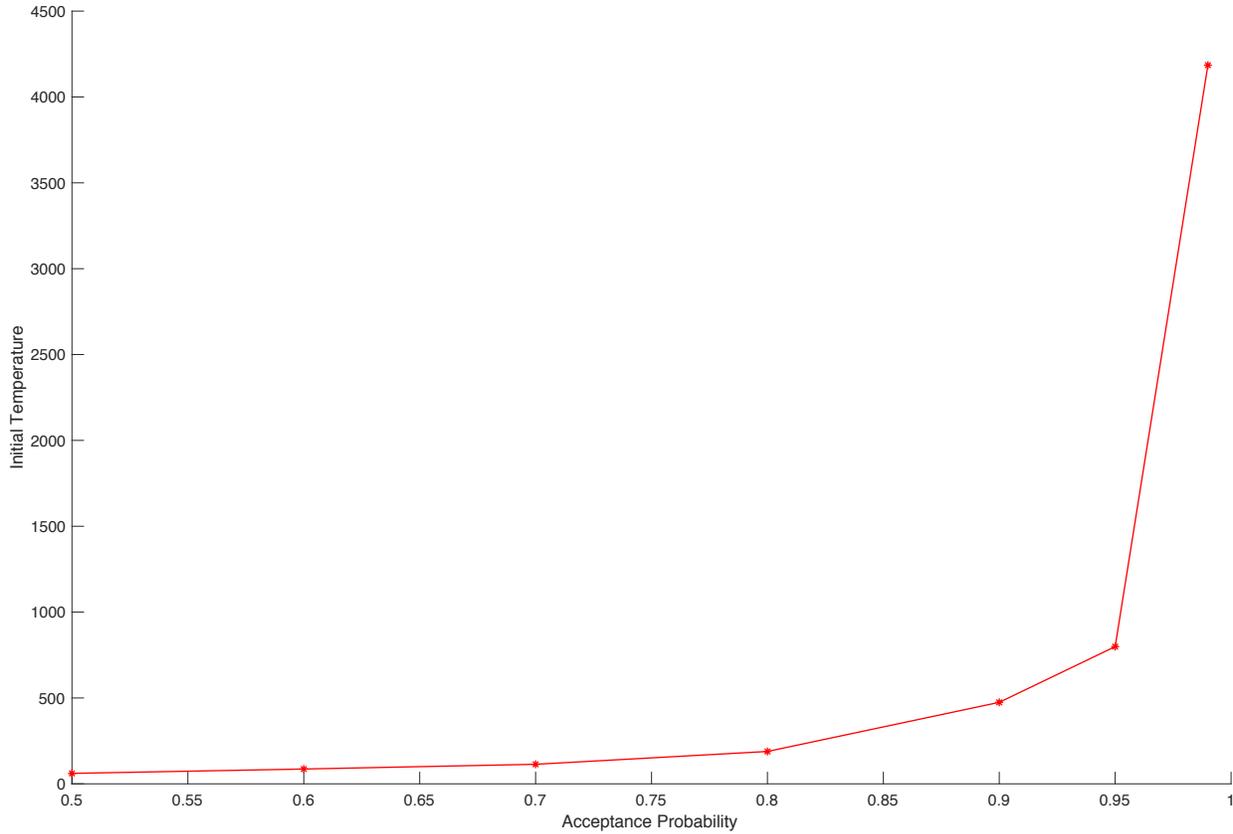

*Figure 40: Initial Temperature with Different Acceptance Probabilities*

The *AP* value does significantly impact the value of the $T_0$, as expected by equation (4.10), and demonstrated by Figure 40. This figure visualizes the difference in $T_0$ as *AP* changes. The value of $T_0$ is taken from a single run of each Monte Carlo simulation, since all $T_0$ values within each simulation will be very similar. From the figure, as the value of *AP* grows, the value of $T_0$ also grows, in an almost exponential manner. Between *AP* = 0.5 and *AP* = 0.9, the value of $T_0$ grows gradually between $0°C$ and $1000°C$. However, the growth of $T_0$ between *AP* = 0.95 and *AP* = 0.99 is very sudden, jumping between almost $1000°C$ to almost $4500°C$. The value of $T_0$ impacts the overall cooling process as the SA algorithm runs. If $T_0$ is not hot enough, there may not be enough of a temperature difference for the cluster configurations to be agitated then cooled. However, if $T_0$ is too hot, the algorithm may take a long run time to reach a cooled state. A high $T_0$ for a system, while possibly taking a long time to run, would not adversely impact the quality of the refined clusters. However, a low $T_0$ value could result in suboptimal solutions. From the results seen in Figure 39, the lowest $T_0$ value produced by $AP = 0.5$ does not significantly change the quality of the solution using the highest $T_0$ value produced by $AP = 0.99$. However, to be conservative, an $AP$ value of 0.9 or higher would be recommended for this SA system.

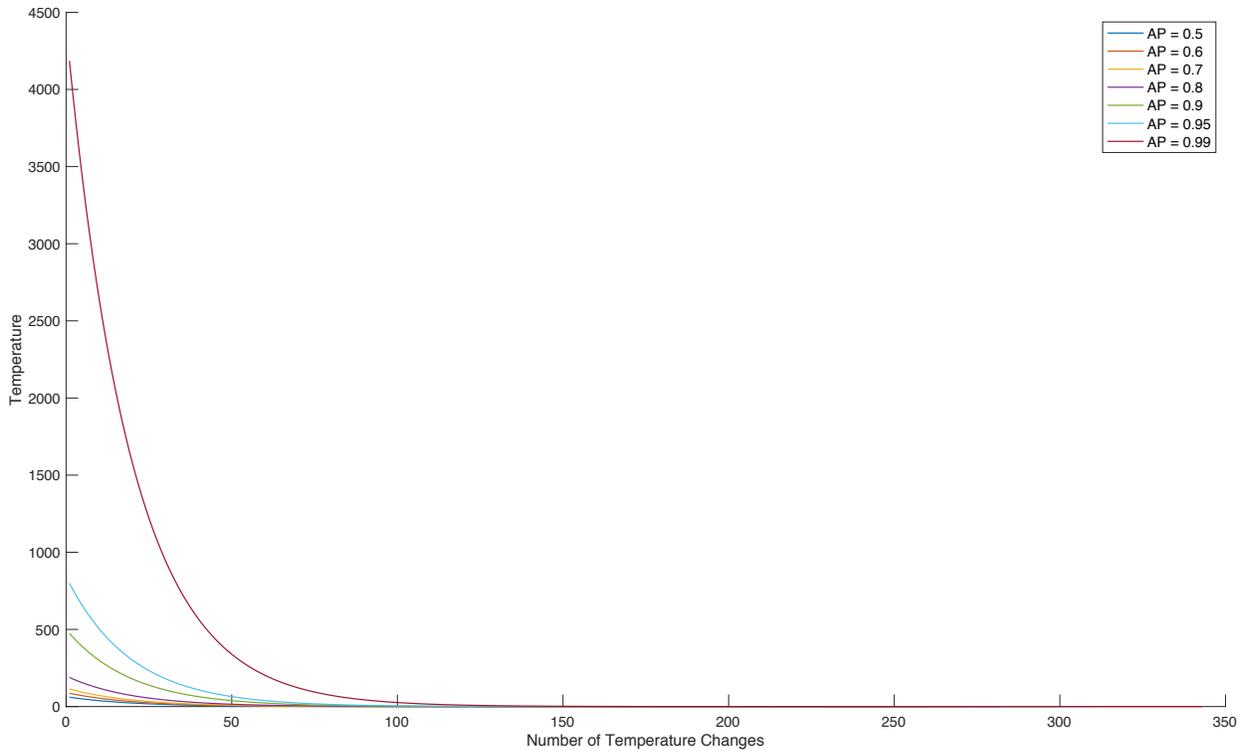

*Figure 41: Temperature Changes with Different Acceptance Probabilities*

In Figure 41, the temperature change of each SA system using different $AP$ values are compared, with the temperature($T$) value taken from one run in each Monte Carlo simulation. In each SA run, $T$, starting at $T_0$, is cooled at a rate controlled by the cooling rate ($\alpha$), seen in equation (4.13). The number of temperature changes is the number of times $T$ is reduced until the minimum temperature ($T_{min}$) is reached. If $T_0$ is large, there are more temperature changes, because it takes longer for $T$ to be cooled to $T_{min}$. For this experiment, the value of $\alpha$ was held consistent between the simulations at $\alpha = 0.95$, but the value of $T_0$ differs depending on the $AP$ value. The simulation using $AP = 0.99$ produced the most dramatic temperature changes, compared to the temperature changes of using other $AP$ values. This is due to the fact that $AP = 0.99$ also produced the largest $T_0$, and the amount of $T$ reduced by $\alpha$ when $T_0$ is large is more than when $T_0$ is smaller. The SA system when using $AP = 0.99$ also took the longest to cool when compared to other SA systems. Therefore, the rate of temperature change and speed to completion of a SA system is correlated to $T_0$, and thus also correlated to $AP$.

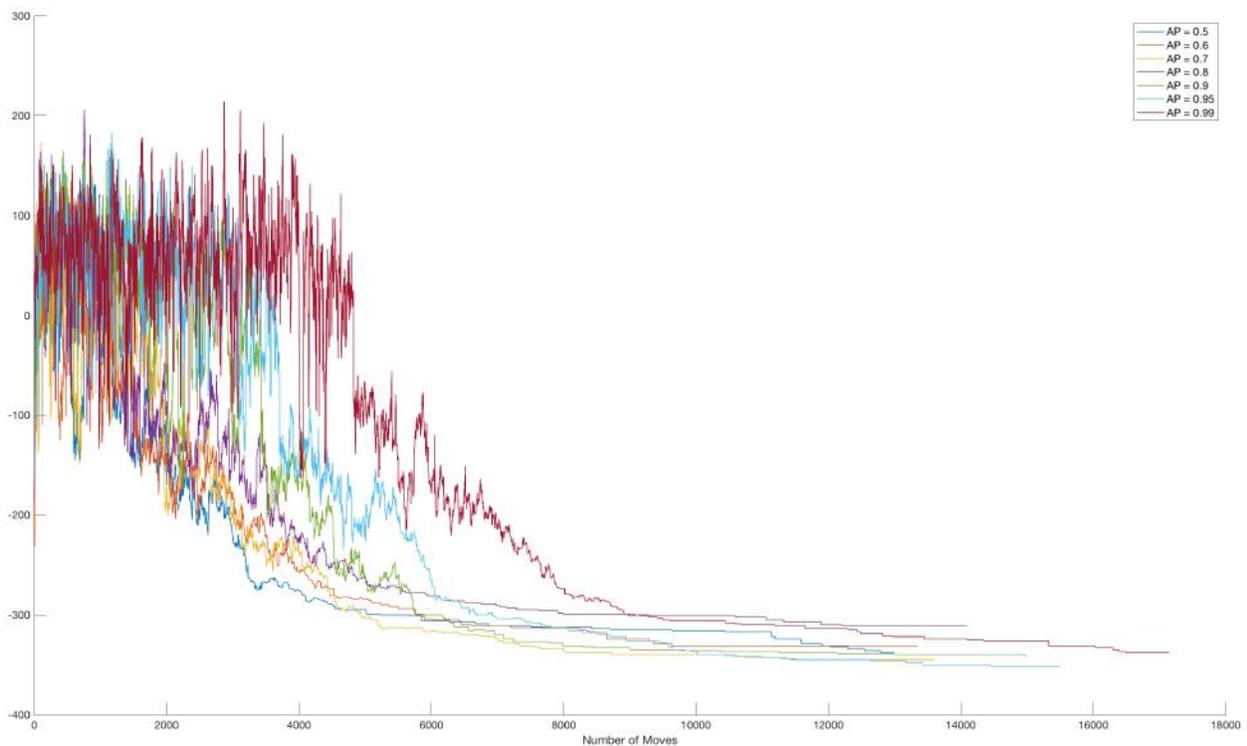

*Figure 42: Cost Value Change with Different Acceptance Probabilities*

Figure 42 visualizes the change in $Cost$ values for each experimental $AP$ value, where each set of $Cost$ values are taken from one run in each Monte Carlo simulation. As opposed to $S_T$, where the goal is to maximize the value, the goal of the SA system is to minimize $Cost$. The relationship between $Cost$ and $S_T$ is outlined in equation (4.7). For all values of $AP$, the change in $Cost$ fluctuates very wildly in the initial stages. However, as the SA system cools, the $Cost$ value is continuously minimized. Eventually, the system terminates at $T_{min}$ with a minimized $Cost$. The $AP$ value changes the number of moves a SA system undergoes, as the number of moves is equal to the number of times $T$ is reduced, multiplied by the number of iterations at each $T$ ($i_T$). Systems with larger $AP$ values will take longer to run, and generate more moves per run. The larger AP values will also result in the system taking longer to minimize $Cost$. For $AP = 0.99$, the system left the wild fluctuation of $Cost$ values and trended towards a more gradual minimization at around 5000 moves. Smaller $AP$ values entered the gradual minimization stage sooner, with less moves made. The value of $AP$ does not appear to impact the final minimized $Cost$ value of the system.

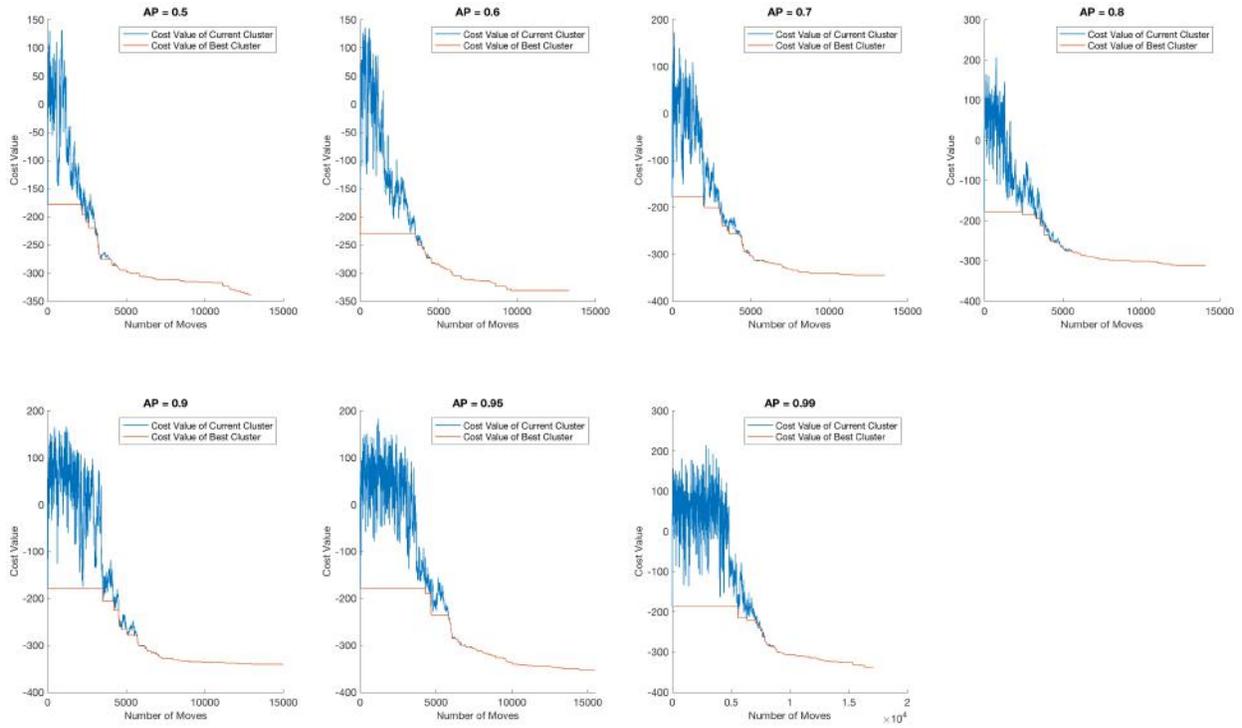

*Figure 43: Cost Value Change of Current and Best Clusters with Different Acceptance Probabilities*

Figure 43 maps the difference between the $Cost$ value of the current clusters and the $Cost$ value of the best clusters found thus far for different $AP$ values. Both sets of $Cost$ values are obtained from one run of each Monte Carlo simulation. The $Cost$ of the current clusters in a system is shown in blue, while the cost of the best clusters in the same system is shown in red. The distinct stages of $Cost$ value changes, seen also in Figure 42, is reflected in this figure. Initially, the $Cost$ value is in a stage of fluctuation, while not improving the best clusters. After a certain number of moves, the $Cost$ of the best clusters begins to decrease, as the $Cost$ of the current clusters leave the fluctuation stage and trend towards being minimized. At some point, the $Cost$ of the current clusters and the $Cost$ of the best clusters merge, where each new solution is accepted if it is better than the currently best solution. These stages align with the behavior of the SA system. While the system is hot, SA allows for the acceptance of almost any solution, leading to the large fluctuation in $Cost$ values. As the system cools, the acceptance criteria become stricter, and eventually will only accept the best solutions. This behavior is consistent across the different values of $AP$ used. However, the values of $AP$ change the duration of each stage, where smaller values of $AP$ lead to a shorter fluctuation stage, and a shorter overall length of run. As $AP$ increases, the duration of the fluctuation stage also increases, and the duration of the SA run also increases. This behavior is consistent with findings from previous figures in this section.

Overall, the value of $AP$ directly and almost exponentially impacts the value of $T_0$ in a SA system. The change in $T_0$ also affect the rate at which $T$ changes in a SA system and the duration of the SA run. In this case, the value of $AP$ does not change the quality of the refined student LCs. However, knowing that a low $T_0$ can lead to suboptimal solutions, it is recommended in this case to use an $AP$ value of 0.9 or larger.

### 5.4.3 Cooling Rate Results

This experiment will examine the relationship between cooling rate ($\alpha$) and temperature ($T$), and the impact $\alpha$ has on a SA run. The details of the experiment are described in Section 4.5.3, and other variables used in the experiment are kept consistent. SA was performed for the following $\alpha$ values, listed in Table 17.

*Table 17: Experimental Values of Cooling Rate*

| Parameter | Experimental Value | | | | | | |
|---|---|---|---|---|---|---|---|
| $\alpha$ | 0.5 | 0.6 | 0.7 | 0.8 | 0.9 | 0.95 | 0.99 |

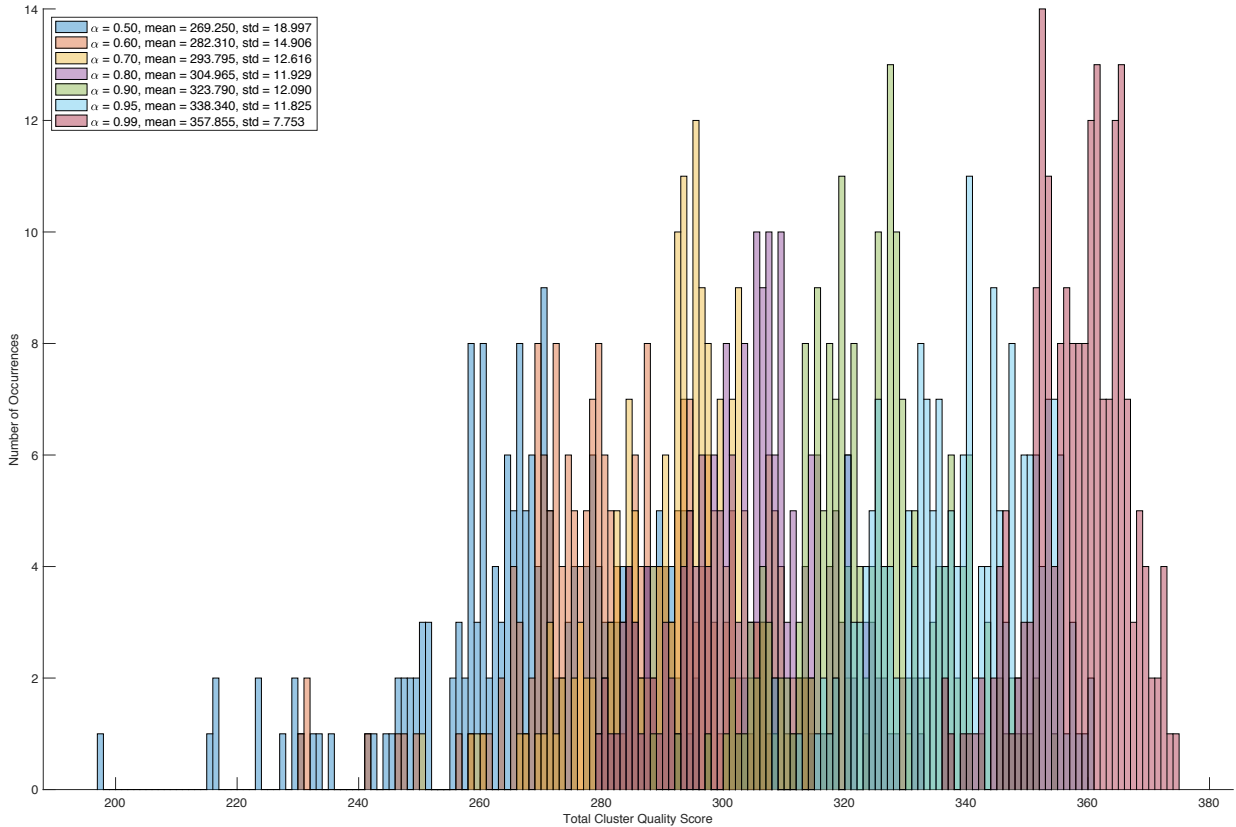

*Figure 44: Histogram of Cluster Refinement Results with Different Cooling Rates*

Figure 44 captures the result of the cluster refinement, where each set of results from a 200-run Monte Carlo simulation using a different $\alpha$ value is displayed, alongside the mean and standard deviation of $S_T$ for each simulation. Each set of refined clusters acts as a potential student LC configuration. The histogram shows a shift in the quality of the clusters, whereas the value of $\alpha$ increases, $S_T$ also generally increases. This trend is also visualized in the following figure.

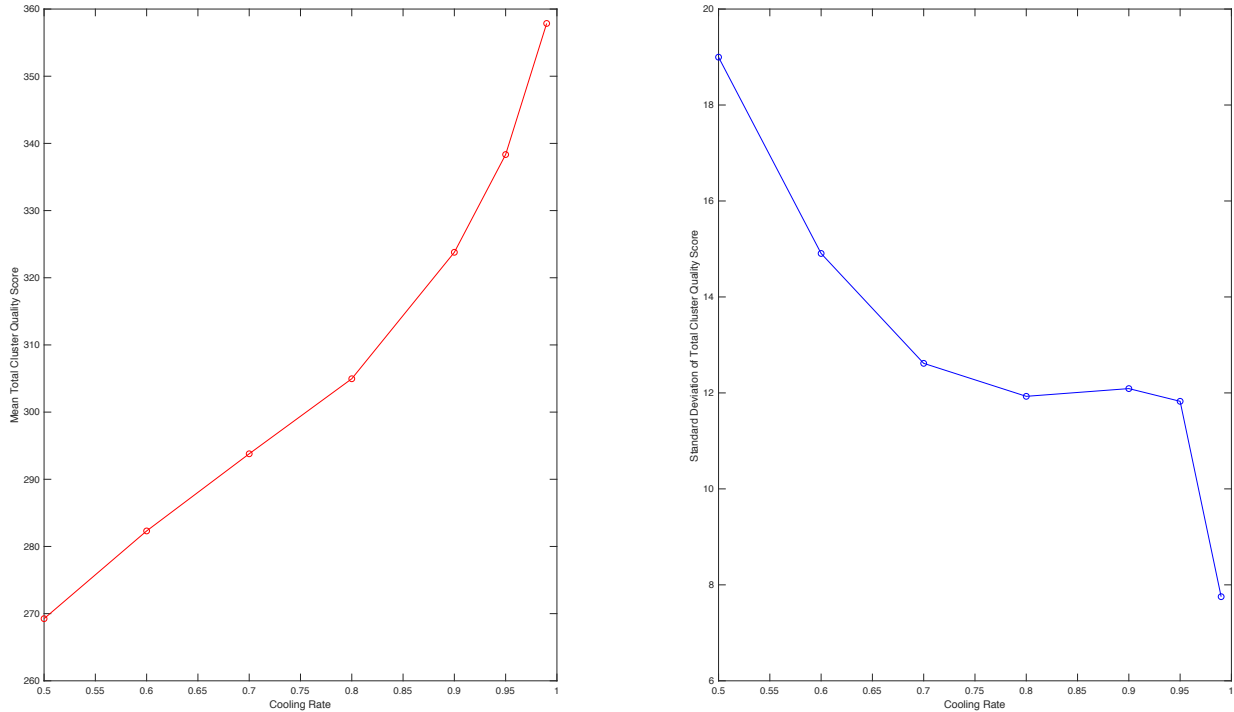

*Figure 45: Mean Total Cluster Quality Score (left) and Standard Deviation of Total Cluster Quality Score (right) with Different Cooling Rates*

Figure 45 compares the mean $S_T$ and the standard deviation of $S_T$ with the set of different $\alpha$ values. Mean $S_T$ is visualized on the left in red, whereas the standard deviation of $S_T$ is visualized on the left in blue. The variable $\alpha$ dictates the speed at which $T$ is cooled during the SA process, where $\alpha$ indicates the percentage of heat that is carried onto the next cycle. For example, for $\alpha = 0.5$, $T$ is reduced by 50%, and the new $T$ is 50% of the current $T$. This figure demonstrates a relatively linear relationship between $\alpha$ and the mean $S_T$ of each Monte Carlo simulation. As the value of $\alpha$ increased, the quality of the refined clusters also increased. In addition, the standard deviation of $S_T$ decreased as $\alpha$ increased, as the clustering solutions became less varied as $\alpha$ increased. This result demonstrated the importance of $T$ cooling speed in producing higher quality clusters. A large $\alpha$ value leads to a slow-cooling SA system, whereas a small $\alpha$ value leads to a fast-cooling system. A slow-cooling system allows the system to remain at higher temperature for a long time, giving the system ample opportunity to search for and find the best solutions. The more $T$ cools, the lower the chance the SA system will accept a "bad" solution. The SA algorithm needs to occasionally accept "bad" solutions, in order to eventually find better solutions. A fast-cooling system would not allow the system ample time to accept "bad" solutions, and therefore the system will not have the chance of finding better solutions later. Therefore, a slow-cooling system, accompanied by a large $\alpha$ value, is necessary to obtain high quality cluster refinements.

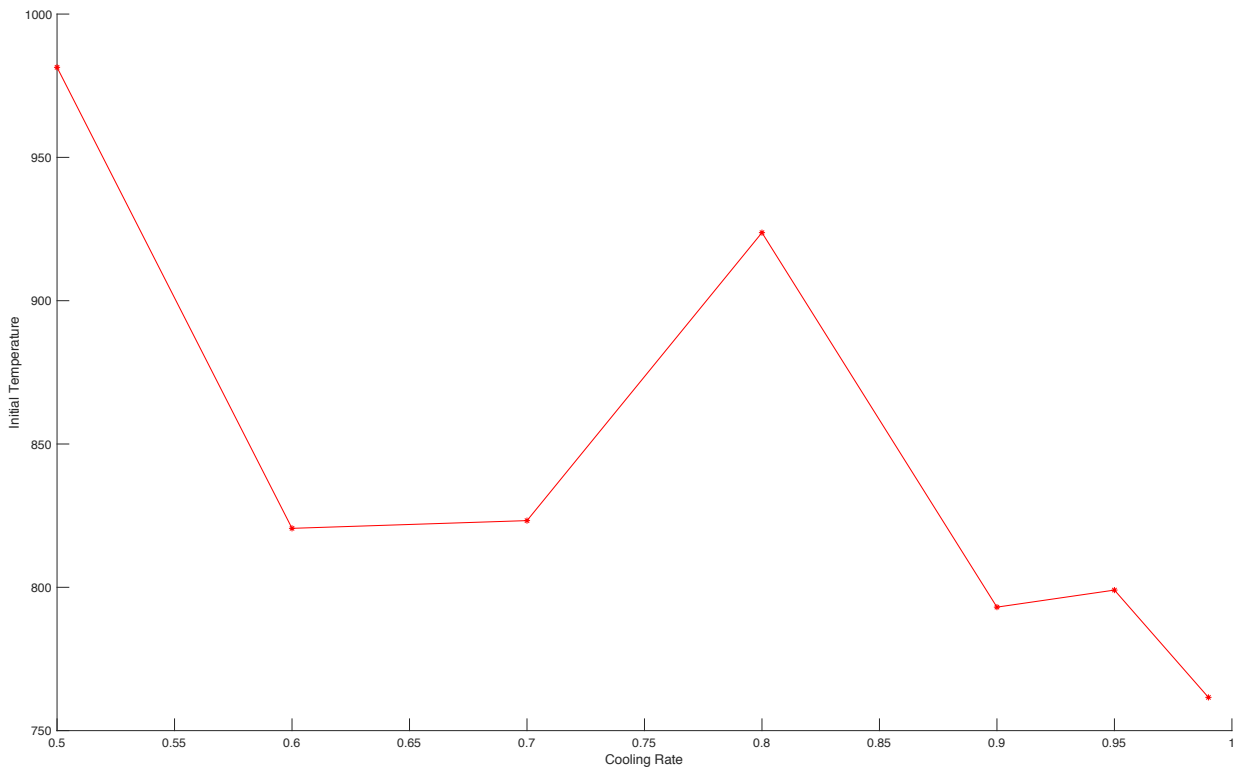

*Figure 46: Initial Temperature with Different Cooling Rates*

Figure 46 compares $T_0$ to different $\alpha$ values, where all $T_0$ values were obtained using $AP = 0.95$. The value of $T_0$ fluctuates between 750 °$C$ and 1000 °$C$, with no discernable pattern. This is due to the fact that $T_0$ and $\alpha$ are independent of each other, and the value of $\alpha$ does not affect the value of $T_0$ at all. The variable $\alpha$ only begins to be used after $T_0$ has already been calculated.

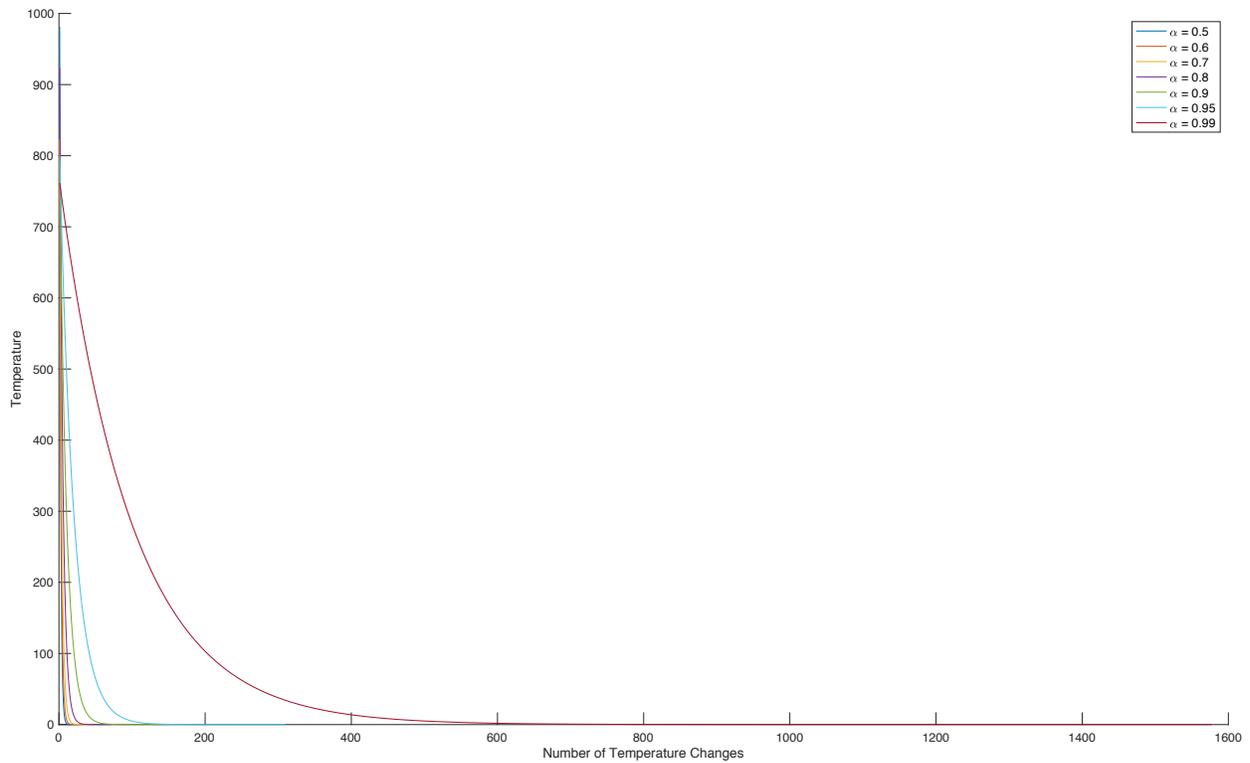

*Figure 47: Temperature Changes with Different Cooling Rates*

Figure 47 compares the rate at which the value of $T$ changes for different $\alpha$ values. As aforementioned, the rate at which $T$ cools, starting from $T_0$, is controlled by the value of $\alpha$. Therefore, if $\alpha$ is small, $T$ cools very rapidly. Conversely, $T$ cools very slowly if $\alpha$ is closer to 1. This fact is visually reflected in this figure, where $T$, when using $\alpha = 0.99$, takes a significantly longer time to cool compared to other $\alpha$ values. The $T$ of the SA system when using $\alpha = 0.99$ have been cooled almost 1600 times before reaching completion. In comparison, the smaller $\alpha$ values are cooled less than 200 times prior to completion. Even for $\alpha = 0.95$, the SA system has been cooled for less than 400 times, less than ¼ the number of times the system using $\alpha = 0.99$ has been cooled. As mentioned previously, a slow-cooling SA system is more preferable because it will yield higher quality clusters. The only drawback to a slow-cooling system is that they take a long time to run, and could become computationally expensive. This figure demonstrates how fast each SA system compares to one another when using different $\alpha$ values.

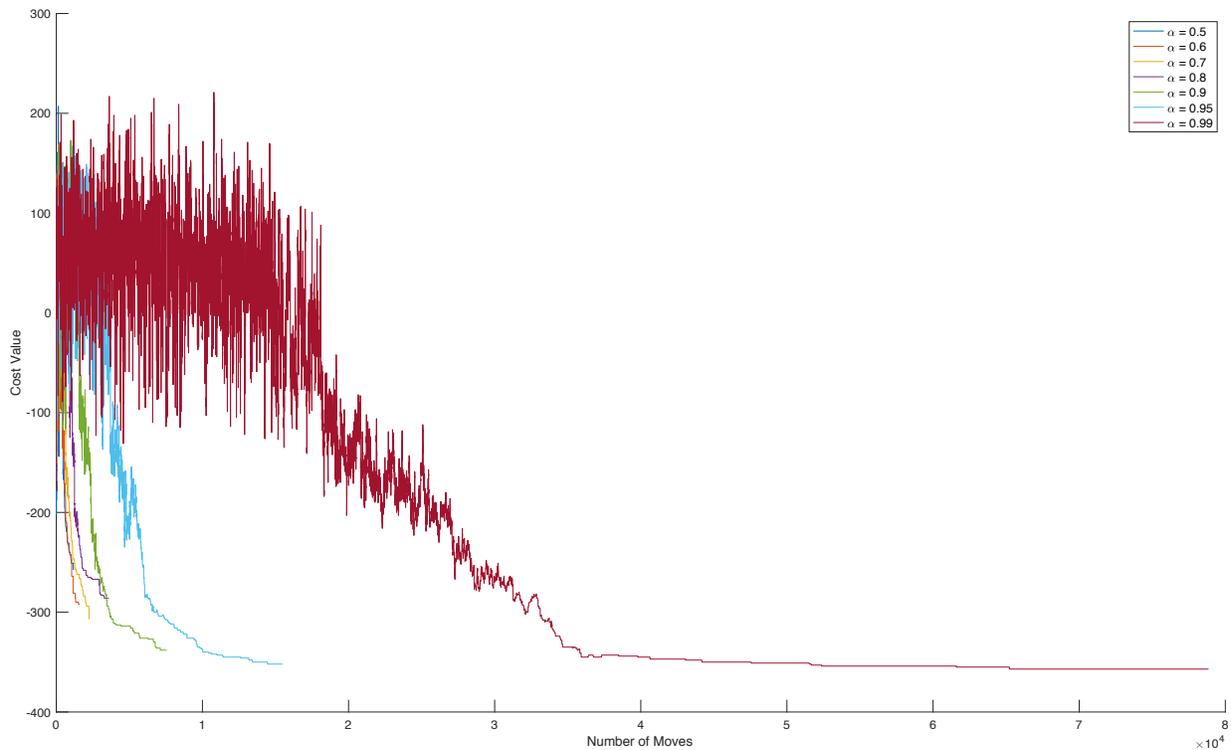

*Figure 48: Cost Value Change with Different Cooling Rates*

Figure 48 compares how the $Cost$ value changed for each system using a different value of $\alpha$. Once again, this figure demonstrated how much dramatically longer the SA system is using $\alpha = 0.99$, when compared to $\alpha = 0.95$ and smaller. Because $\alpha$ impacts how long the SA system runs, $\alpha$ also impacts how long the $Cost$ value will wildly fluctuate before beginning to be gradually minimized. For $\alpha = 0.99$, this process took a much longer time than smaller $\alpha$ values. However, unlike $AP$, $\alpha$ has a distinct impact on the final minimized $Cost$ value of each SA system. Smaller $\alpha$ values lead to a system that ends very quickly, with a $Cost$ value that barely reaches -300. Larger $\alpha$ values far surpass that value, with some $Cost$ solutions reaching almost -350. Since the objective of the SA algorithm is to minimize $Cost$, the latter with larger $\alpha$ values are more preferable. For all of the large $\alpha$ values tested, an $\alpha$ value of 0.95 or more should suffice. While it is clear that $\alpha$ = 0.99 produced the highest cluster quality, $\alpha$ = 0.95 was able to produce a more suboptimal, but still comparable, solution. Also, $\alpha$ = 0.95 is a reasonable compromise since it is a less computationally expensive operation due to its faster runtime.

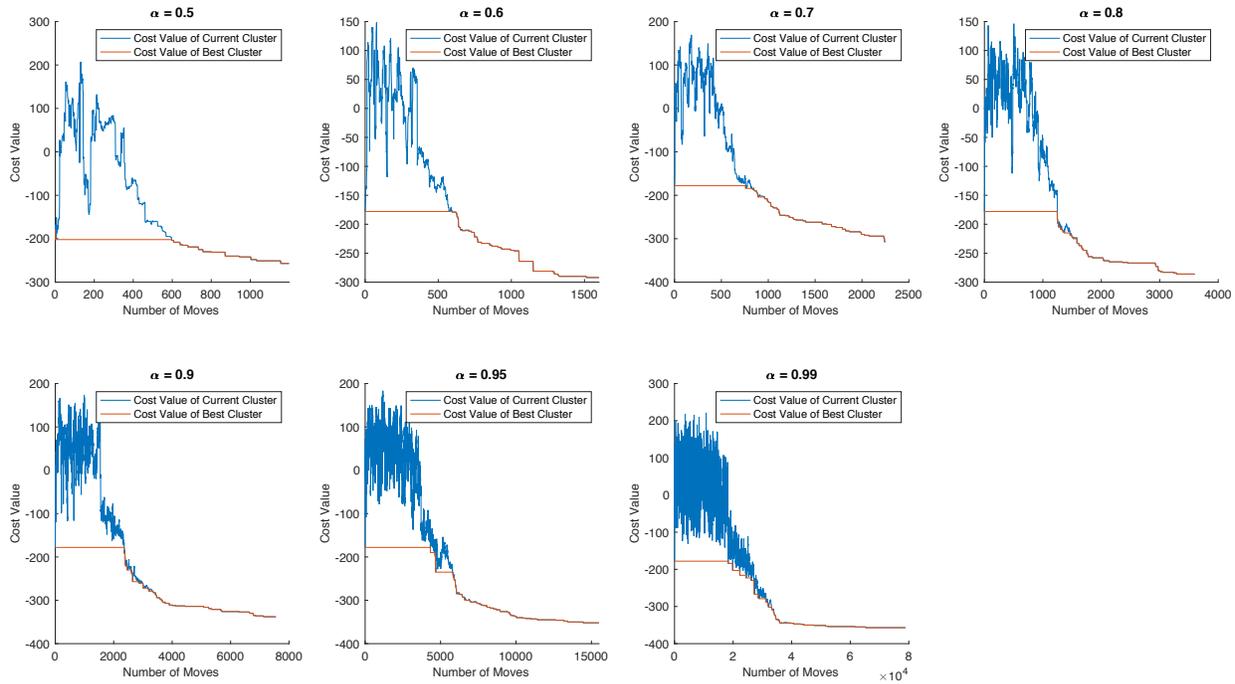

*Figure 49: Cost Value of Current and Best Clusters with Different Cooling Rates*

Figure 49 compares the $Cost$ values of the current and best clusters for each $\alpha$ value. This figure reiterates the relationship between $\alpha$ and the refined cluster quality. For $\alpha$ = 0.5, the $Cost$ value of the best clusters changed very little throughout the entire run. In addition, there are not many $Cost$ value changes for the current solution when $\alpha$ = 0.5. As the value of $\alpha$ increases, the duration of the SA run becomes longer, as indicated by the increasing number of moves. At the same time, the $Cost$ value of the best clusters are also being minimized more. This figure visualizes the distinct difference in how the $Cost$ values of both the current and best solutions in a SA system using different $\alpha$ values.

Overall, $\alpha$ has a direct impact on both the runtime of the SA algorithm and the quality of the final solution. A large $\alpha$ value will cool $T$ of the SA system slowly, thus leading to more opportunities to accept "bad" solutions prior to accepting better solutions. By allowing the system to have this longer runtime, the SA algorithm would have longer to search for and find the global optimum. Therefore, large $\alpha$ values directly correlate to better cluster refinement results, thus better student LCs refinement results. For this case, $\alpha \geq 0.95$ is recommended.

### 5.4.4 Minimum Temperature Results

This experiment will examine the impact of the minimum terminating temperature of a SA system ($T_{min}$) on the quality of the final LC configurations. The details of this experiment are outlined in Section 4.5.4. The $T_{min}$ values used in this experiment are listed in Table 18, while all other variables are kept consistent.

*Table 18: Experimental Value of Minimum Temperature*

| Parameter | Experimental Value | | | | | | | |
|---|---|---|---|---|---|---|---|---|
| $T_{min}$ | 10 | 5 | 1 | 0.1 | 0.01 | 0.001 | 0.0001 | 0.00001 |

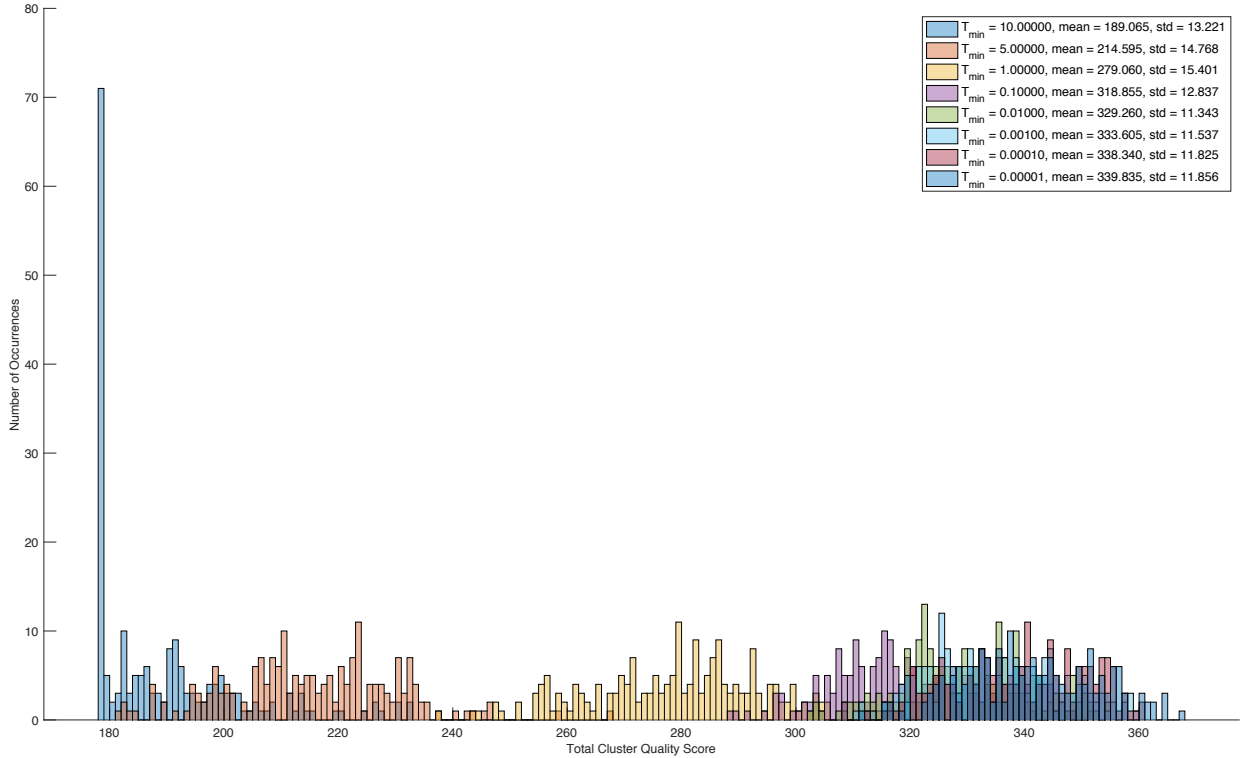

*Figure 50: Histogram of Cluster Refinement Results with Different Minimum Temperatures*

In Figure 50, the cluster refinement results from the 200-run Monte Carlo simulations using different $T_{min}$ values are showcased, where every set of solution is a potential student LC configuration. The mean and standard deviation of $S_T$ for each Monte Carlo simulation is also represented. From the figure, it can be seen that as $T_{min}$ decreases, the overall quality of the clusters improves. For $T_{min} = 10$, there was an overwhelming number of solutions with $S_T = 178$, which was the $S_T$ of the initial input cluster. Therefore, a large number of solutions have not been refined when $T_{min} = 10$ is used. For the $T_{min}$ values at 0.01 or smaller, the cluster refinement results generally overlap. The cluster quality change at smaller $T_{min}$ values appear to be minimal.

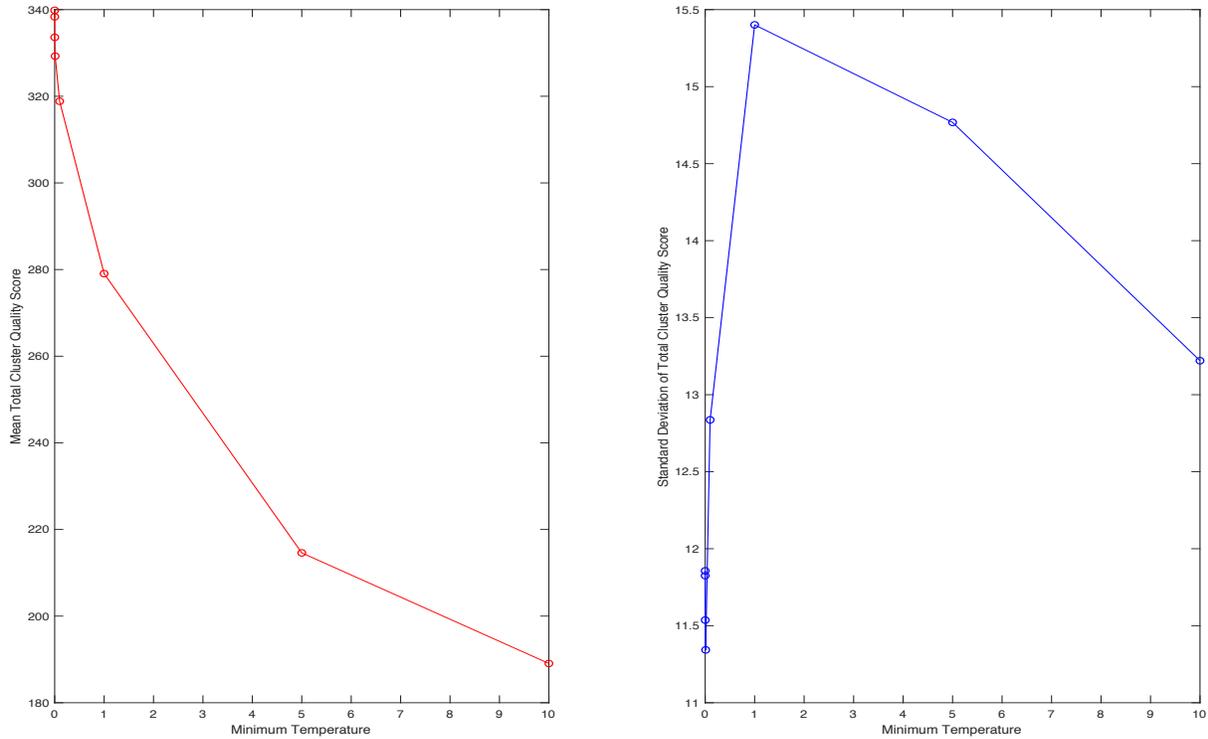

*Figure 51: Mean Total Cluster Quality Score (left) and Standard Deviation of Total Cluster Quality Score (right) with Different Minimum Temperatures*

Figure 51 displays $T_{min}$ compared to the mean $S_T$ and standard deviation of $S_T$ for the Monte Carlo simulation results in Figure 50. The mean $S_T$ is displayed on the left in red, whereas the standard deviation of $S_T$ is displayed on the right in blue. The mean $S_T$ plot showed the same trend observed in the previous histogram, where the quality of the cluster refinement result improves as $T_{min}$ decreases. Initially, there is a large improvement in cluster quality as $T_{min}$ decreases, especially between $T_{min} = 5$ and $T_{min} = 1$. As $T_{min}$ approaches 0.01 and smaller, the cluster quality change becomes much less significant, improving from $S_T = 329$ to $S_T = 340$. However, there are less correlation between the $T_{min}$ and the standard deviation of $S_T$. Standard deviation of $S_T$ increase between $T_{min} = 10$ and $T_{min} = 1$, and it decreases following $T_{min} < 1$. For this experiment, the most varied solutions in the Monte Carlo simulation are at $T_{min} = 1$, and the variation decreases sharply for $T_{min} < 1$. From the combined results in this figure, having $T_{min} \leq 0.01$ generally produces high quality and preferable solutions.

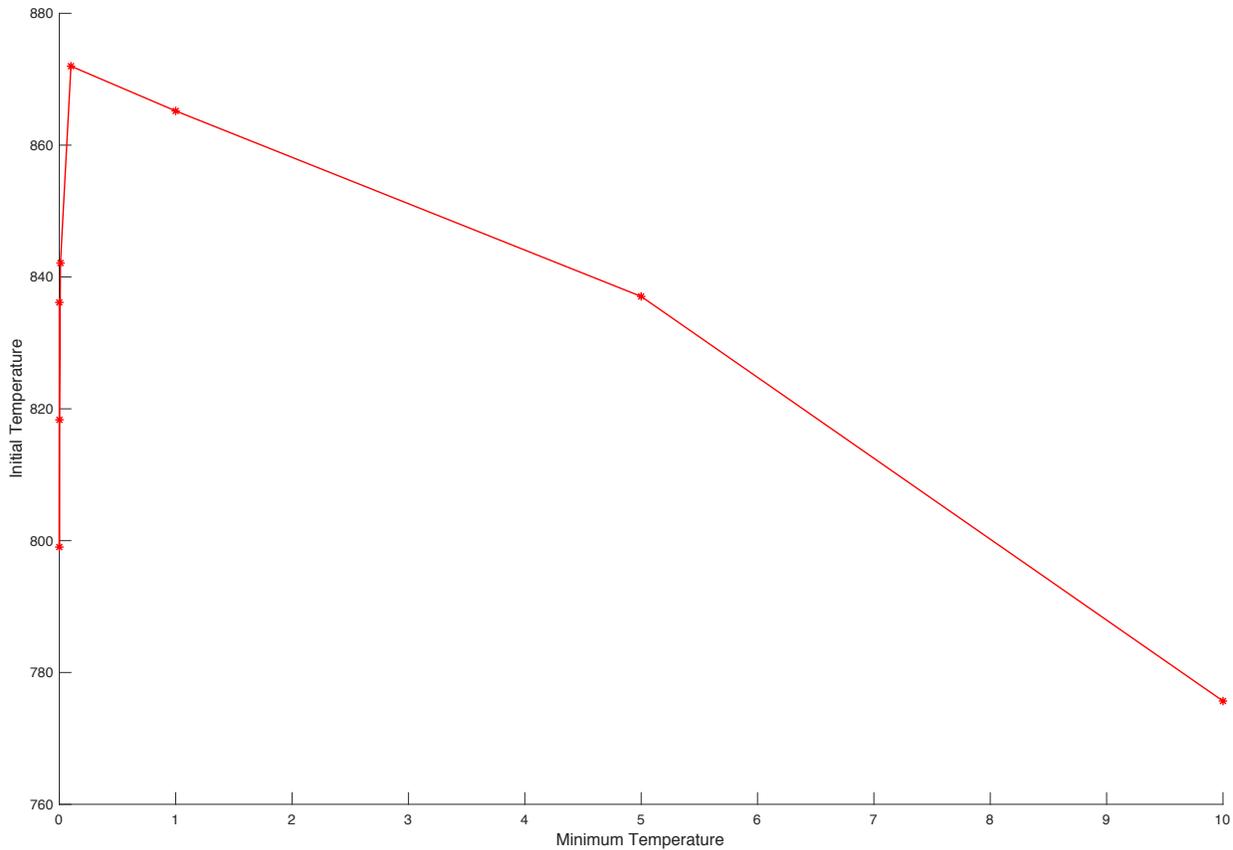

*Figure 52: Initial Temperature with Different Minimum Temperatures*

Figure 52 compares the value of $T_0$ with the different $T_{min}$ values in the experiment. There is no discernable correlation between $T_0$ and $T_{min}$. The variable $T_0$ is computed at the very beginning of a SA run, while $T_{min}$ is only used to determine the completion of a SA run, by checking whether the $T$ of the system has reached $T_{min}$.

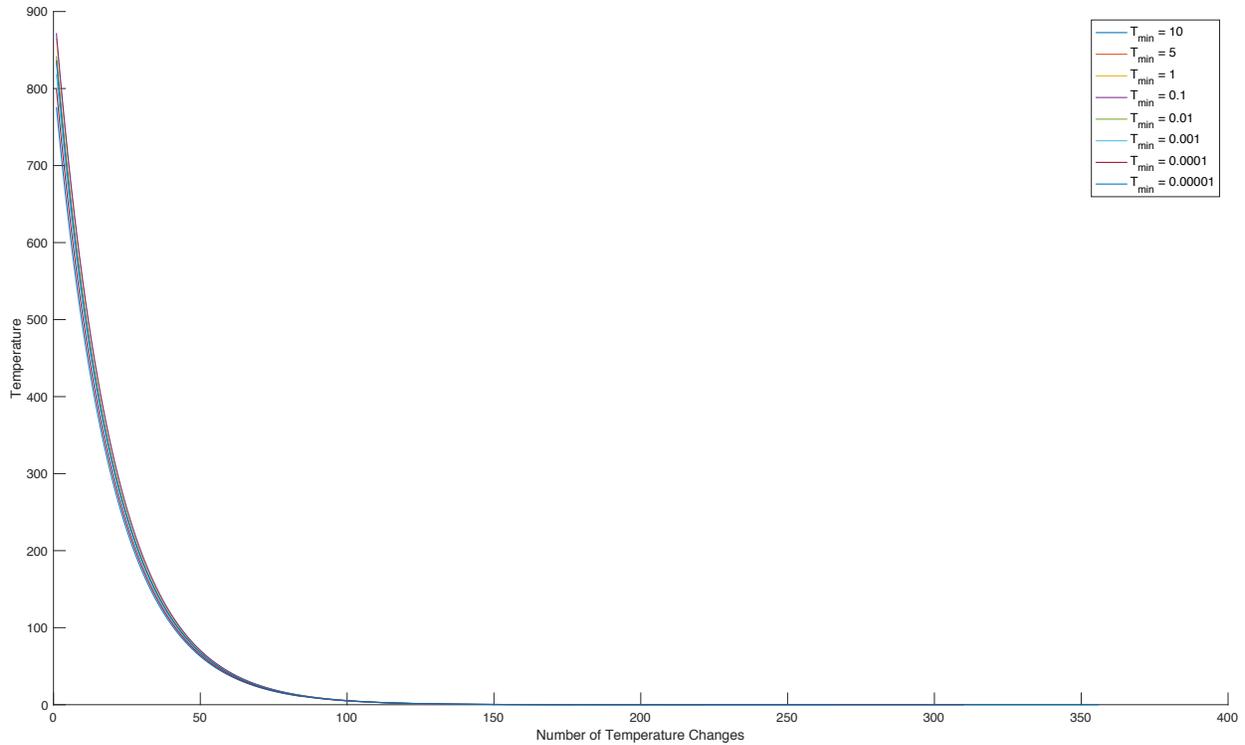

*Figure 53: Temperature Changes with Different Minimum Temperatures*

Figure 53 compares how the value of $T$ changes for each SA run with different $T_{min}$ values. From the figure, it can be seen that the rate at with $T$ changes is almost identical despite the different $T_{min}$ values. The rate at which $T$ changes is controlled by the variables $\alpha$ and $T_0$, neither of which are related to $T_{min}$. In this figure, $\alpha$ is held consistent across the different SA runs, and $T_0$ is generated stochastically using the same $AP$ value. The slight difference in the rate $T$ changes between the different runs comes from the different $T_0$ values, shown in Figure 52. If $T_0$ was the same for each run, the rate at which $T$ changes would have also been the same for all the runs. Therefore, $T_{min}$ does not affect the rate at which $T$ changes for a given SA system.

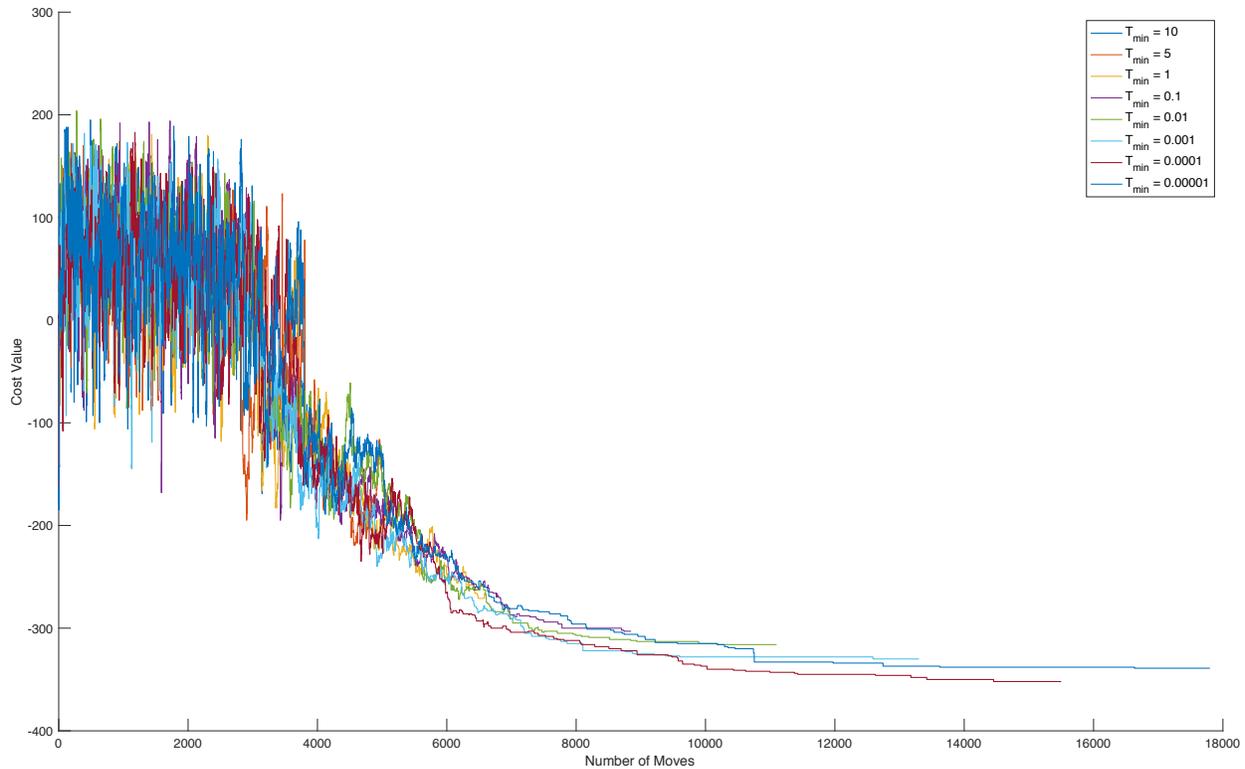

*Figure 54: Cost Value Change with Different Minimum Temperatures*

Figure 54 visualizes the $Cost$ changes of the SA runs using different $T_{min}$ values. From the figure, all of the $Cost$ value fluctuations occurred within the same range. The main factor of differentiation in these results is the duration of each run. As $T_{min}$ decreases, the duration of the SA run increases. The difference in the duration does not appear to be very large for the smaller $T_{min}$ values. This is due to the fact that $T_{min}$ directly controls when the SA run would end. The terminating condition for a SA run is when the $T$ of the system is cooled to below $T_{min}$. Therefore, a large $T_{min}$ would terminate the run early, whereas a smaller $T_{min}$ would keep the system running longer at a very cooled stage, close to 0°C. In addition, it is also observed here that for $T_{min} \leq 0.01$, the $Cost$ value of the system is generally minimized in the same range. However, due to the overlap in data, it is difficult to observe values where $T_{min} > 0.01$. Each $Cost$ change will be broken down in the following figure.

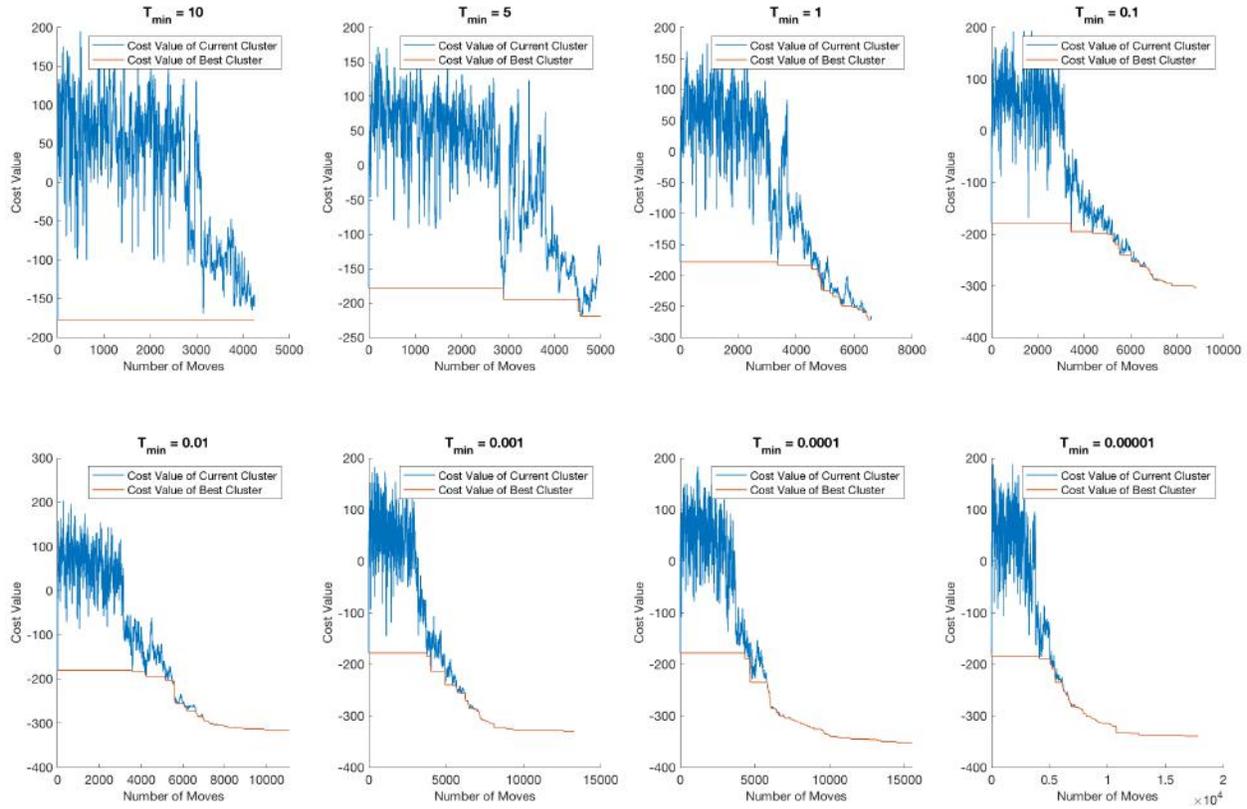

*Figure 55: Cost Value of Current and Best Clusters with Different Minimum Temperatures*

Figure 55 visualizes the change in $Cost$ value of the current and the best cluster for different $T_{min}$ values separately. From this figure, the impact of $T_{min}$ on the quality of the final solution can clearly be seen. For $T_{min}$ = 10, the $Cost$ value of the best clusters never changed from the initial to the final solution. In fact, the SA algorithm using $T_{min}$ = 10 only demonstrated the wild fluctuation stage of the search, and has not entered the $Cost$ minimization stage. For that SA run, the system was terminated too early due to the high $T_{min}$ value, and was not able to find the global optimum. As $T_{min}$ decreases, the $Cost$ value of the best clusters is further minimized, and the duration of the SA run is increased. In addition, the different stages of how the best $Cost$ value changed became more pronounced, where the $Cost$ minimization stage was demonstrated more as $T_{min}$ decreased. The change in the best $Cost$ value was especially significant between $T_{min}$ = 5 to $T_{min} = 0.1$. From these results, it appears that for this SA system, there exist a region for $T_{min}$ where the minimization of the best $Cost$ value happens rapidly. Assessing the figure, this region for $T_{min}$ exists between $10 > T_{min} > 0.1$. Once $T_{min} \leq 0.1$, the minimization of the best $Cost$ slows down significantly. The trade-off for smaller $T_{min}$ is a longer, and thus more computationally expensive, runtime.

Overall, $T_{min}$ directly controls the duration in which SA runs, and thus have a direct impact on the quality of the final LC solution. Between $10 > T_{min} > 0.1$, the quality of the final solution improves rapidly. When $T_{min} \leq 0.1$, the quality still continues to grow, but much less significantly. The trade-off when using a very small $T_{min}$ value is a longer and more computationally expensive SA run. Therefore, $T_{min} \leq 0.1$ is recommended for this particular SA system.

### 5.4.5 Maximum Number of Iterations at Each Temperature Results

Once the $T$ has been cooled in a SA run, a set number of perturbation moves are made while keeping $T$ constant. That set is the number of iterations at each $T$ ($i_T$). This experiment explores the impact of the value of $i_T$ on the outcome of the SA run. The details of this experiment are described in Section 4.5.5. Variables $AP$, $\alpha$, and $T_{min}$ are kept constant, while the value of $i_T$ used are displayed in Table 19.

*Table 19: Experimental Value of Number of Iterations at Each Temperature*

| Parameter | Experimental Value | | | | | | | | | |
|---|---|---|---|---|---|---|---|---|---|---|
| $i_T$ | 10 | 20 | 30 | 40 | 50 | 60 | 70 | 80 | 90 | 100 |

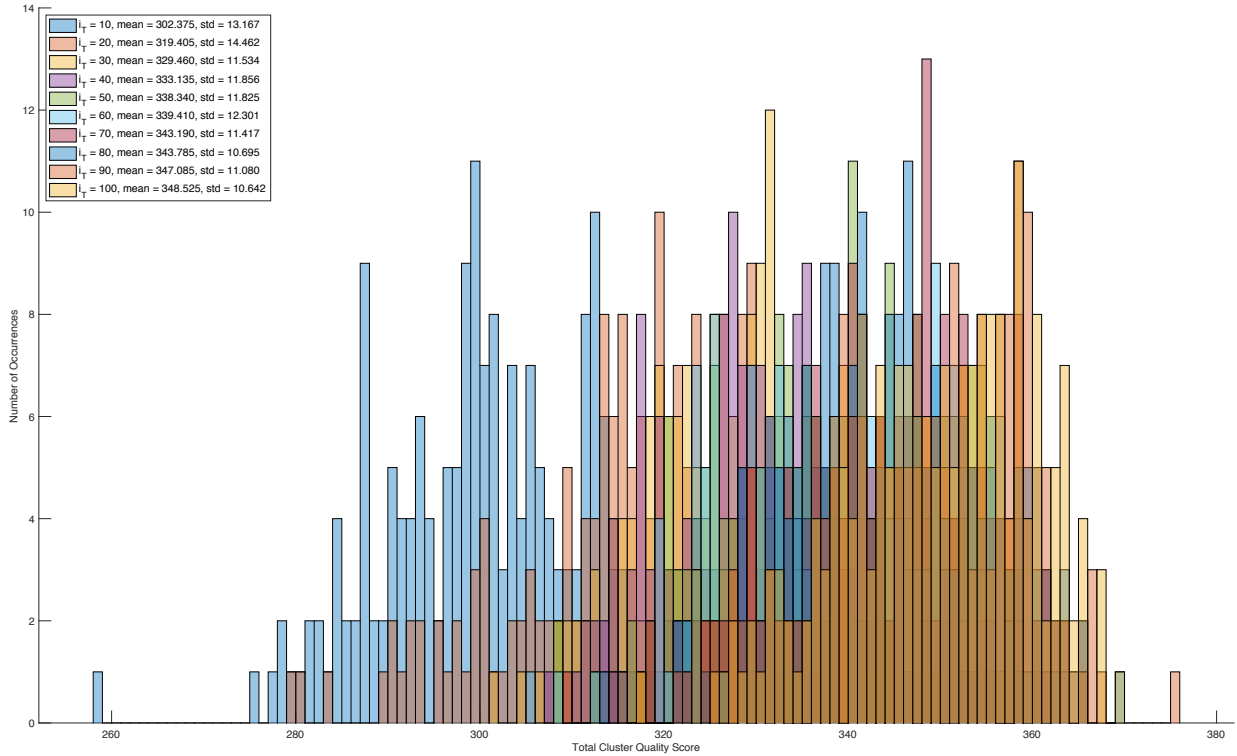

*Figure 56: Histogram of Cluster Refinement Results with Different Number of Iterations at Each Temperature*

Figure 56 displays the cluster refinement results from the 200-run Monte Carlo simulations using different $i_T$ values as a histogram. Every set of clusters produced by a single SA run can act as student LCs. The mean and standard deviation of $S_T$ for each simulation are also displayed. From the figure, it can be seen that the cluster results generally tend to increase in quality as the value of $i_T$ increases. There appear to be large increases between $i_T = 10$ and $i_T = 20$. The increases in cluster quality becomes smaller for larger $i_T$ values, as indicated for the larger amount of overlap in the individual histograms representing those simulations.

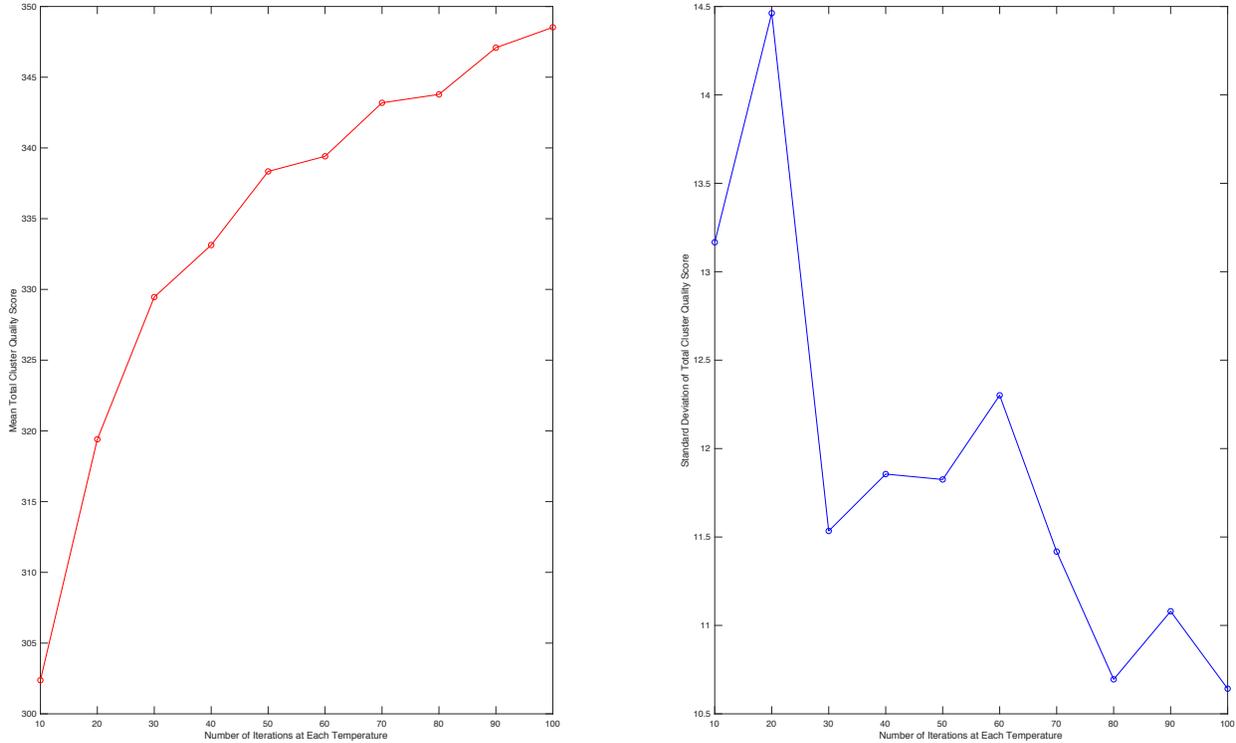

*Figure 57: Mean Total Cluster Quality Score (left) and Standard Deviation of Total Cluster Quality Score (right) with Different Number of Iterations at Each Temperature*

Figure 57 visualizes the mean and standard deviation of $S_T$ against different $i_T$ values. The mean $S_T$ is visualized left in red, whereas the standard deviation of $S_T$ is visualized right in blue. From the figure, it is apparent that as $i_T$ increases, the mean $S_T$ of the simulation also increases. The change in cluster quality for low $i_T$ values are larger than the change in quality when $i_T$ is higher. For $10 \leq i_T \leq 50$, $S_T$ improved by almost 40. However, for $50 \leq i_T \leq 100$, $S_T$ improved by only around 10. The standard deviation of $S_T$ also generally decreases as the value of $i_T$ increases. Overall, larger $i_T$ values lead to an increase in quality and a decrease in variability in the final cluster solution.

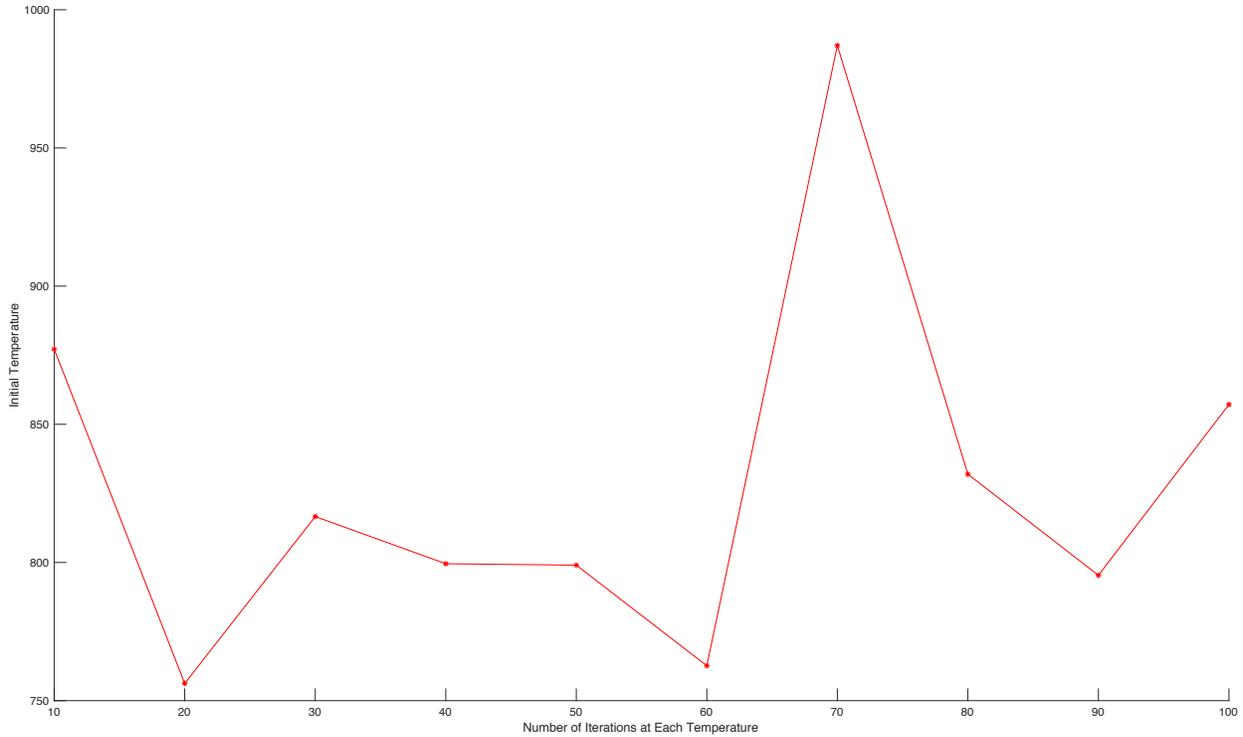

*Figure 58: Initial Temperature with Different Number of Iterations at Each Temperature*

Figure 58 compares the value of $T_0$ with different values of $i_T$. The range of $T_0$ varies between 750 °C and 1000 °C. There is no visible correlation between $T_0$ and $i_T$, as $i_T$ is a variable set prior to the SA run, that controls how many moves are being made at each $T$, whereas $T_0$ is independently calculated at the beginning of a run.

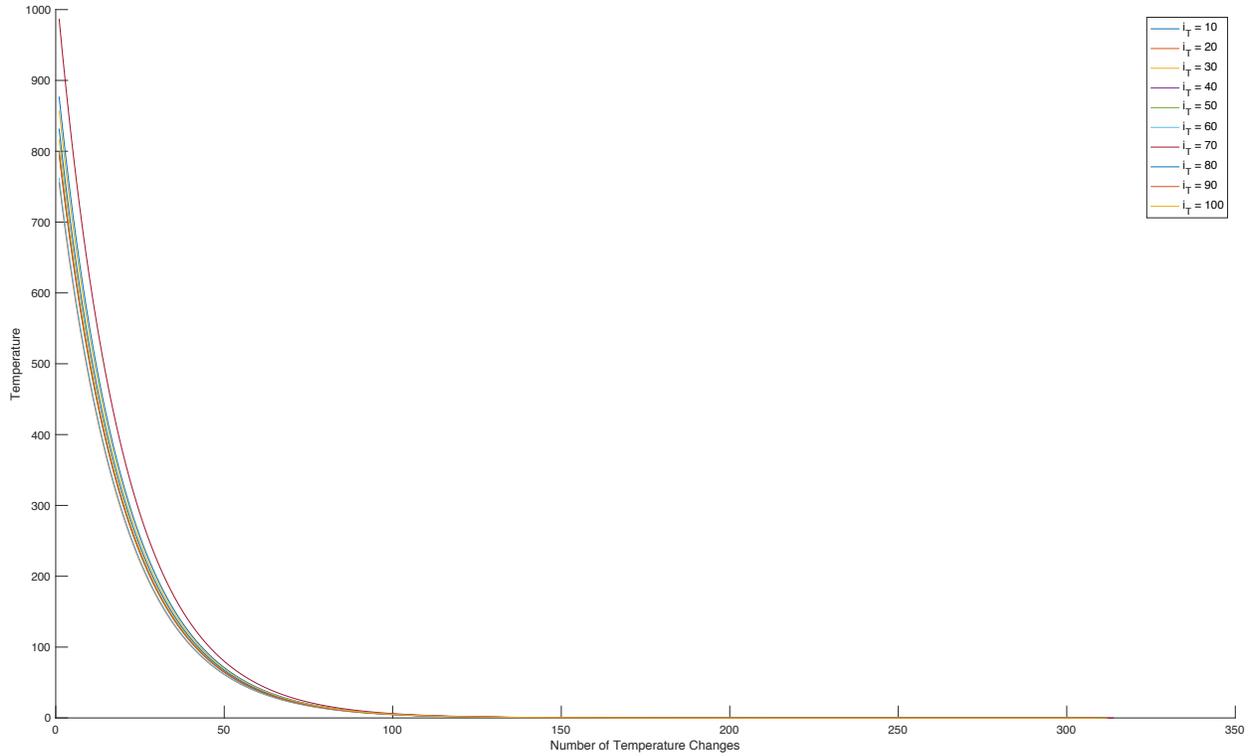

*Figure 59: Temperature Changes with Different Number of Iterations at Each Temperature*

Figure 59 compares the rate at which $T$ changes for each SA run using a different $i_T$ value. Much like in Figure 53, the difference in the rate of $T$ changes is very similar across the SA runs. The rate at which $T$ changes is controlled by both $\alpha$ and $T_0$. Since it was established in the previous figure that $i_T$ does not impact the value of $T_0$, $i_T$ therefore also does not impact the speed at which $T$ changes. Since $\alpha$ is held consistently across this experiment, the stochastically generated value of $T_0$ is what causes the slight variation in the rate at which $T$ is changing. If the same $T_0$ is used across all of the SA runs, the resulting change in $T$ would also be the same.

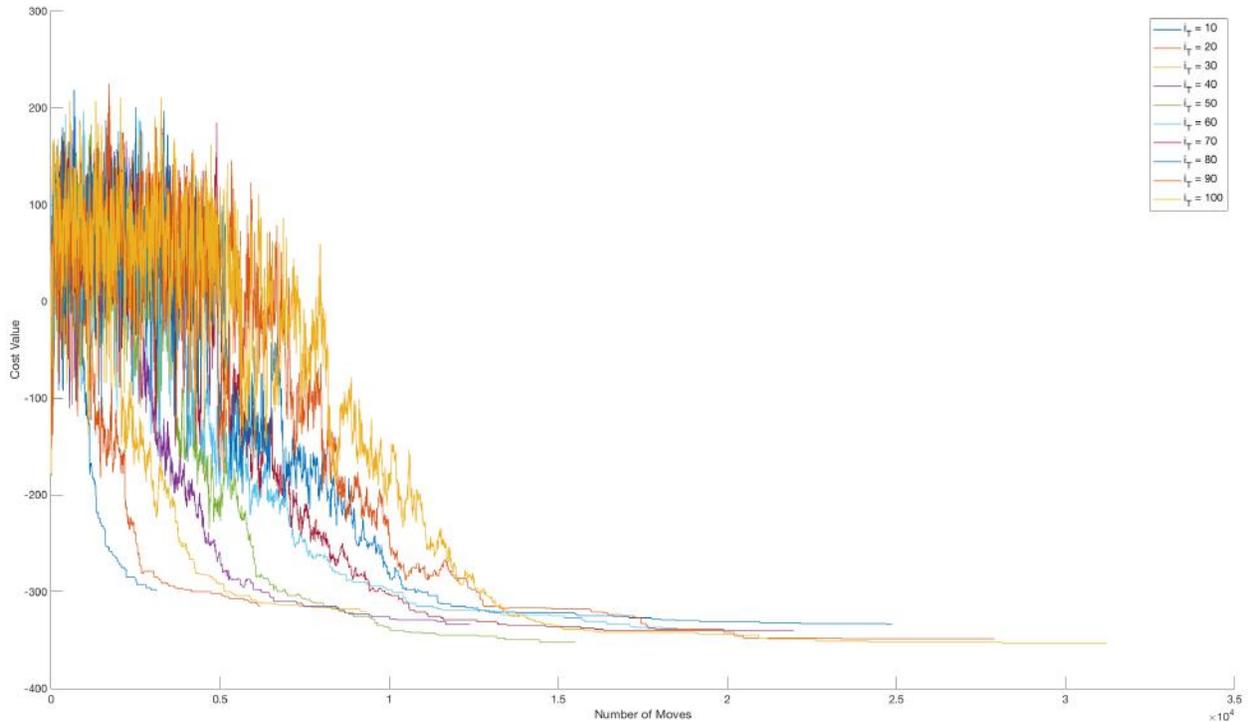

*Figure 60: Cost Value Changes with Different Number of Iterations at Each Temperature*

Figure 60 visualizes the $Cost$ value changes for SA runs using different values of $i_T$. From the figure, the impact of $i_T$ on how $Cost$ values are minimized throughout the SA run is visualized. For $i_T$ with lower values, the SA run terminated very early with a final $Cost$ value that has not been fully minimized. As $i_T$ increases, the duration of the SA run also increases, and the $Cost$ is minimized more and more. The number of moves is equal to $i_T$ multiplied by the number of $T$ reductions in a SA system. If we assume the number of times $T$ is reduced is the same across the different SA runs, then for every increase in 10 iterations for $i_T$, the number of moves is increased by 10 multiplied by the number of $T$ reductions. For example, $i_T = 50$ yields five times the number of moves compared to $i_T = 10$, and $i_T = 100$ will yield a SA run that is 10 times longer than $i_T = 10$. However, the number of $T$ reductions is not the same across the different SA runs. In this experiment, the number of times $T$ changes is controlled by $\alpha, T_0$ and $T_{min}$. Both $\alpha$ and $T_{min}$ are variables held constant across the different SA runs, but $T_0$ differs slightly for each run, as seen in Figure 58. Therefore, the number of times $T$ changes for each SA run is very similar, barring the difference in $T_0$. Overall, $i_T = 100$ will still produce a SA run that is almost 10 times longer than using $i_T = 10$.

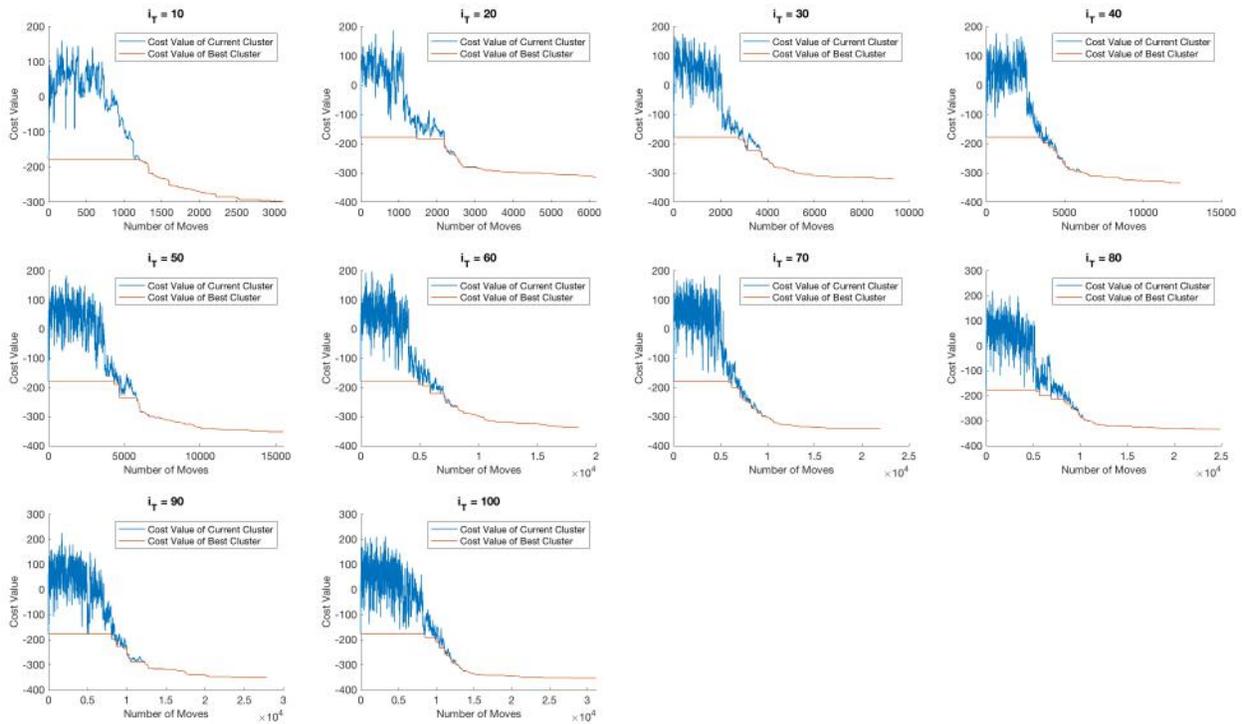

*Figure 61: Cost Value of Best and Current Clusters with Different Number of Iterations at Each Temperature*

Figure 61 compares the $Cost$ values of both the best and current clusters using different $i_T$ values. From this figure, it can be observed that as the value of $i_T$ increases, the $Cost$ value of the best clusters becomes more minimized, and the duration of the SA run increases. Both of these facts are reflected in the results from previous figures. It should be noted that even for $i_T = 10$, the $Cost$ value has moved past the fluctuation stage and entered the minimization stage. Compared to Figure 55, where a high $T_{min}$ prevents the SA algorithm from even entering the $Cost$ minimization stage, it appears that $i_T$ has less impact on the quality of the final solution compared to $T_{min}$. The value of $i_T$ also seems to have less impact on the quality of the final solution than the value of $\alpha$, where a small value of $\alpha$ significantly reduces the amount in which $Cost$ is minimized, as per Figure 49. However, $i_T$ still impacts the quality of the final clusters, as a large $i_T$ produces clusters with more minimized $Cost$ values, thus higher quality clusters.

Overall, $i_T$ changes the duration of a SA run, where an increase in the value of $i_T$ leads to a longer SA run. As a direct result of $i_T$ increasing, the quality of the final clusters also increases. The cluster quality when $10 \leq i_T \leq 50$ increases much faster than when $50 \leq i_T \leq 100$. The final clusters refined by SA can therefore be used as student LCs. While $i_T$ appears to impact the SA system less than variables $\alpha$ and $T_{min}$, it is still important to use a reasonably large $i_T$ value. A value of $i_T \geq 50$ is recommended, since it is a compromise between having high quality final solutions without overextending a computationally expensive SA run.

### 5.4.6 Simulated Annealing Learning Community Refinement Results

As evident from the previous experiments, SA is capable of refining the LCs produced by the BC algorithm to produce LCs with higher quality. The BC algorithm was able to produce clustering solutions with $S_T = 178$, and SA improved upon these results by producing refined LCs with $S_T$ consistently above 300. The SA experimental results recommended the following values for SA algorithmic parameters for optimal LC refinement:

- $N \geq 600$
- $1 > AP \geq 0.9$
- $1 > \alpha \geq 0.9$

- $0.1 \geq T_{min} > 0$
- $i_T \geq 50$

Using these recommendations, a student LC solution refined by SA is tabulated and visualized.

*Table 20: Simulated Annealing Learning Community Refinement Results*

| LC ID | Learning Community Membership | | | | | | | | | |
|---|---|---|---|---|---|---|---|---|---|---|
| 1 | 10 | 40 | 51 | 57 | 67 | 68 | 71 | 76 | 77 | 80 |
| 2 | 20 | 42 | 49 | 55 | 60 | 63 | 69 | 75 | 79 | 81 |
| 3 | 19 | 22 | 26 | 27 | 28 | 38 | 61 | 65 | 70 | 73 |
| 4 | 2 | 21 | 39 | 48 | 62 | 64 | 66 | 72 | 74 | 78 |
| 5 | 1 | 7 | 14 | 18 | 29 | 36 | 47 | 50 | 53 | 59 |
| 6 | 12 | 16 | 35 | 37 | 43 | 45 | 46 | 54 | 56 | 58 |
| 7 | 5 | 6 | 9 | 13 | 15 | 24 | 30 | 31 | 32 | 41 |
| 8 | 3 | 4 | 8 | 17 | 23 | 25 | 33 | 34 | 44 | 52 |
| 9 | 11 | | | | | | | | | |

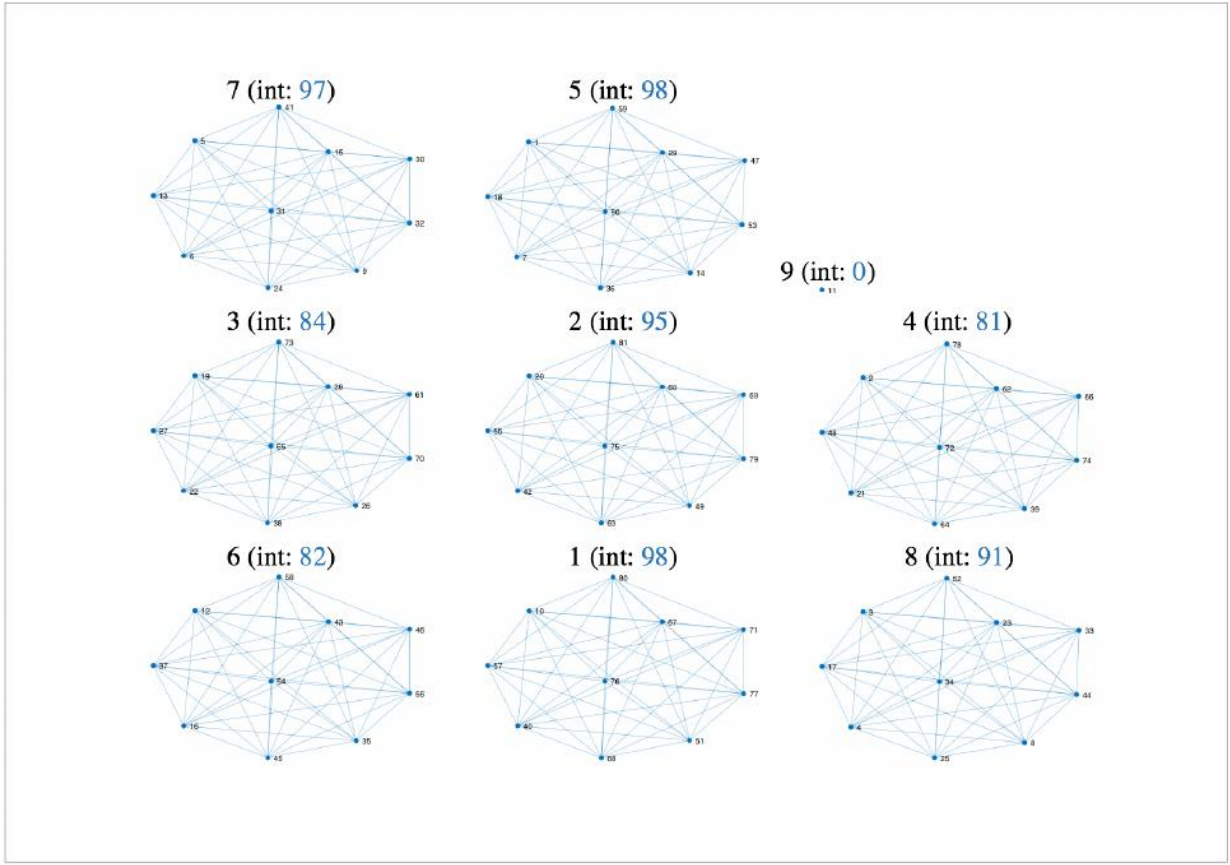

*Figure 62: Simulated Annealing Learning Community Refinement Results*

The preceding Table 20 tabulates the LC configuration as refined by SA, and the preceding Figure 62 illustrates these refined LCs. The SA parameter values used are displayed in Table 21.

*Table 21: Simulated Annealing Parameter Value Used in Learning Community Refinement*

| Parameter | Value |
| --- | --- |
| $N$ | 600 |
| $AP$ | 0.95 |
| $\alpha$ | 0.95 |
| $T_{min}$ | 0.0001 |
| $i_T$ | 50 |

Since the SA algorithm preserves the maximum cluster size set by the initial input LCs, the refined LCs are the same size as the LCs produced by the BC algorithm. There are eight full LCs and one individual serving as a single LC. The number of internal connections within each LC is indicated in blue, alongside the ID of the LC. The refined clusters display a very consistent range of internal connections, between 81 to 98. This is especially notable when compared to the initial input LCs produced by the BC algorithm, visualized in Figure 35. The LCs produced by the BC algorithm showed a much larger range of internal connections, between 41 to 110. The SA algorithm refined the student LCs by balancing out the internal connection among the LCs. Prior to the refinement, there are LCs with very strong connections mixed with clusters with weak connections. After refinement, all LCs have around the same strong connections internally.

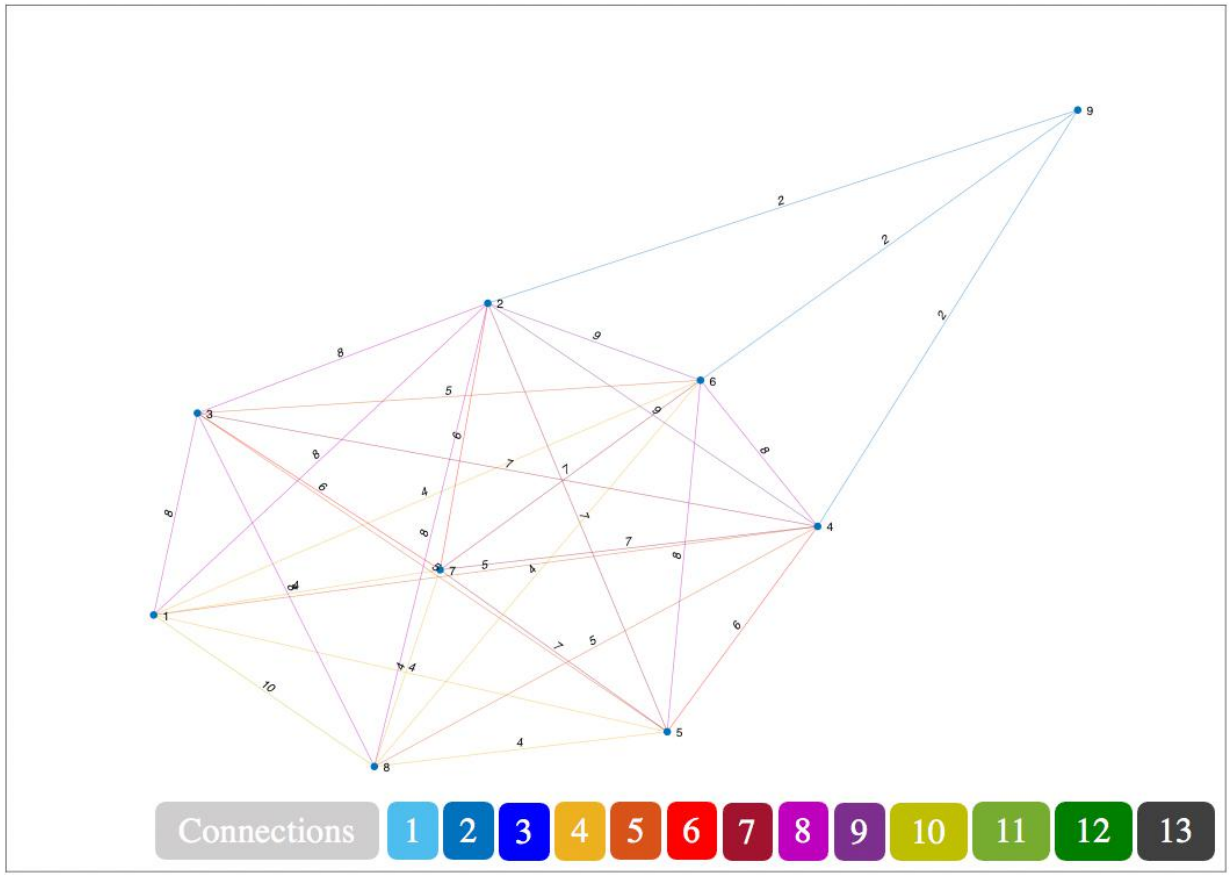

*Figure 63: External Connections between Refined Learning Communities*

The preceding Figure 63 illustrates the external connections remaining between the refined LCs, where the different colours indicates the different number of remaining connections between LCs. These external connections have been minimized as much as possible, per the goal of the SA algorithm, but there are as many as 10 connections unsevered between LCs. Overall, there has been a reduction in the external connections when compared to the LCs produced by the BC algorithm, seen in Figure 36. This may be better illustrated by the following figures.

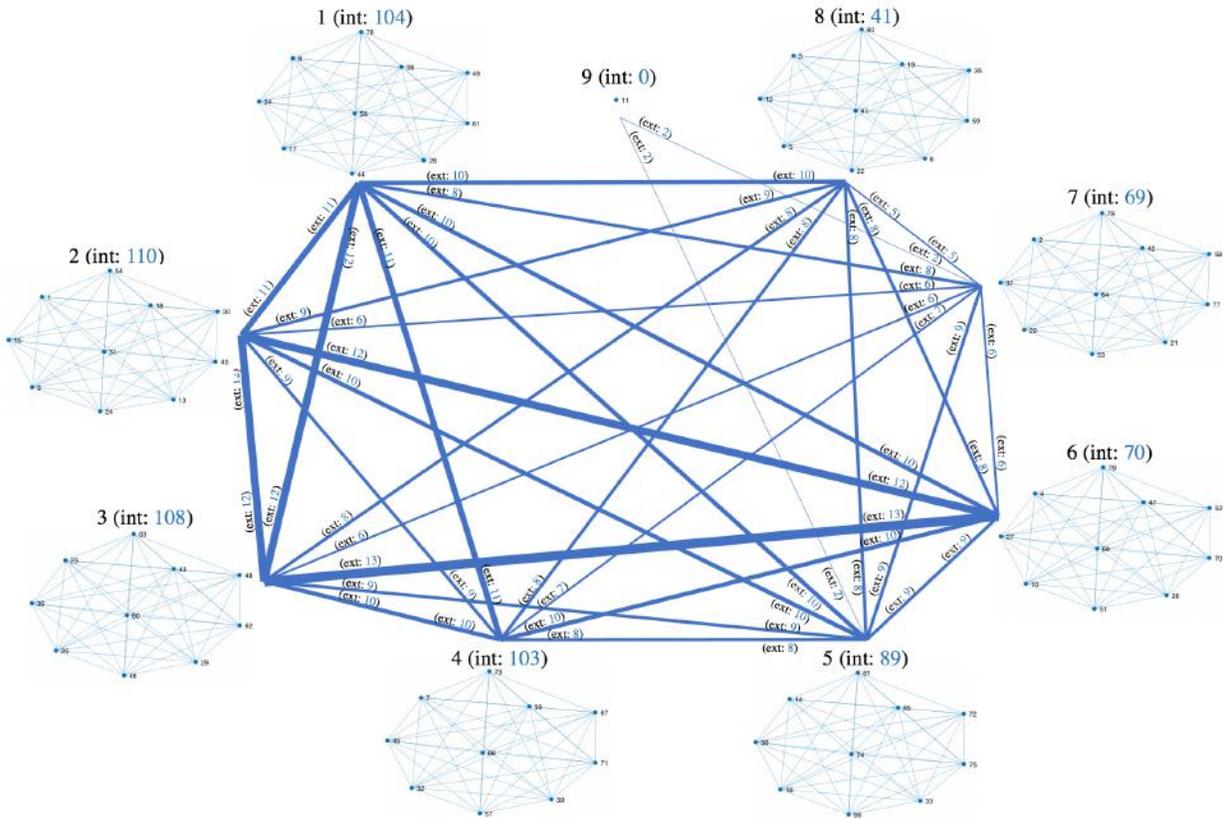

*Figure 64: Connections of Clusters Created by the Best Choice Algorithm*

Figure 64 visualizes the student LC network of the results produced by the BC algorithm, using a maximum cluster size of ten, and the $d_l$ function in equation (4.3). This figure combines the results from Figure 35 and Figure 36, and serves as the initial input LC used for the previous SA experiments. Within the figure, each LC is labeled with a LC ID and the number of internal connections within the LC. Lines connecting LCs represents how many external connections exist between two LCs. Each line is labeled with the number of external connections, and the thickness of the line is proportional to the number of external connections, with a thicker line representing more connections between two LCs. From the figure, it can be seen that there are a lot of pairs of LCs with a significant number of external connections between them, with LC pairs having up to thirteen external connections between them. This set of LCs is one of the best results produced by the BC algorithm, but it can be improved upon using SA, by rearranging the LC configurations in order to balance the strength of internal connections and reduce the number of external connections between LCs.

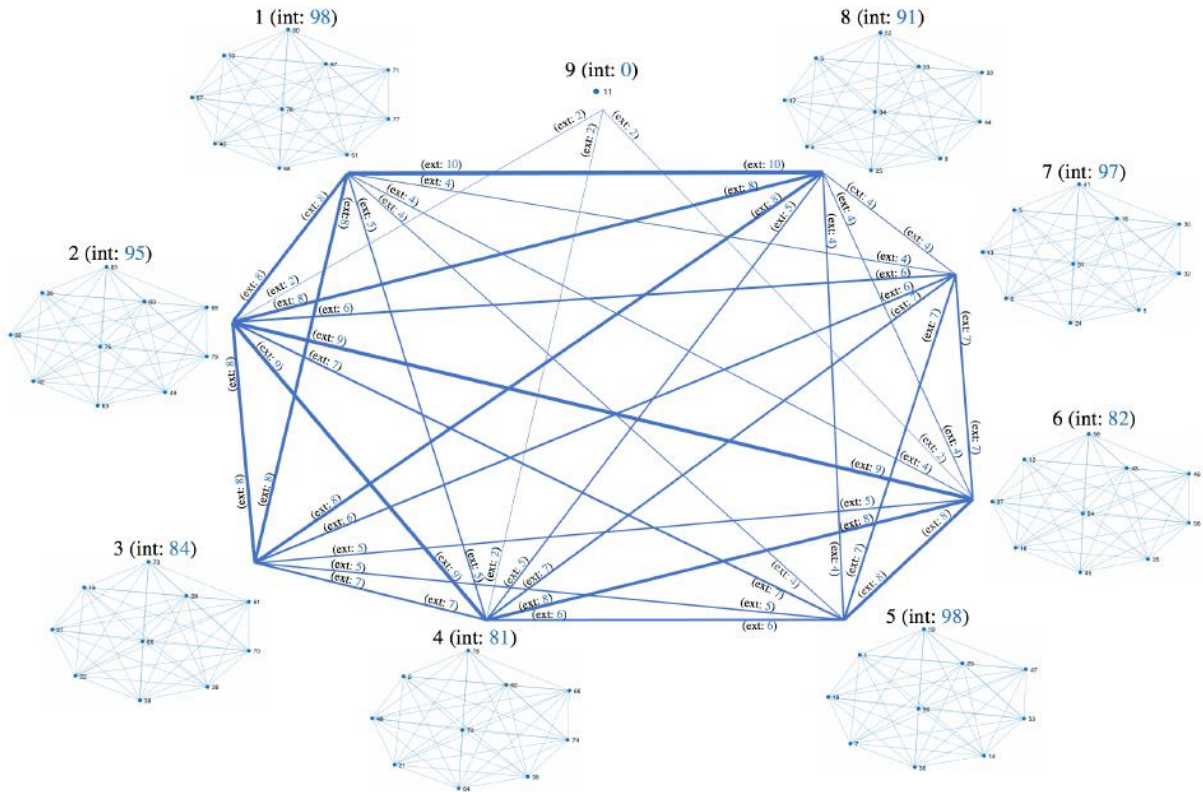

*Figure 65: Connections of Clusters Refined by the Simulated Annealing Algorithm*

Figure 65 visualizes the LC network refined by the SA algorithm, using the parameters listed in Table 21. This figure combines the results from Figure 62 and Figure 63. The LC configurations post-refinement is largely different compared to the initial input LCs in Figure 64, and improvements made to the network can be clearly seen by examining the external connections between LCs. Unlike Figure 64, the external connections between any pair of LCs in Figure 65 are much smaller on average. There are typically five to nine external connections between any two pairs of LCs, as opposed to ten to thirteen connections between a pair of LCs prior to refinement. As mentioned earlier, the internal connections of each LC are more balanced, ranging from 81 to 98, as opposed to 41 to 110 in the LCs prior to refinement. Through the SA cluster refinement process, the LCs in the network became more balanced internally, and less connected externally. The $S_T$ increased from 178 to 352 during the refinement process. The evolution of the LC quality from Figure 64 to Figure 65 demonstrated SA's ability to refine LCs created by the BC algorithm, by showing the improvement in the quality of the resulting solution.

## 5.5 Learning Community Creation and Refinement for Large Datasets

Previous sections demonstrated the steps necessary to process student enrollment data, and how VLSI algorithms can be used to create and refine student LCs. Together, these steps form the basis of the LC Creation and Refinement Framework, introduced in Figure 8. This framework should be able to create student LCs without distinguishing the students' faculty or department affiliation, thus usable on a larger dataset than the 3rd year ENEL Fall 2020 enrollment dataset used thus far.

### 5.5.1 Fall 2020 and Winter 2021 Student Enrollment Network

The LC Creation and Refinement Framework will be tested on a large enrollment dataset obtained from the University of Calgary Office of the Registrar. This dataset contains both Fall 2020 and Winter 2021 student enrollment information for 3939 students across five different faculties, and was fully anonymized by the Office of the Registrar. Specifically, the dataset contains students in the following years and programs, as listed in the following table.

*Table 22: Fall 2020 and Winter 2021 Student Enrollment Dataset*

| Semester | Year in Program | Major | Faculty |
|---|---|---|---|
| Fall 2020 and Winter 2021 | 1 | Business (BUSI and GENL) | Haskayne School of Business |
| | 2 | | |
| | 3 | | |
| | 4 | | |
| | 1 | Economics (ECON) | Faculty of Arts |
| | 2 | | |
| | 3 | | |
| | 4 | | |
| | 1 | Political Sciences (POLI) | |
| | 2 | | |
| | 3 | | |
| | 4 | | |
| | 1 | Natural Sciences (NTSC) | Faculty of Sciences |
| | 2 | | |
| | 3 | | |
| | 4 | | |
| | 1 | Nursing (BNDE – Direct Entry for High School Applicants, BNTR – Transfer | Faculty of Nursing |
| | 2 | | |
| | 3 | | |
| | 4 | | |

| | | Students, BNDH – Degree Holder) | |
| | 2 | Electrical Engineering (ENEL) | |
| | 3 | | |
| | 4 | | |
| | 2 | Software Engineering (ENSF) | Schulich School of Engineering |
| | 3 | | |
| | 4 | | |
| | 2 | Mechanical Engineering (MENG) | |
| | 3 | | |
| | 4 | | |
| | 2 | Energy Engineering (ENER) | |
| | 3 | | |
| | 4 | | |

Table 22 provides an overview of the Fall 2020 and Winter 2021 enrollment dataset, including the students' year in program, major, and faculty. This dataset offers a cross-section of the undergraduate programs offered at the University of Calgary, within five major faculties at this university. The dataset also contains both the unique student ID and their course enrollment, the information necessary for the enrollment network modeling stage. Prior to modeling this data for LC creation and refinement, the dataset will be separated into the Fall and the Winter semesters. The student enrollment network would differ between the Fall and the Winter semesters, as the courses they are enrolled in changes, thus the students they are in contact with also changes.

Once the dataset is separated into Fall and Winter semesters, they enter the enrollment network modeling stage by generating the *Adjacency Matrix (A)* and *Connectivity Matrix (C)*. Both the Fall and the Winter student enrollment networks will be processed as Fully Dense Networks. The following figures visualizes both matrices for both the Fall and Winter datasets.

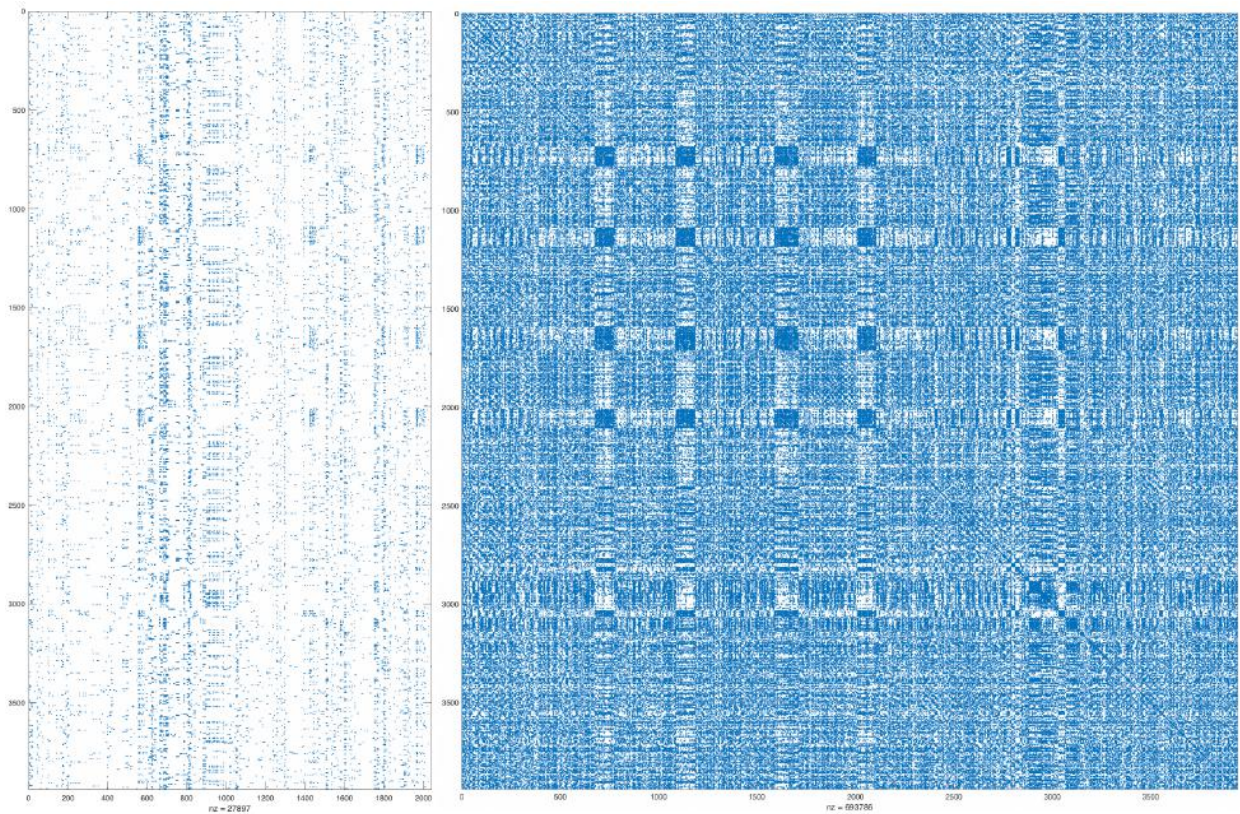

*Figure 66: Adjacency Matrix (left) and Connectivity matrix (right) of Fall 2020 Student Enrollment Network (Fully Dense Network)*

Figure 66 visualizes the *A* Matrix (left) and *C* Matrix (right) of the Fall 2020 student enrollment network. In this semester, 3939 students are enrolled in over 2000 courses together, as seen in the *A* Matrix. From the *C* Matrix, it can be observed that there are pockets of student connections that are denser than other connections. It could be inferred that those students belong in the same program and year. Overall, compared to the *A* and *C* matrices of the 3rd year ENEL Fall 2020 student enrollment network, shown in Figure 9, this figure highlights the difference in scale and complexity of this dataset.

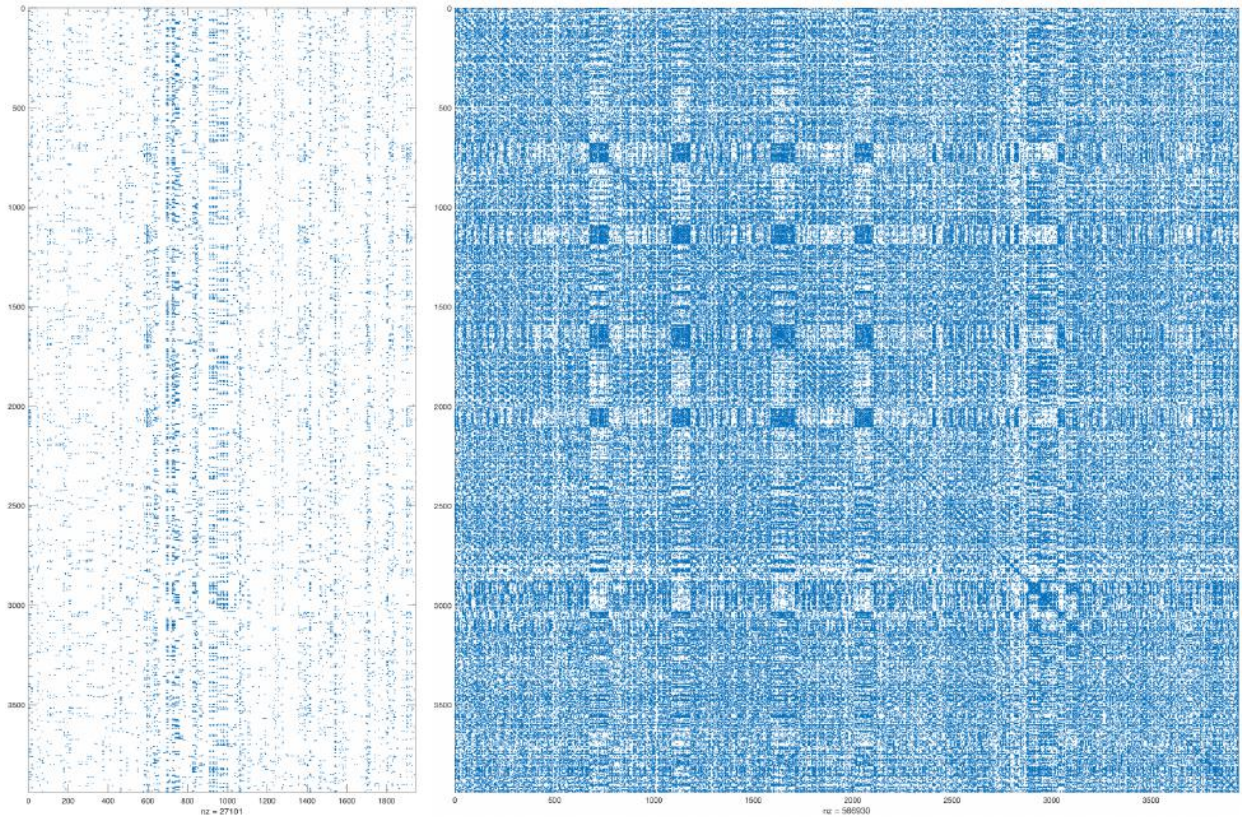

*Figure 67: Adjacency Matrix (left) and Connectivity matrix (right) of Winter 2021 Student Enrollment Network (Fully Dense Network)*

Figure 67 visualizes the *A* Matrix (left) and *C* Matrix (right) or the Winter 2021 student enrollment network. Visually, this figure is very similar to Figure 66, where students are still very connected, sometimes in denser pockets. The 3939 students are enrolled in only over 1800 courses in the Winter 2021 semester, compared to over 2000 in the Fall 2020 semester. The presence of less courses in the semester implies there would also be fewer overall connections between the students in this semester. Both the Fall and Winter enrollment networks will be visualized in the following figures.

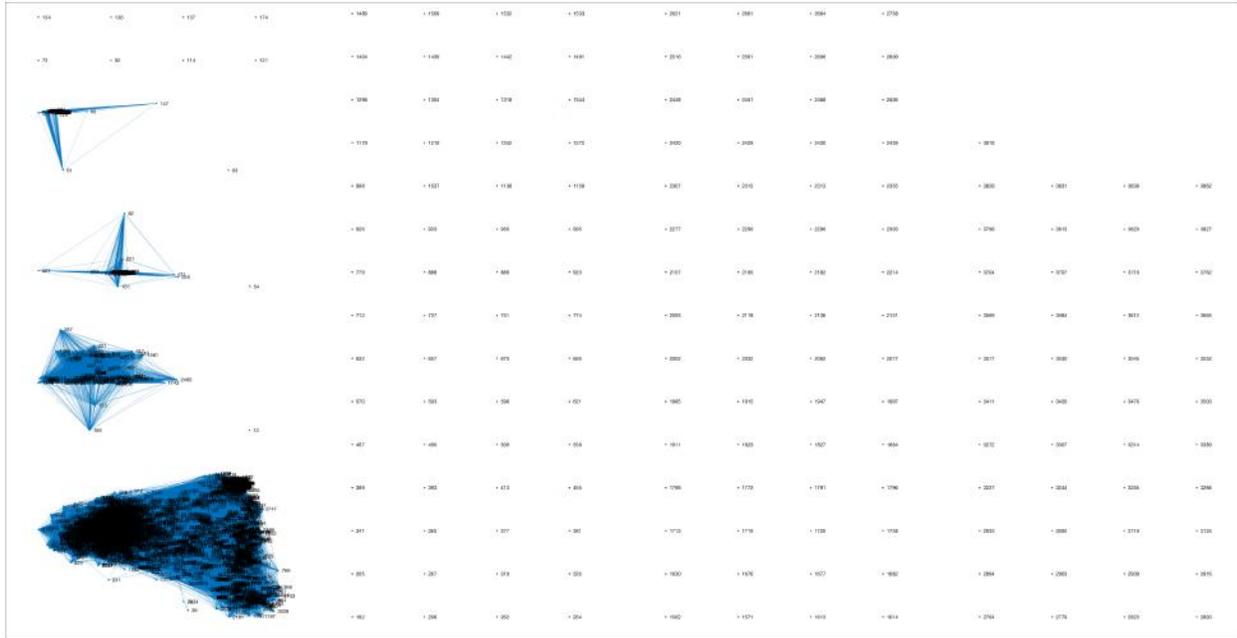

*Figure 68: Connectivity Graph of Fall 2020 Student Enrollment (Fully Dense Network)*

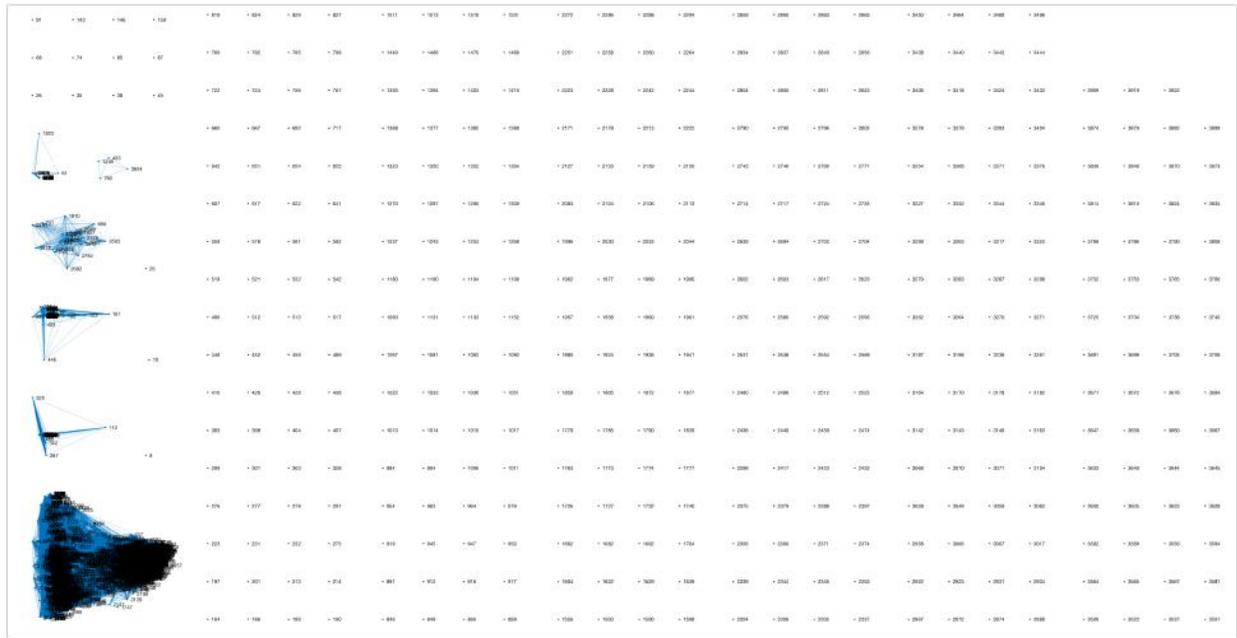

*Figure 69: Connectivity Graph of Winter 2021 Student Enrollment (Fully Dense Network)*

Figure 68 and Figure 69 visualizes the Fall 2020 and Winter 2021 student enrollment networks respectively. Both figures appear similar to each other, where there are different clusters of smaller student networks nested within the larger network. For both semesters, there are one very dense network and several smaller independent networks. The smaller independent networks are all different cohorts in the undergraduate nursing program, between the three routes of direct entry from high school (BNDE), degree holder (BNDH), and transfer (BNTR). Students in these nursing programs are not taking courses in any other programs, and therefore have formed their own smaller enrollment networks. The large dense network includes students in all other programs and majors, except nursing. For both the Fall and the Winter semesters, there are a collection of students not connected to any other students. These students are not taking courses in the given semester, but are enrolled in courses in the other semester. In the Fall 2020 semester, there are 176 students not taking courses, but are taking courses in the Winter 2021 semester. Conversely, there are 414 students taking courses in the Fall 2020 semester that are not taking courses in the Winter 2021 semester. This fact aligns with the observation that there are also less courses being taken in the Winter 2021 semester. The following FIGURE visualizes these nested student enrollment networks for both the Fall 2020 and Winter 2021 semesters, without the students not enrolled in courses.

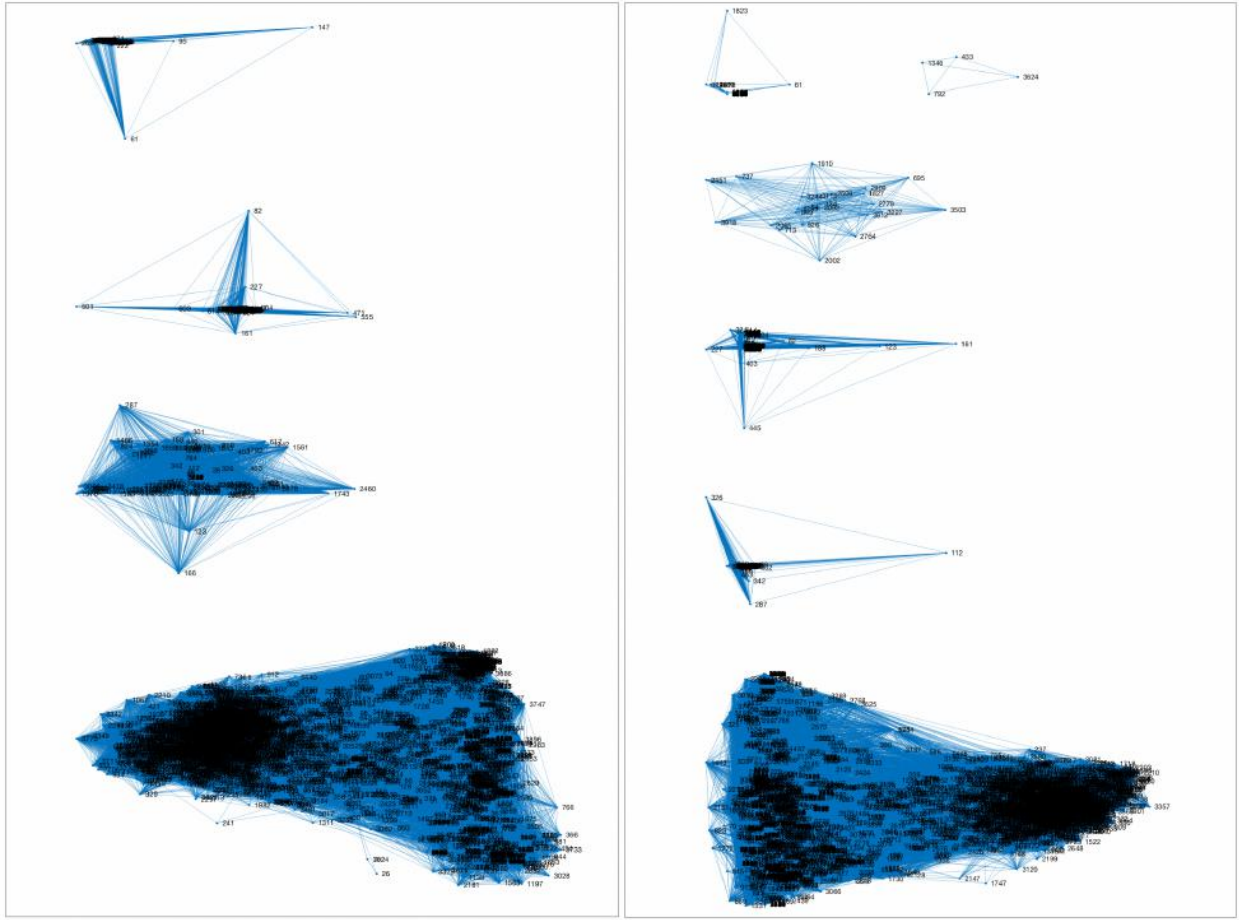

*Figure 70: Enlarged Connectivity Graph of Fall 2020 (left) and Winter 2021 (right) Student Enrollment (Fully Dense Network)*

### 5.5.2 Optimal Learning Community Creation Results for Large Datasets

Following the network modeling stage, the Fully Dense Fall 2020 and Winter 2021 student enrollment networks are used to create LCs using the BC algorithm. The maximum cluster size is set to ten for both the Fall 2020 and the Winter 2021 networks, and the $d_l$ function will be used to generate the LCs, as per recommendation in Section 5.3. The following figures visualizes the LCs created with the BC algorithm.

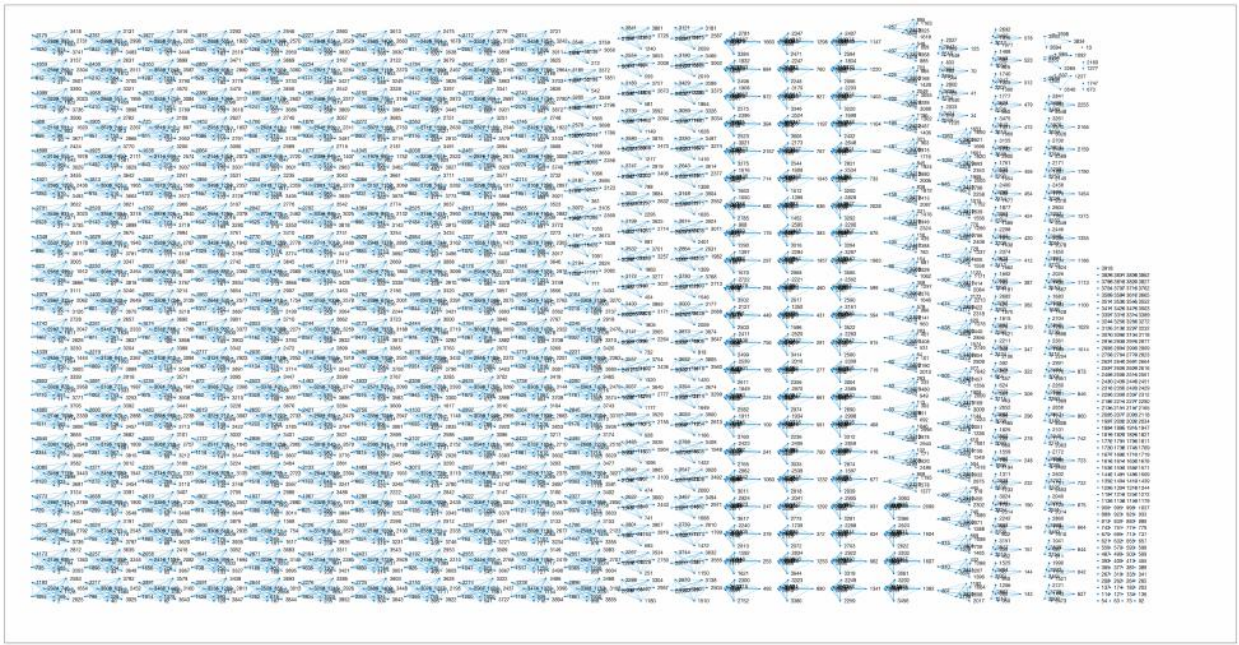

*Figure 71: Optimized Learning Community Creation Results for Fall 2020 Student Enrollment Network*

Figure 71 depicts the set of LCs created by the BC algorithm for the Fall 2020 student enrollment network. Overall, 427 LCs were created, with sizes ranging from two to ten students. The remaining students are unclustered due to their lack of course enrollment during the semester. The quality of the resulting LCs is $S_T$ = -5175. Since LC quality is reflected by maximizing the value of $S_T$, the resulting $S_T$ value indicates that there are many unsevered connections for each LC.

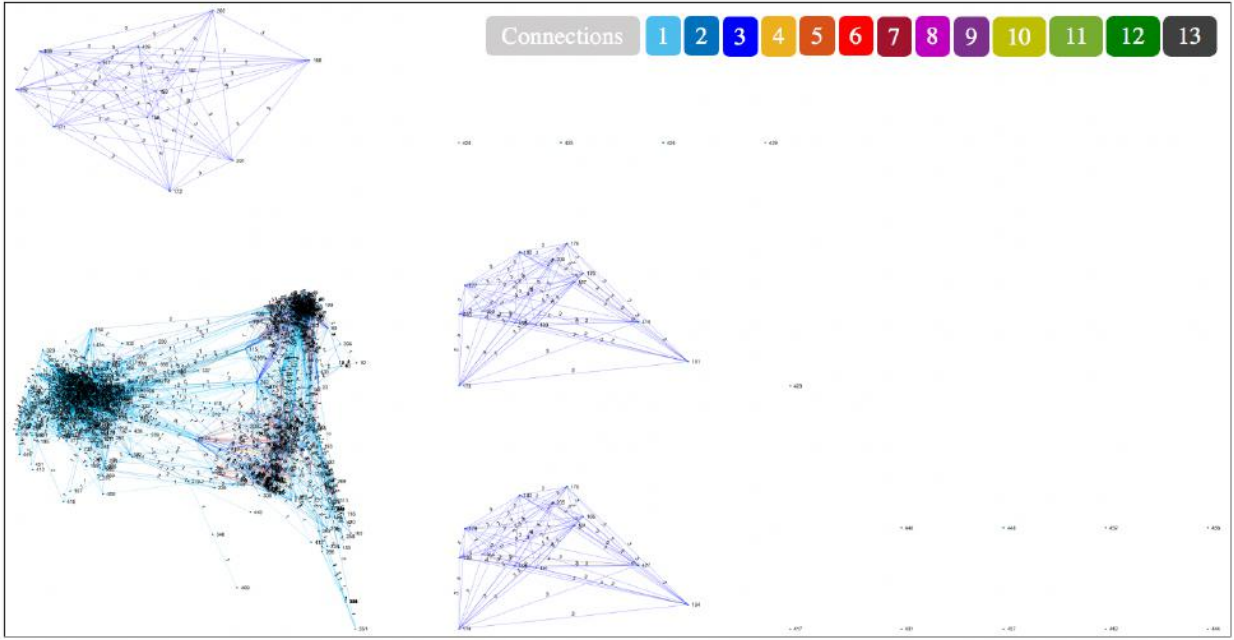

*Figure 72: External Connections between Learning Communities for Fall 2020 Student Enrollment Network*

Figure 72 visualizes the external connections between the LCs created for the Fall 2020 student enrollment network, where the number of connections between LCs are indicated by different colours. From the figure, it could be seen that there remain several collections of LCs, including one large and densely connected group of LCs. While the LCs are still connected externally, many LC pairs are connected by one to three connections, which is not a lot. The issue lies in the fact that many LCs are connected to many other LCs, and the total number of external connections adds up. From these results, the LCs created should be further refined using SA to increase the overall quality of the LCs.

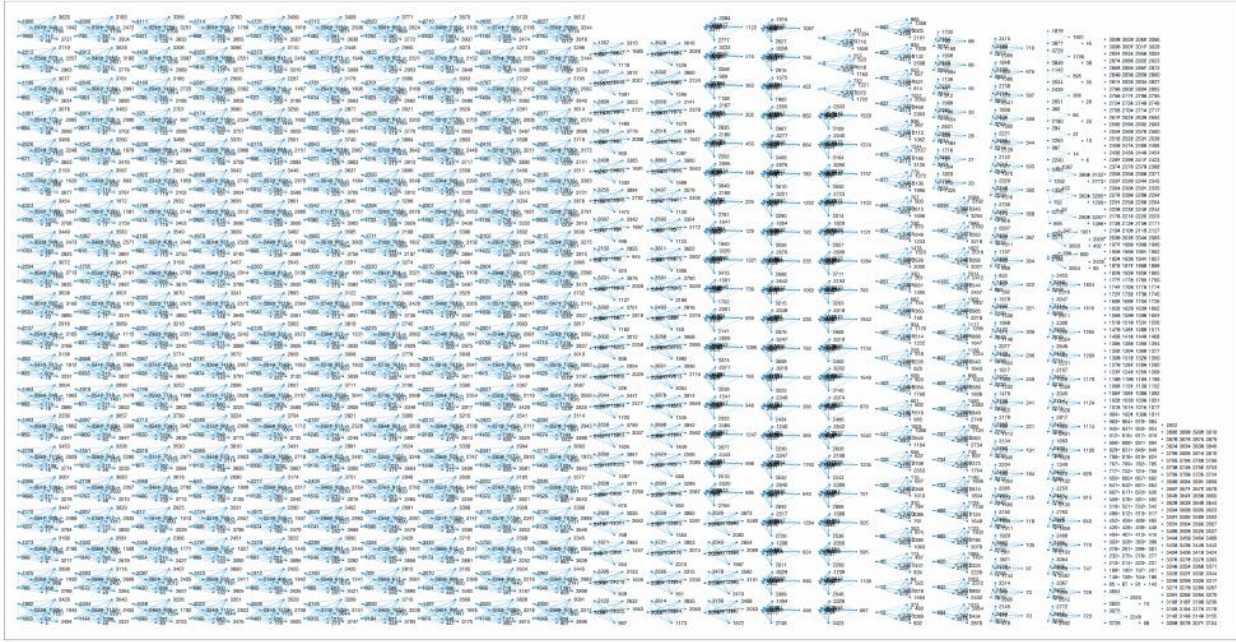

*Figure 73: Optimized Learning Community Creation Results for Winter 2021 Student Enrollment Network*

Figure 73 depicts the LCs created using the BC algorithm for the Winter 2021 student enrollment network. Overall, 410 LCs were created, with sizes ranging from two to ten. The remaining students are unclustered due to their lack of course enrollments in this semester. Compared to the Fall 2020 student enrollment network in Figure 71, less LCs were created. This is possibly due to the fact that there were also less students enrolled in courses in the Winter 2021 semester. The quality of the LCs for this network is $S_T$ = -5387, which is not ideal given the goal to maximize $S_T$. Also, the quality of the LCs created for this network is lower than the LCs for the Fall 2020 network.

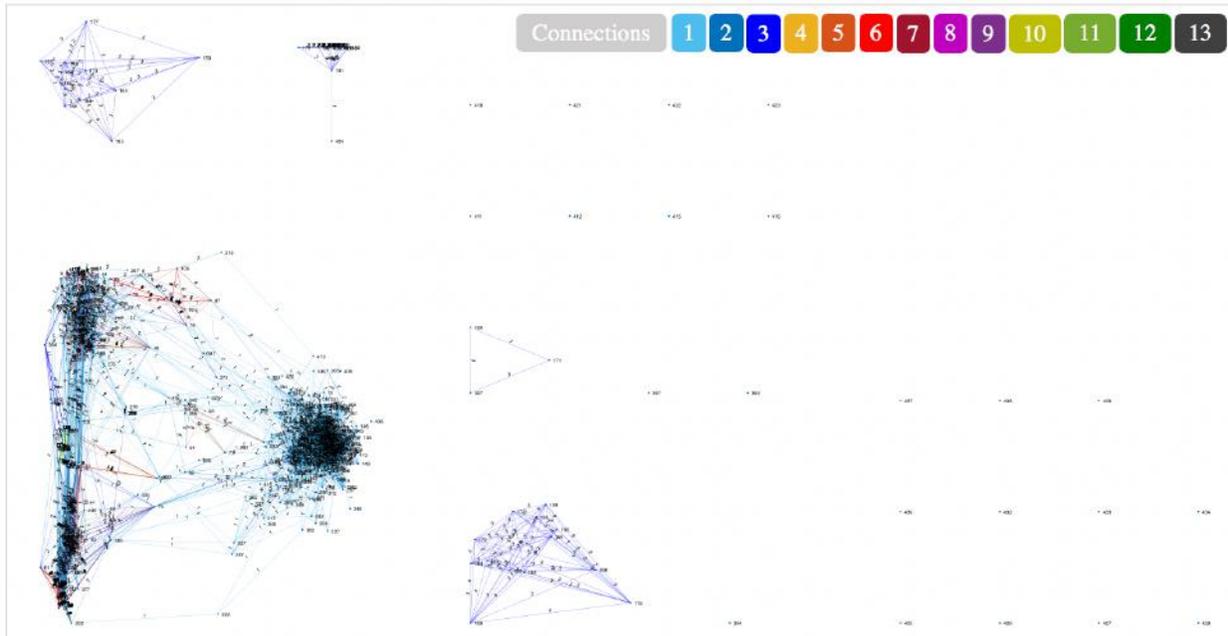

*Figure 74: External Connections between Learning Communities for Winter 2021 Student Enrollment Network*

Figure 74 visualizes the external connections between LCs created for the Winter 2021 student enrollment network. Much like Figure 72, LCs in the Winter enrollment network are still externally connected in several collections, and one very densely connected group of LCs. Many of the LC pairs are connected by one to three external connections, but most individual LCs are connected to many other LCs. Overall, the BC algorithm created student LCs for the very large enrollment networks in the Fall 2020 and Winter 2021 semesters. However, the quality of these LCs leaves a lot to be desired. These LCs will serve as the initial inputs for the refinement stage using SA.

### 5.5.3 Stochastic Learning Community Refinement Results for Large Datasets

Following the LC Creation stage using the BC algorithm, the created LCs are refined with the SA algorithm during the LC Refinement stage. SA preserves the maximum cluster size as well as the number of LCs created. Therefore, the refined LCs will still have a maximum cluster size of ten.

For the Fall 2020 student enrollment network, there will be 427 LCs post-refinement, whereas there will be 410 LCs post-refinement for the Winter 2021 network. For both networks, SA will refine the input LCs using the parameter values listed in Table 21, as per recommendation in Section 5.4. The following figures visualizes the LCs refined by SA.

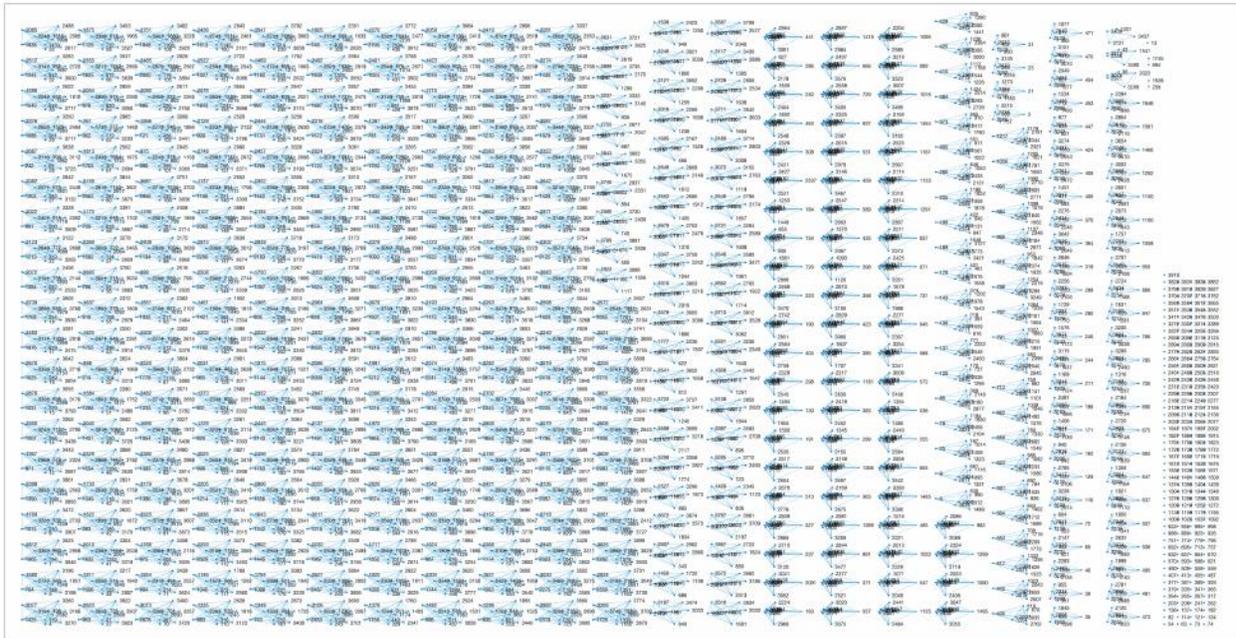

*Figure 75: Stochastic Learning Community Refinement Results for Fall 2020 Student Enrollment Network*

Figure 75 visualizes the refined LCs following the SA algorithm for the Fall 2020 student enrollment network. The number of LCs remains the same as Figure 71, where these LCs were first created. Visually, the refined LCs appears very similar to the LCs prior to refinement. However, the quality of these LCs has significantly improved since they were first created by the BC algorithm. For the LCs post-refinement, $S_T$ = -2048. The value of $S_T$ increased by 3127 during the refinement process, significantly increasing the quality of the LCs.

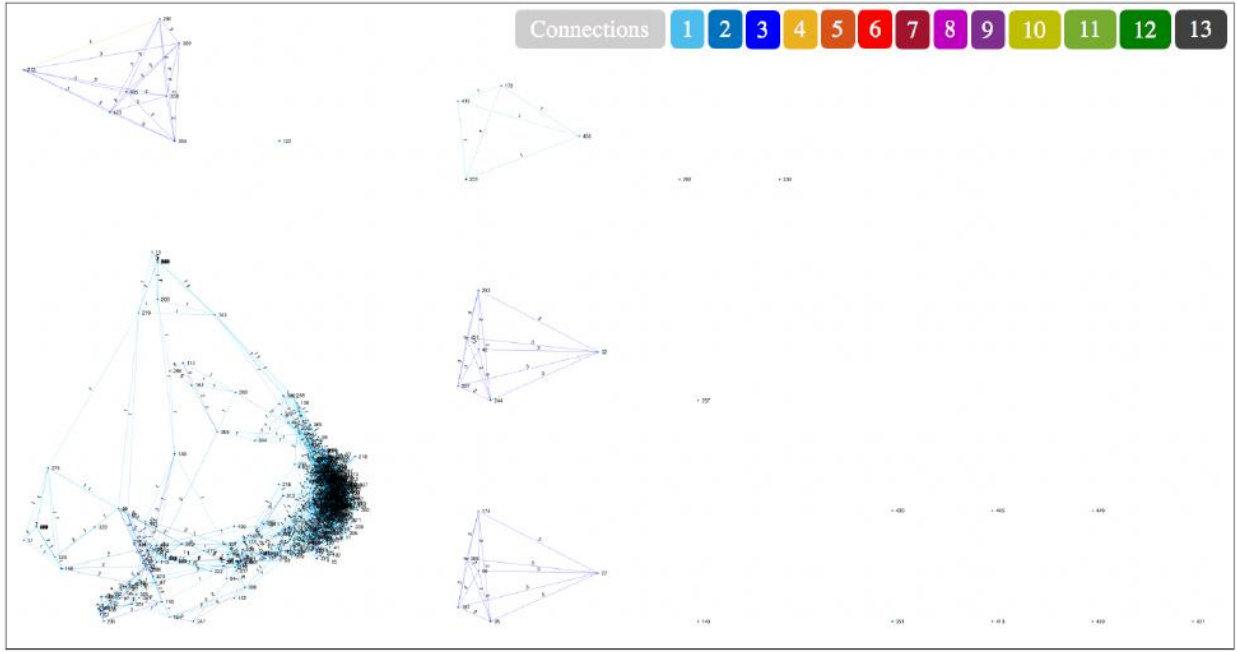

*Figure 76: External Connections between Learning Communities for Fall 2020 Student Enrollment Network*

Figure 76 visualizes the external connections between LCs post-refinement, for the Fall 2020 student enrollment network, where the connections are indicated by different colours. Compared to the LCs' external connections pre-refinement in Figure 72, the LCs are much less densely connected in this figure. For pairs of LCs, they are mostly still minimally connected, with one to three connections. While SA has significantly improved the quality of the LCs for the Fall 2020 student enrollment network, there is still a group of tightly connected LCs that the refinement process was not able to remove.

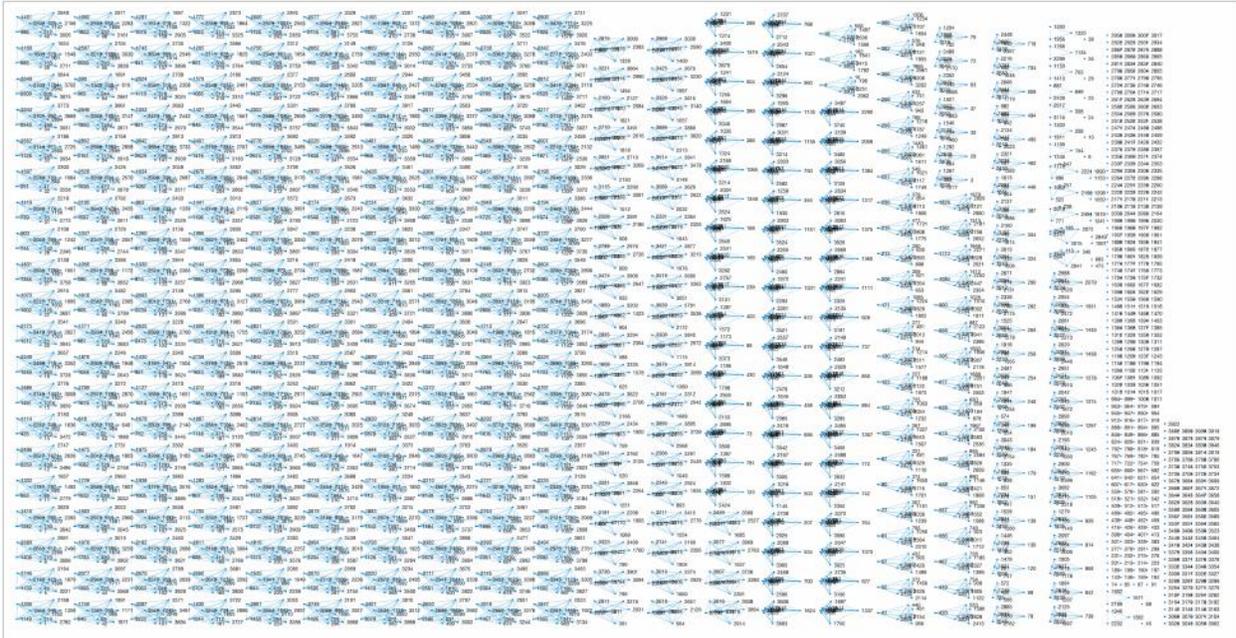

*Figure 77: Stochastic Learning Community Refinement Results for Winter 2021 Student Enrollment Network*

Figure 77 depicts the LCs post-refinement using the SA algorithm, for the Winter 2021 student enrollment network. Using the LCs created by the BC algorithm in Figure 73 as a starting point, SA preserved the number of LCs and the maximum cluster size during the refinement process. Visually, it is difficult to distinguish the post-refinement LCs from the LCs when they were first created. However, the quality of these LCs has dramatically improved, where $S_T$ = -839. Compared to the pre-refinement LCs, whose $S_T$ = -5387, the SA refinement process increased the LC quality by 4548. This is also a much larger increase in quality when compared to the refined LCs for the Fall 2020 network.

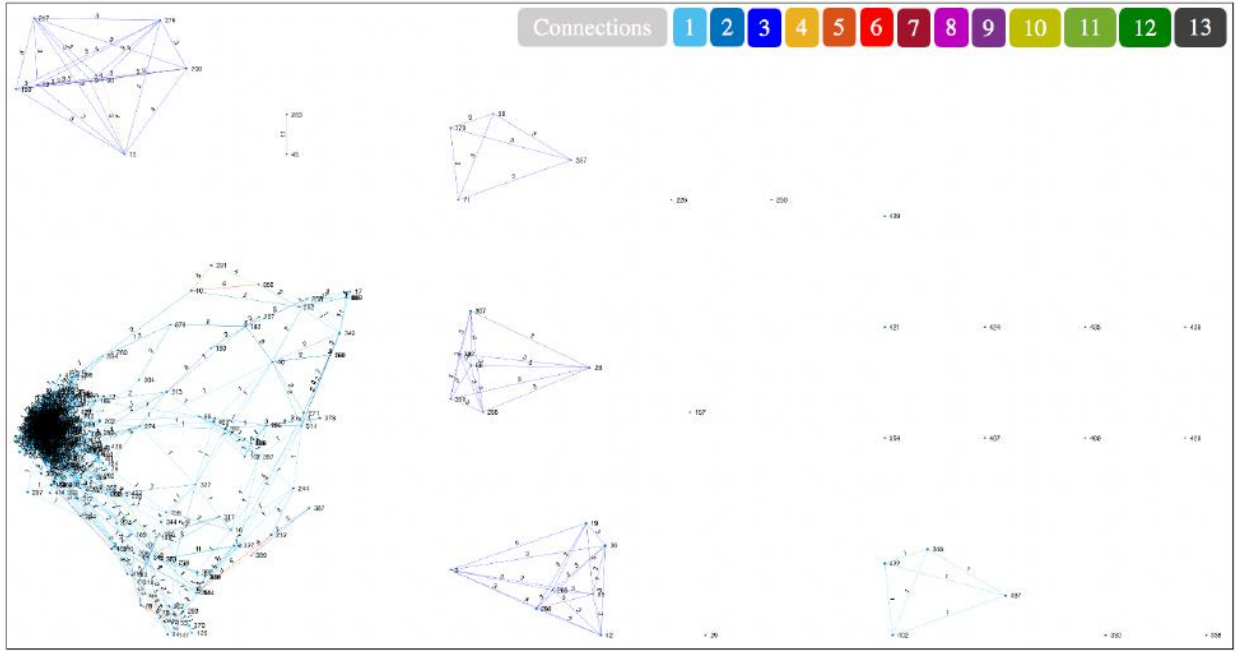

*Figure 78: External Connections between Learning Communities for Winter 2021 Student Enrollment Network*

Figure 78 visualizes the external connections between the refined LCs for the Winter 2021 student enrollment network. Compared to the external connections between the pre-refined LCs in Figure 74, this figure depicts fewer external connections, and there are on average between one to three connections between most pairs of LCs. However, much like the external connections between the refined LCs for the Fall 2020 network in Figure 76, there is still one group of densely connected LCs that the SA refinement process was not able to remove. Overall, the refinement stage was able to significantly improve the quality of the LCs, but there remains some room for improvement.

## 5.6 Discussion of Results

The results in this chapter demonstrated the potential application of modified VLSI algorithms in creating and refining LCs to support online and blended learning in a group environment. Section 5.2 demonstrated how to convert a student enrollment dataset into a network similar to a circuit netlist. This was possible because students enrolled in a university is inherently similar to a circuit. Students connected by the mutual courses they are enrolled in forms an enrollment network similar to how cells in a circuit are connected by nets that forms the circuit netlist. A major assumption in this thesis is that these two types of networks are equivalent, therefor the VLSI algorithms designed for the circuit netlist can be applied onto the enrollment network. However, students enrolled in a university are also not similar to a circuit netlist in many ways. Individual circuit components are designed for a specific function, and are set to always perform that function. Therefore, a circuit netlist is inherently predictable, because all components will do what they are designed for if there are no malfunctions. However, students are not inherently predictable, because they can form and break connections independent from their course enrollment. Even if they are enrolled in a course, they have the option to not participate or drop out at any point, because students are not governed by their course enrollment schedule. From purely an academic scheduling perspective, it is possible to model students into a circuit-like network based on their enrollment. However, the full extent of the student network in a university, including their relationships and interactions outside of courses, cannot be fully comprehended from simply a course enrollment dataset.

Section 5.3 demonstrated the capabilities and the drawbacks of using VLSI clustering algorithms for student LC creation, where the BC algorithm demonstrated the most potential in this application. Algorithmically, it is possible to cluster densely connected student enrollment networks into clusters that has relatively little cross-LC connections. The clustering process also takes into account the diversity in scheduling between students, grouping together students that would already spend a lot of time together by the virtue of being enrolled in the same courses. A major drawback of the BC algorithm is that there are still many external connections between the created LCs. Therefore, the LC creation process must be paired with a refinement process to further optimize the final LC configuration. VLSI algorithms were chosen for student LC creation because these types of algorithms should translate well from circuit applications to dense circuit-like student network applications. Clustering techniques beyond VLSI, such as unsupervised machine learning, were considered, but ultimately deemed unnecessary due to the available dataset and size of the problem. For the majority of the research duration, only the ENEL dataset was made available. The size of the ENEL dataset, comprising of only 81 students, is very small, and could result in an overfitted model if machine learning was applied. Furthermore, the general scope of clustering several thousands of students is a small problem that can be accomplished without the computing power of machine learning.

Section 5.4 demonstrated the capabilities and drawbacks of using SA to refine the LC results created using the VLSI clustering algorithms, and the BC algorithm in particular. The experimental results demonstrated that SA is capable of increasing the quality of the initial input LCs. This algorithm pairs well with the VLSI algorithms used to create the LCs because it does not alter the fundamental parameters around which the LCs were constructed around. The VLSI algorithms performed clustering based on students' course schedules and a maximum cluster size. The modified SA algorithm searches through alternative LC configurations by swapping students in different LCs. The swapping mechanic preserves the maximum cluster size of the LCs. In addition, the modification uses the same measurement of LC quality as the VLSI clustering algorithms, by evaluating the quality of each LC using $S_T$. Therefore, the LCs refined by Simulated Annealing would adhere to the original clustering parameters set out before the creation of the LCs. One downside of using SA is that it is a stochastic algorithm, meaning that it would not consistently produce the same result. Therefore, the algorithmic parameters of SA must be carefully tuned, to ensure the final results produced by different runs of the algorithm are comparable. Furthermore, the problem of separating students into LCs, similar to most partitioning problems, is considered NP-hard. This problem is as hard as the hardest NP problem, where the problems could be verified in polynomial time, but it is unsure if it could be solved in polynomial time. When applied to this problem, the SA algorithm runs in polynomial time to go as close to the global minimum as possible. However, the algorithm cannot guarantee whether it found the global minimum, or just a very good local minimum, due to the nature of this problem. Hence, the other drawback of SA is that it is difficult to guarantee the best possible LC configuration will be found in reasonable time, because the algorithm does not go through every possible option for a solution.

Section 5.5 demonstrated the LC Creation and Refinement Framework on two very large student enrollment networks. The framework is indeed capable of creating and refining student LCs based on large enrollment datasets without having to process the dataset into faculty- or program-level student course enrollments. In a scenario where a large number of LCs must be created at once for students enrolled across many different programs, this framework is capable of handling the task. However, the quality of the resulting LCs created for these large networks leaves a lot to be desired, especially when compared to the quality of the LCs created for only one program, the 3rd year ENEL Fall 2020 enrollment network. The major assumption in the application of the framework to these large networks is that the parameters used during both the BC and the SA algorithms stayed constant from the experimental results obtained for the ENEL network. Perhaps, the algorithmic parameters must be re-tuned for the large networks. While the framework could be applied on large enrollment networks, this broad-stroke student LC creation is not very practical from an education perspective. Different faculties and programs have different approaches to their education, and how they structure content delivery. Student LCs should be intentionally designed to suit the content delivery approach, and therefore LCs across different programs should not be created using the same set of design parameters. The limitation of the proposed framework is that it does not allow different design parameters to be applied partially to a multi-program enrollment dataset. The current proposed framework is best suited to create tailored student LCs for students enrolled within the same, or very similar, programs.

During the pandemic, this framework offers academic institutions the ability to create student LCs that allows for both online and blended learning. LCs could be brought on-campus for in-person laboratory-based design exercises with a paired TA. Between the contained nature of the cohorts and the limited cross-cohort connections, it would be easier to perform contact tracing should the need arise. During the online lecture components, students within each LC can engage in active learning and support each other through peer-tutoring sessions. Leveraging these LCs can facilitate a sense of community and promote teambuilding during group design work.

Looking beyond the pandemic, this proposed framework raised major implications. From a technical perspective, VLSI algorithms are shown to have more applications beyond circuit design, and can be adaptable to human-centric situations. Following the logic that human-based networks can be represented as graphs, many VLSI algorithms can be applied to segment, rearrange, or connect human-centric networks, opening up many new potential applications. Furthermore, this framework can be tailored to accommodate new content delivery models. In a blended learning setting, student LCs can be created based on whether the students prefer online or in-person lectures, and when they will be present for hands-on tutorial or laboratory sessions. For an online course, student LCs cohorts can be created based on their time zone, allowing for student support regardless of their physical location. In fact, future improvements of this framework can explore LC clustering parameters beyond student scheduling. Student LCs can be optimized for factors such as common interest, physical location, or even values, goals, and other factors that emphasizes the students' sense of community. In addition, this framework has applications outside of creating student LCs, as community-based learning can benefit beyond undergraduate education. For example, participants of an academic conference can be clustered together based on their mutual interests in certain sessions, allowing for further networking opportunities and deeper level discussions within each group of participants.

## 5.7 Summary

This chapter highlighted the experimental results of the LC Creation and Refinement Framework first introduced in Chapter 4. The 3$^{rd}$ year ENEL Fall 2020 student enrollment network was visualized as both a Fully Dense and Sparse networks. The following Table 23 summarized the major outcomes of the HC, MHC, and BC clustering algorithms.

*Table 23: Summary of Results for Hyperedge Coarsening, Modified Hyperedge Coarsening, and*

*Best Choice Algorithms*

| Algorithm | Experiment | Result | Cluster Quality | Assessment |
|---|---|---|---|---|
| HC | Create LCs from the ENEL network (Fully Dense and Sparse). | Density of the network does not affect result | One major LC created, most students were unclustered. | Algorithm not suited for creating LCs from densely connected networks |
| MHC | Create LCs from the ENEL network (Fully Dense and Sparse). | Density of the network does not affect result. | Five major LCs created, with several small LCs and unclustered students. | Algorithm performed better than HC, but produced unevenly sized clusters, and not suited for creating LCs. |
| BC | Optimal Maximum Cluster Size | Clusters of size nine or ten is the optimal balance between LC size and quality. | Smaller LCs (6 and seven) showed poor quality, but the quality of the | Algorithm performed best of the three, being able to create balanced |

| | | | LCs dramatically improved from size eight and above. | LCs of specific sizes. For this student connectivity network, maximum cluster size of nine or ten using $d_l$ for the Fully Dense Network would produce ideal results. |
|---|---|---|---|---|
| | Optimal Number of Monte Carlo Simulation Runs | Monte Carlo simulation is not needed for $d_l$, $d_n$ required at least 100 simulations and should be used for Sparse Network | The same LCs are produced using $d_l$, and quality varied when $d_n$ is used (Sparse Network performed better than Fully Dense Network) | |

Between the three VLSI clustering algorithms highlighted in Table 23, BC delivered a balanced set of clusters that has limited external connections between LCs. Using $d_l$ and a maximum cluster size of nine or ten yields the most balanced LCs for this Fully Dense student enrollment network. The LCs created using the BC algorithm will be used as the initial input during the SA-based LC refinement process.

The following table summarizes the experimental findings from fine-tuning SA algorithmic parameters.

*Table 24: Summary of Results for Simulated Annealing*

| Algorithm | Experiment | Result | Impact to Cluster Quality | Impact to Temperature |
|---|---|---|---|---|
| SA | Number of Initial Perturbations for Average Initial Cost | N ≥ 600 is recommended to allow the system to reach a steady state when computing for the average initial cost value. | N has no direct impact to cluster quality. | The N value directly impacts $T_0$ of the system. |
| | Initial Acceptance Probability | AP ≥ 0.9 is recommended, as low $T_0$ can lead to suboptimal solutions. | AP has no direct impact on cluster quality. | AP directly and almost exponentially impacts $T_0$, which also affect the rate T cools. |
| | Cooling Rate | α ≥ 0.95 is recommended, since a small α can lead to the system not accepting "bad" solutions. | Cluster quality improved significantly as the α increased. | α controls the rate at which T cools, where smaller α value cools the system faster. |

|  | Minimum Temperature | $T_{min} \leq 0.1$ is recommended, since a large $T_{min}$ will lead to the system terminating early. | $T_{min}$ has a direct impact on cluster quality, where it rapidly improved between $10 > T_{min} > 0.1$. | The value of $T_{min}$ does not affect T of the system. |
|---|---|---|---|---|
|  | Maximum Number of Iterations at Each Temperature | $i_T \geq 50$ is recommended, as a compromise between cluster quality and a computationally expensive SA run. | $i_T$ directly impacts cluster quality, where it grew significantly between $10 \leq i_T \leq 50$. | The value of $i_T$ does not affect T of the system. |

By carefully selecting its parameters, SA is capable of refining LCs produced by the BC algorithm. The refined LCs showed marked improvement from the initial input LCs, thus proving SA to be a viable method of refining existing LCs while preserving its original design parameters.

The entire LC Creation and Refinement Framework is tested on the Fall 2020 and Winter 2021 student enrollment network, where the results showed that the framework is capable of creating student LCs from a large and multi-program enrollment dataset. However, the current design of the framework is best suited for creating student LCs for single-program enrollment networks, where the LC design parameters are tailored to the content delivery model of that particular program.

# Chapter 6 : Conclusion

This thesis presented a methodology for the optimized creation and refinement of student LCs, based on VLSI algorithms. Due to COVID-19, academic institutions have largely been conducting courses online, leading to possible feelings of isolation in students. As academic institutions re-open post-pandemic, there could be a transitional phase between fully online and fully face-to-face teaching, leading to some models of blended learning. During this phase, it may still be necessary to prevent large amounts of student-student contact and transmission. LCs are designed to address these issues, to support both online and blended learning while its static nature of small groups of students can limit additional transmission. However, due to the complexity of large student networks, LCs can be difficult to design.

In this thesis, VLSI clustering and optimizing algorithms have been modified for the purpose of student LC creation. Students are represented as a network, connected by the common courses they are enrolled in. Student enrollment data are modelled into this type of network. The goal of the clustering algorithms is to create LCs of a specific size, while maximizing the number of internal connections within a LC, and minimizing the number of connections between LCs. This optimization criteria were chosen to increase the contacts within each LC to support student learning, while decrease external contacts to limit disease transmission in the pandemic setting. The goal of the optimization algorithm is to further refine the LCs by the same criteria, using a global optimum search. Three clustering algorithms, Hyperedge Coarsening (HC), Modified Hyperedge Coarsening (MHC), and Best Choice (BC), were modified to operate on the student enrollment network. The global optimization algorithm chosen to refine the clustering results was Simulated Annealing (SA).

Clustering results showed that BC was the best clustering algorithm of the three used, capable of consistently producing high quality LCs based on the Fall 2020 3rd year Electrical Engineering (ENEL) dataset. The results also examined the impact of different maximum cluster sizes and the number of Monte Carlo simulations on the resulting clusters. A cluster size of nine or ten yielded the most balanced clusters for the dense ENEL dataset. SA was able to further fine-tune the clustering results, yielding LCs with much higher scoring. Results also show the impact of altering the five major SA parameters: Number of Initial Perturbations (N), Acceptance Probability (AP), Cooling Rate ($\alpha$), minimum temperature ($T_{min}$), and the number of iterations at each temperature ($i_T$), on the resulting clusters. Finally, all components of the LC creation and refinement framework, including the enrollment network modeling, the BC clustering algorithm, and the SA refinement algorithm, are connected and applied to a large Fall 2020 and Winter 2021 student enrollment dataset. Results showed the framework is capable of handling large student enrollment datasets from students in multiple faculties. The implications of the research presented in this thesis showed that a VLSI-based LC creation and refinement framework is capable of creating high quality LCs, with possible applications in realms beyond higher education.

The remainder of this chapter is organized as follows: Section 6.1 highlights the contributions made by the research presented in this thesis. In Section 6.2, the assumptions and limitations made in this thesis are discussed. Finally, Section 6.3 proposes the future work to be done for this research.

## 6.1 Summary of Contribution

The following research contributions were made in this thesis:

- Proposed a method using Adjacency and Connectivity matrices to represent student enrollment data as a network of students, connected by their mutually enrolled courses

- Modification and implementation of HC, MHC, and BC on student enrollment networks to create student LCs

- Experimental result demonstrating the impact of maximum cluster size and number of Monte Carlo simulation on BC clustering results

- Clustering result comparison demonstrating BC as the best clustering algorithm for student LC creation

- Modification and implementation of SA on student LCs to further refine the quality of the LCs

- Experimental result demonstrating the impact of N, AP, $\alpha$, $T_{min}$, and $i_T$ on SA refinement results

- Proposed a framework to create and refine student LCs, and demonstrated the operationality of the framework on a large Fall 2020 and Winter 2021 student enrollment dataset.

## 6.2 Limitations

The purpose of this research is to develop a VLSI-based method for student LC creation and refinement. In this research, several assumptions and compromises were made that acts as limitations to this project. A major assumption in this research was that student enrollment networks are equivalent to circuit netlists, and therefore VLSI algorithms can directly be applied to student enrollment networks. In reality, students are connected in a network with far more complex relationships, and capturing that network with only enrollment data is an oversimplification. VLSI clustering algorithms were chosen for this task because they work well with circuit netlists that the student enrollment network is modelled after. However, there are many clustering algorithms outside of VLSI applications, and the design decision to primarily focus on VLSI algorithms is a limitation in of itself. Due to the complexity of the problem presented, stochastic algorithms were chosen for both the LC creation and refinement stages. While consistently good solutions are achievable through careful fine-tuning of parameters, stochastic algorithms are limited by the fact that they typically do not produce the same LCs given the same inputs. The variability in solutions can be difficult when the framework is used in practice. The use of stochastic algorithms, and SA in particular, is capable of finding very good solutions, but can never guarantee the solution to be the best student LC configuration due to the nature of the problem. Furthermore, the entire LC Creation and Refinement Framework is limited by its ability to only take in one set of clustering and refinement parameters. When this framework is applied on a large multi-program enrollment dataset, the design parameters cannot be tailored and applied partially to different programs.

The algorithms used in this framework were proposed and modified with the context of COVID-19 in mind, in order to maximize the internal LC connections, while minimizing the external LC connections. A major limitation of this research is that in the future, once COVID-19 is entirely beyond concern, the clustering criteria proposed may no longer be the best criteria for LC creation. In that case, further research is needed to re-examine what are the best clustering criteria for student LCs, and what are the best metrics to evaluate the resulting LCs. Based on literature, LC clustering criteria that can support student learning includes external factors, such as access or function of the group, or internal factors, such as relationships, values, goals, or common interests. LCs can also be clustered for time zone to support students in various physical locations during online learning.

## 6.3 Future Work

This thesis provided a framework of using VLSI algorithms to create and refine student LCs. In the future, there are multiple directions in which the work in this thesis can be extended. One possible research is to extend the VLSI clustering and refinement algorithms for group creation beyond student LCs. This can include creating groups of people in different professional contexts or conference settings. Another extension is to explore clustering criteria beyond course connection, and focus on creating LCs based on students' relationship, goals, vision, common interests, or any additional factors that can further facilitate learning. For example, in an engineering design classroom, LCs can be created for students who have a common vision and idea for their final project. As mentioned previously, LCs can also be created based on time zone to support accessing online learning from different physical locations.

The clustering algorithms used in this thesis can also be benchmarked against non-VLSI clustering algorithms, such as k-means or unsupervised clustering using machine learning. In particular, the large Fall 2020 and Winter 2021 dataset could be used in training the machine learning models. If the clustering algorithm can be improved, it would reduce or eliminate the need for the SA refinement stage. The framework can also be expanded to accept multiple clustering criteria, for better application on very large enrollment networks. Finally, student LCs created by the proposed methodology should be studied in an online or blended classroom, to evaluate the efficacy of created LCs in a learning environment, as well as the impact these created LCs have on students. LCs have been created based on the best scheduling match between students, but that does not indicate the group will work cohesively on a personal level. Classroom observations should be performed for LCs created using this framework, and the performance of students in these LCs should be benchmarked against students in randomly created LCs. This study should assess whether LCs can benefit student learning in online and blended classrooms, and whether algorithmically created LCs would benefit student learning more than randomly created LCs.